%% file: bare_jrnl_new_sample4.tex
\documentclass[lettersize,journal]{IEEEtran}
\usepackage{amsmath,amssymb,amsfonts}
\usepackage{algorithmic}
\usepackage{array}
\usepackage[caption=false,font=normalsize,labelfont=sf,textfont=sf]{subfig}
\usepackage{textcomp}
\usepackage{stfloats}
\usepackage{url}
\usepackage{soul}
\usepackage{verbatim}
\usepackage{graphicx}
\usepackage{cite}
\usepackage{wrapfig}
\usepackage{supertabular}
\usepackage{caption}
\usepackage{subcaption}
\usepackage{enumitem}
\usepackage{xurl}
\usepackage{lipsum}
\usepackage{footnote}
\usepackage{xcolor}
\usepackage{chngcntr}
\counterwithin*{subsubsection}{subsection}

\newlist{CC}{enumerate}{1}
\setlist[CC]{label=\textbf{C\arabic*},ref={C\arabic*}}

 \usepackage[linesnumbered,ruled]{algorithm2e}
 \usepackage{mathtools}
\usepackage{comment}

\renewcommand{\footnoterule}{%
    \kern -3pt
    \hrule width \columnwidth
    \kern 2.6pt
}

\makeatletter
\long\def\@makefntext#1{%
    \parindent 1em\noindent
    \hb@xt@ 1.8em{\hss\@makefnmark}#1}
\makeatother

\usepackage{blindtext}
\usepackage{tikz}
\usetikzlibrary{arrows, automata, backgrounds, calendar, chains, matrix, mindmap, patterns, petri, shadows, shapes.geometric, shapes.misc, spy, trees}

\usepackage{bchart}

\usepackage[utf8]{inputenc}
\usepackage{textcomp}

\usepackage{pgfplots}
\pgfplotsset{width=10cm,compat=1.9}

\usepackage[export]{adjustbox}
\usepackage{multirow}
\usepackage{afterpage}

\ifCLASSINFOpdf

\else

\fi

\begin{document}

\title{DRACO: Data Replication and Collection Framework for Enhanced Data Availability and Robustness in IoT Networks}

\author{Waleed Bin Qaim, \IEEEmembership{Member,~IEEE}, Öznur~Özkasap, \IEEEmembership{Senior Member,~IEEE},\\ Rabia Qadar, \IEEEmembership{Graduate Student Member,~IEEE},
Moncef Gabbouj, \IEEEmembership{Fellow,~IEEE}%
\thanks{Manuscript submitted for review: \today. This work has been mainly conducted while the first author was studying at Koç University and as part of the graduate thesis research.}%
\thanks{Waleed Bin Qaim and Moncef Gabbouj are with the Unit of Computing Sciences, Faculty of Information Technology and Communication Sciences, Tampere University, 33720 Tampere, Finland (e-mails: waleed.binqaim@tuni.fi, moncef.gabbouj@tuni.fi).}%
\thanks{Rabia Qadar is with the Unit of Electrical Engineering, Faculty of Information Technology and Communication Sciences, Tampere University, 33720 Tampere, Finland (e-mail: rabia.qadar@tuni.fi).}%
\thanks{Öznur~Özkasap is with the Department of Computer Engineering, Koç University, Istanbul, Turkey (email: oozkasap@ku.edu.tr).}}

% The paper headers
\markboth{Journal of \LaTeX\ Class Files,~Vol.~XX, No.~X, October~2025}%
{W. Bin Qaim \MakeLowercase{\textit{et al.}}: DRACO: Data Replication and Collection Framework for Enhanced Data Availability and Robustness in IoT Networks}

\IEEEpubid{0000--0000/00\$00.00~\copyright~2025 IEEE}
% Remember, if you use this you must call \IEEEpubidadjcol in the second
% column for its text to clear the IEEEpubid mark.

\maketitle

\begin{abstract}

The Internet of Things (IoT) bridges the gap between the physical and digital worlds, enabling seamless interaction with real-world objects via the Internet. However, IoT systems face significant challenges in ensuring efficient data generation, collection, and management, particularly due to the resource-constrained and unreliable nature of connected devices, which can lead to data loss. This paper presents DRACO (Data Replication and Collection), a framework that integrates a distributed hop-by-hop data replication approach with an overhead-free mobile sink-based data collection strategy. DRACO enhances data availability, optimizes replica placement, and ensures efficient data retrieval even under node failures and varying network densities. Extensive ns-3 simulations demonstrate that DRACO outperforms state-of-the-art techniques, improving data availability by up to 15\% and 34\%, and replica creation by up to 18\% and 40\%, compared to greedy and random replication techniques, respectively. DRACO also ensures efficient data dissemination through optimized replica distribution and achieves superior data collection efficiency under varying node densities and failure scenarios as compared to commonly used uncontrolled sink mobility approaches namely random walk and self-avoiding random walk. By addressing key IoT data management challenges, DRACO offers a scalable and resilient solution well-suited for emerging use cases.

\end{abstract}


%%%%%%%%%%%%%%%%%%%%%%%%%%%%%%%%%%%%%%%%%%%%%%%%%%%%%%%%%%%%%%%%
\section{Introduction}
%%%%%%%%%%%%%%%%%%%%%%%%%%%%%%%%%%%%%%%%%%%%%%%%%%%%%%%%%%%%%%%%

\IEEEPARstart{T}{he} Internet of Things (IoT) represents a network of interconnected devices that generate, process, and exchange data to drive automation and enhance efficiency across diverse systems. These devices are typically equipped with sensors for data generation, computing units for processing, and communication modules for data transmission, enabling seamless integration of physical and digital systems. IoT aims to enable smart decision-making while minimizing human intervention, making it a cornerstone technology across several domains \cite{sinche2019survey}.

\textcolor{black}{IoT technology powers a wide range of applications, including healthcare, smart cities, industrial automation, surveillance, intelligent transportation systems, environmental, habitat monitoring, and beyond \cite{jannu2023advanced, hayyolalam2021edge, ahmed2023aitel, hayyolalam2021edgewc, hui2017major, ahmed2016internet, bonola2016opportunistic, farahani2018towards, rahmani2018exploiting}. 
By facilitating real-time connectivity and control, IoT enables faster and more informed decision-making, optimizes resource utilization, enhances operational efficiency, supports predictive maintenance, and delivers tailored solutions, providing significant benefits across industries through automation and data-driven insights \cite{walia2023ai}.}

\textcolor{black}{IoT devices are capable of monitoring various physical phenomena in their vicinity such as temperature, humidity, noise, and pressure \cite{feng2014coverage}. Data generated by these devices is either transmitted directly to the Internet or aggregated at a central gateway. The gateway node serves as a hub for processing network data and is typically connected to the Internet, enabling remote access and control \cite{yaqoob2017internet, gubbi2013internet}. IoT networks may deploy different data collection strategies tailored to specific applications. For instance, gateways may remain stationary to continuously receive data streams or act as mobile units that visit devices intermittently to retrieve data. Fixed gateways require devices to route data across the network, often involving significant communication overhead and energy consumption. In contrast, mobile gateways reduce routing requirements, conserve device energy, and extend network lifespan \cite{yun2010maximizing, wang2015improving}.}

\IEEEpubidadjcol

\textcolor{black}{The choice of gateway type depends on application requirements. Real-time applications, such as surveillance and emergency response, benefit from static gateways for consistent data availability \cite{lee2009wireless}. Conversely, delay-tolerant applications, such as environmental or habitat monitoring, often use mobile gateways for efficient data collection \cite{merrett2010wireless}. By adapting the data collection mechanisms to specific needs, IoT systems offer flexible, scalable, and efficient solutions for diverse scenarios, enhancing their applicability across various domains \cite{al2015internet}.}

\textcolor{black}{IoT devices generate vast volumes of valuable data, necessitating efficient data management techniques \cite{zhu2014survey}. However, the resource-constrained and often unreliable nature of IoT devices poses challenges, including data loss and mismanagement \cite{younis2014topology}. To address these issues, several data management techniques have been proposed \cite{sun2021data, ma2013data, abu2013data, ahmed2012techniques}. One such technique is data replication, which allows creating redundant copies of data across multiple devices in the network \cite{qaim2018state, milani2017systematic, hara2006data}. Data replication technique enhances fault tolerance, improves data availability, optimizes memory utilization, and facilitates efficient data retrieval and query handling \cite{martins2006survey}.}

\textcolor{black}{In this paper, we propose DRACO (Data Replication and Collection), a comprehensive framework designed to enhance data resilience and retrieval efficiency in IoT networks through intelligent replication and collection strategies. The main contributions of this paper are as follows:}

\begin{CC}
	\item
    \textcolor{black}{We design a fully distributed hop-by-hop data replication mechanism within DRACO that enables sensor nodes to autonomously create and manage data replicas, thereby improving data resilience in the face of node failures without requiring centralized coordination.}

	\item
    \textcolor{black}{We conduct a detailed analysis of the impact of replication on data availability under varying node failure rates and examine how node density influences the replication footprint, providing valuable insights for resource-aware deployment planning.}

	\item
    \textcolor{black}{We introduce an intelligent mobile sink-based data collection strategy that leverages replication to minimize the number of node visits required during data collection, significantly reducing communication overhead.}

	\item
    \textcolor{black}{We analyze how the interaction of data replication, node density, and failure rates affects the performance of the proposed data collection mechanism, offering a holistic understanding of system dynamics under realistic conditions.}

	\item
    \textcolor{black}{We implement DRACO in Network Simulator-3 (ns-3) and perform extensive simulations to benchmark its performance against state-of-the-art techniques, demonstrating substantial improvements in data availability, replica creation, and collection efficiency.}
    
    \end{CC}

\textcolor{black}{The remainder of this paper is organized as follows. Section \ref{sec2} reviews related work. Section \ref{sec3} presents the system overview. Section \ref{sec4} details the proposed data replication technique and its performance analysis. Section \ref{sec5} introduces the data collection mechanism and presents experimental results. Finally, Section \ref{sec6} concludes the paper with key findings and highlights future research directions.}

\section{Related Work}
%%%%%%%%%%%%%%%%%%%%%%%%%%%%%%%%%%%%%%%%%%%%%%%%%%%%%%%%%%%%%%%%
\label{sec2}
%A number of data dissemination and data collection approaches have been proposed in the literature using data replication \cite{sun2021data, zhu2014survey, abu2013data, ma2013data}.
%In this section, we present some of the related works targeting efficient data dissemination and collection in IoT networks incorporating data replication.
%and  by highlighting their pros and cons. \par

In the following subsections, we provide an overview of relevant solutions available in the literature, targeting efficient data dissemination and collection in IoT networks incorporating data replication. Moreover, in the final subsection, we discuss how this work aligns with the existing landscape of solutions.

\subsection*{Data Dissemination:}

\textcolor{black}{A hybrid message replication scheme for IoT networks is proposed in \cite{rathi2024energy} that combines piggybacking and individual frame transmission to balance energy efficiency and latency. While effective for applications with high delay tolerance and small message sizes, the reliance on piggybacking introduces delays in latency-critical use cases, limiting its applicability for ultra-low-latency scenarios. Additionally, in large-scale IoT surveillance systems, a low-complexity data replication scheme \cite{GONIZZI201516} ensures data availability by selecting replica nodes based on available memory space. This approach enhances storage capacity, robustness against node failures, and resilience to memory overflow in scenarios with infrequent data retrieval.}

DEEP, a proactive data dissemination and storage scheme for Wireless Sensor Networks (WSNs), employs probabilistic data forwarding and replication to reduce communication overhead \cite{vecchio2010deep}. By enabling a mobile sink to visit a small subset of nodes, DEEP achieves a reasonable data gathering efficiency while minimizing overhead, making it suitable for uncontrolled sink mobility scenarios. Similarly, ProFlex \cite{maia2013distributed}, designed for heterogeneous WSNs with mobile sinks, leverages distributed replication among selected nodes to reduce overhead while maintaining high data collection efficiency. By incorporating data correlation, ProFlex further minimizes replication costs, offering robust performance under varying failure and message loss conditions.

Grid-based replication strategies, such as ADR \cite{chen2016efficient}, address the query hotspot problem by dynamically creating and removing replica nodes near overloaded nodes. This approach effectively balances energy consumption and query latency, outperforming similar techniques in scalability and responsiveness to dynamic query loads. Similarly, for improving service reliability, the service-oriented approach (SOA) \cite{neumann2010redundancy} incorporates a scoring-based replica selection and recovery mechanism. SOA ensures reliable data retrieval and service continuation with better storage efficiency and energy utilization compared to randomized and bounded flooding techniques.

\subsection*{Data Collection:}

\textcolor{black}{Traditionally, data collection in IoT networks relied on a static data collection point/sink node concept, requiring all nodes to forward data along established paths \cite{zhou2018survey}. This approach incurred high routing overhead and created energy hotspots near the sink, where nodes closer to the sink were mostly engaged in relaying data from the rest of the network to the sink node, often becoming single points of failure in the network.}

To address these challenges, mobile sink nodes have been introduced, offering advantages such as reducing routing overhead and enabling data collection in sparse or even partially disconnected networks \cite{ali2022data}. Instead of relaying data to a fixed point, nodes store data locally for collection by the mobile sink. However, in large networks, it is inefficient for the sink to visit every node, necessitating data aggregation to maximize data collection by visiting a small subset of nodes \cite{xu2021minimizing}.

\textcolor{black}{The sink's mobility trajectory significantly affects data collection efficiency and can follow one of two approaches: controlled or uncontrolled mobility. In controlled mobility, the sink follows a predetermined path, visiting specific nodes (rendezvous points) where network data must be aggregated \cite{zhang2017energy}. While effective, this method retains some routing overhead, as nodes must forward data to these rendezvous points \cite{zhang2017energy, xing2008rendezvous}. In contrast, uncontrolled sink mobility completely eliminates routing overhead where nodes do not need to relay all network data to a common point or some distinguished nodes. Instead, an efficient data dissemination approach is thus required to distribute data across the network, enabling the sink to collect most of the network data by randomly visiting only a fraction of nodes~\cite{temene2022survey}.}

\begin{table*}[t]
\centering
\label{tab:comparison}
\renewcommand{\arraystretch}{1.2}
\begin{tabular}
{|p{0.9cm}|p{2.6cm}|p{2cm}|p{2.5cm}|p{8cm}|}
\hline
\textbf{[Ref.]} & \textbf{Replication Strategy} & \textbf{Collection Method} & \textbf{Routing Overhead} & \textbf{Remarks} \\
\hline

\cite{rathi2024energy} & Piggybacked and individual-frame replication & Fixed sink & High (for urgent messages); Low (for delay tolerant ones) & 
Optimized for small message sizes; not ideal for ultra-low-latency use cases\\
\hline

\cite{GONIZZI201516}& Greedy node selection & Mobile sink & Low & Biased replica distribution; benchmarked in simulations \\
\hline

\cite{vecchio2010deep} & Probabilistic flooding & Mobile sink with self-avoiding random walk & Low & Early WSN model; lacks intelligent sink decision-making; benchmarked in simulations \\
\hline

\cite{maia2013distributed} & Distributed, correlation-aware & Mobile sink with random walk & Medium & Handles message loss; relies on partial routing; benchmarked in simulations \\
\hline

\cite{chen2016efficient} & Grid-based dynamic replication & Fixed sink & High & Addresses query hotspot; not mobility-resilient \\
\hline

\cite{neumann2010redundancy} & Scoring-based redundancy & Fixed sink & High & Reliable service continuation; suited for static topologies \\
\hline

% Random Replication & Random neighbor selection & Mobile sink & Medium & Benchmarked in simulations; lacks replica placement control \\
% \hline
\textbf{DRACO} & \textbf{Hop-by-hop, point-to-point} & \textbf{Intelligent mobile sink} & \textbf{None} & \textbf{Fully distributed; routing-free; replica-aware adaptive collection} \\
\hline
\end{tabular}
\caption{\textcolor{black}{Comparison of DRACO with representative data dissemination and collection techniques in IoT networks}}
\label{comptable}
\end{table*}

\subsection*{Contributions and novelties of this work:}

% In comparison to the related works, we propose an efficient data dissemination approach coupled with an intelligent data collection mechanism for homogeneous IoT-based WSNs. The proposed approach does not incorporate any multihop routing mechanism, rather the network data is replicated among network nodes in a hop by hop fashion using point to point connections. The proposed approach is fully distributed where the nodes do not need to know and maintain the whole network knowledge rather each node communicates with its immediate neighbors only which completely eliminates the need for a proper routing structure. Furthermore, an intelligent data collection mechanism is proposed where the sink node has no prior knowledge of the network nodes and the available data items. Thus, it is allowed to freely visit random nodes in the network with some induced intelligence to ensure maximum data collection efficiency. We present a detailed simulation-based analysis of the proposed data replication and data collections mechanisms to validate its applicability as well as analyze the network performance with extensive experiments.

\textcolor{black}{While prior solutions offer partial improvements in data availability or collection efficiency, they often rely on the routing infrastructure or lack adaptability to node failures and dynamic sink behavior. In contrast, DRACO introduces a fully distributed, hop-by-hop replication strategy paired with an intelligent, routing-free data collection mechanism, offering enhanced data availability and resilience with reduced overhead. Table~\ref{comptable} presents a comparative summary of representative data dissemination and collection schemes relevant to IoT networks, alongside our proposed framework, DRACO. The comparison highlights differences in replication strategies, data collection methods, and routing overhead.}

\textcolor{black}{In comparison to the related works, DRACO includes an efficient data dissemination approach coupled with an intelligent data collection mechanism for IoT networks. The DRACO framework does not incorporate any multi-hop routing mechanism; rather, the network data is replicated among network nodes in a hop-by-hop fashion using point-to-point connections. Moreover, our proposed approach is fully distributed where nodes do not need to know and maintain the whole network knowledge, rather each node communicates only with its immediate neighbors, eliminating the need for a proper routing structure.}

\textcolor{black}{The proposed DRACO framework is implemented and extensively evaluated using Network Simulator-3 (ns-3)~\cite{NS3}. ns-3 is a C++ based discrete-event network simulator. It is open-source software, widely used by the research community, which provides realistic models for different networking protocols and standards (e.g., Zigbee, LTE, Wi-Fi, etc.), an active development community, and extensibility. This allows us to capture realistic communication dynamics, which are crucial for accurately assessing the performance of the proposed framework.}

\textcolor{black}{Comparative simulation results show that, compared to another state-of-the-art technique, the proposed data replication algorithm within the DRACO framework, improves data availability and average replicas created in the network with a maximum gain of approximately 15\% and 18\%, respectively. Moreover, our technique also improves the quality of data dissemination by achieving a better replica spread in the network in terms of distance traveled by successive replicas from the source node.}

\textcolor{black}{Furthermore, an intelligent data collection mechanism is proposed where the sink node has no prior knowledge of the network nodes and the available data items. Thus, it is allowed to freely visit random nodes in the network with some induced intelligence to ensure maximum data collection efficiency. Hence, with the proposed intelligent data collection mechanism, we are able to achieve higher data collection efficiency compared to other state-of-the-art data collection techniques. We present a detailed simulation-based analysis of the proposed DRACO framework 
%data replication and data collection mechanisms
to validate its applicability as well as analyze the network performance with extensive experiments.}

%%%%%%%%%%%%%%%%%%%%%%%%%%%%%%%%%%%%%%%%%%%%%%%%%%%%%%%%%%%%%%%%
\section{System Overview}
%%%%%%%%%%%%%%%%%%%%%%%%%%%%%%%%%%%%%%%%%%%%%%%%%%%%%%%%%%%%%%%%
\label{sec3}

\textcolor{black}{In this section, we present our system preliminaries including system model, details of the node failure model, and simulation setup. We also describe the performance metrics used in this paper to evaluate and compare the performance of the proposed solution. Table~\ref{parameters} lists all the notations used throughout this paper.}

\begin{table}[t]
    \centering
    \renewcommand{\arraystretch}{1.2}
        \caption{\textcolor{black}{List of the main notations}}\label{parameters}
    \begin{tabular}{| c | p{6.65cm} | }
      		\hline\hline	\textbf{Notation} & \textbf{Description} \\
      		\hline\hline
        $A$ & Area of the sensing field (in square meters) deploying IoT devices \\ \hline
        $CN$ & 1-hop neighbors of previous replica nodes for a data item \\ \hline
        $CR$ & Communication radius of the sink node \\ \hline
        $C1, C2$ & Diagonal coordinates of the sensing field \\ \hline
        $D_{i}$ & Data item generated by node $i$ in the network \\ \hline
        $D_{c}$ & Set of collected data items by the mobile sink node \\ \hline
        $H$ & Hop count in terms of the number of hops traversed by a data item \\ \hline
        $i$ & Node ID associated with each IoT device in the system  \\ \hline
    $M$ & Maximum number of sites to visit by the mobile sink node \\ \hline    
  	$N$ & Total number of IoT devices deployed \\ \hline
    $N_{Att}$ & Neighbors' attribute table \\ \hline
    $NoN$ & Number of one-hop neighbors for each node in the system \\ \hline
    $NoN_{sink}$ & Number of nodes available around the mobile sink node at a site \\ \hline
    $PR$ & Previous replicas of a data item \\ \hline
    $PV$ & Previously visited nodes by a data item \\ \hline
    $PVS$ & Previously visited sites by the sink node \\ \hline
       $R$ & Replication degree \\ \hline
       $RC$ & Best replica candidate node for a data item \\ \hline
      $RM$ & Remaining memory space available at a node \\ \hline
      $S$ & Number of sites already visited by the mobile sink node \\ \hline
      $X$ & Coordinates of a site to be visited by the mobile sink node \\ \hline
      $\alpha$ & Radius of the circular communication range of each node \\
 \hline\hline
        \end{tabular}
\end{table}

\subsection*{System Model:}

The system model consists of a set of $N$ IoT devices, or nodes, deployed following a uniform random distribution within a square field of area $A$. Each node is uniquely identified by a node ID ${i}$ where \(i=\{1, 2, ... N\}\). These nodes periodically sense their environment, generating data items at fixed intervals, referred to as the sensing interval. The nodes are assumed to be homogeneous in terms of processing power, memory capacity, and communication capabilities. Each node is equipped with a fixed-size buffer to store data items. This buffer holds not only data items generated by the node itself but also replicas of data items from other nodes, subject to memory constraints.

To manage resources, nodes continuously monitor and share information about their available resources. Key attributes include remaining memory $RM$ and the number of neighboring nodes $NoN$. Nodes periodically broadcast this information to their neighbors, maintaining an updated neighbor attribute table $N_{Att}$ that lists neighboring nodes and their attributes. This information is utilized during the data replication phase, where each node selects the most suitable neighbor to host replicas of its data items.

The replication degree $R$ determines the number of replicas to be created for each data item. For example, a replication degree of 2 ensures that two copies of every data item are generated within the network. The node that generates a data item is referred to as the \textit{data owner}, while nodes that host replicas of data items are termed \textit{replica nodes}. To maximize the spread of replicas across the network, each replica node appends the list of its one-hop neighbors, denoted as $CN$, to the data message. The number of hops traversed by a data item in terms of nodes is tracked as the hop count $H$. Nodes holding replicas of a specific data item are collectively referred to as previous replicas $PR$, and the nodes visited by a data item during its journey are called previously visited nodes $PV$. During the replication process, each node uses this information ($R$, $RM$, $PR$, $PV$, and $CN$) in order to identify the optimal candidate among its neighbors to act as a replica node for a particular data item.

% The system also consists of a mobile sink node that visits the sensor field to collect sensed data from the sensor nodes. This mobile sink node is considered to be more powerful as compared to other ordinary sensor nodes in the system in terms of memory and processing power with no significant resource limitations. The mobile sink node is assumed to have no knowledge about the sensor nodes in the system and their locations. It utilizes the sensor field diagonal coordinates denoted as $C1, C2$ to know the boundaries of the sensor field. The mobile sink node generates random coordinates within the boundary of the sensor field to visit nodes for data collection where it discovers the sensor nodes in its communication radius denoted as $CR$ through advertising its presence and then selects the best node to collect data items (details in the data collection section). The sink also keeps track of the previously visited sites denoted as $PVS$ and nodes to avoid repetition. The set of collected data items by the sink node is denoted as $D_{c}$.

The system also includes a mobile sink node tasked with collecting sensed data from the IoT devices. This mobile sink node is considered to be significantly more powerful as compared to other IoT devices in the system in terms of memory and processing power without significant constraints. The sink node is assumed to have no prior knowledge of the nodes in the system and their locations within the field. Instead, it relies on the diagonal coordinates of the field, denoted as $C1, C2$ to determine the boundaries of the area it needs to cover. 

To collect data, the mobile sink node generates random coordinates within the field boundaries and moves to these locations. Upon reaching a location, it discovers nodes within its communication radius $CR$ by broadcasting its presence. It then selects the most suitable node within this range to gather data items (details provided in the data collection section). To avoid redundant visits, the sink maintains a record of previously visited sites $PVS$ and nodes. The set of collected data items by the sink node during its traversal is denoted as $D_{c}$.

\subsection*{Node Failure Model:}
% To analyze the performance of the proposed data replication and collection mechanisms for system robustness under node failure scenarios, we implement a random node failure model to simulate node failures. In this model, sensor nodes can randomly fail any time during network operation. A failed sensor node no longer remains part of the system. Thus, a failed node does not participate in data generation or replication. Moreover, a failed node is assumed to leave the network forever and does not return to the system, thus resulting in loss of data items it holds not replicated on any other node in the system. \par

To evaluate the performance of the proposed data replication and collection mechanisms under node failure scenarios, we implement a random node failure model to simulate failures during network operation. In this model, nodes can fail unpredictably at any time and permanently leave the system. Once a node fails, it no longer participates in data generation or replication. Consequently, any data items stored exclusively on the failed node are lost, highlighting the importance of effective replication in preserving data and ensuring system robustness.

\subsection*{Simulation Setup:}

\textcolor{black}{The simulation setup consists of 100 nodes (unless otherwise stated) randomly deployed within a 100 m × 100 m square area. The nodes are assumed to be homogeneous in terms of characteristics and resources, each equipped with a fixed buffer size and configured to generate data items periodically. The sensing interval for each node is randomly assigned from a range of [1–4] seconds. The transmit power of all the nodes is set to -20dBm. The physical layer is configured with default settings, i.e., a constant propagation delay and a log-distance propagation loss model. At the start of the simulation, each node is aware only of its own identity and resources. Nodes discover their immediate neighbors through periodic broadcast messages, and they maintain a neighbor attribute table ($N_{Att}$) containing the node ID, MAC address, number of neighbors, and remaining memory of each neighbor. This information is shared via resource advertisement messages broadcasted every 10 seconds.
The simulation runs for 400 seconds, corresponding to the maximum sensing interval. Nodes communicate using either broadcast messages for resource advertisements or unicast messages via MAC addresses for data dissemination, eliminating routing overhead. The simulation is divided into two phases: the data dissemination phase and the data collection phase.
During the data dissemination phase, nodes generate and share data with their neighbors at regular intervals. In the subsequent data collection phase, a mobile sink node traverses the sensor field to gather data. The sink node is assumed to make as many visits as necessary to collect the desired amount of network data, which can be adjusted based on application requirements.
Table \ref{tab:systemparams} summarizes the main system parameters used in the simulations.}

%%%%%%%%%%%%%%%%%%%%%%% Finalized Main system parameter%%%%%%%%%%%%%%
% \begin{table}[b]
% \setlength{\tabcolsep}{6pt}
% \resizebox{\columnwidth}{!}
% {
%         \scalebox{0.2}{
%         \begin{tabular}{l l}
%         \hline
%         %\textbf{Parameter} & \textbf{Value} \\ 
%         %\hline
%         Number of nodes & 100 \\
%         %Buffer size per node & 100 data units \\
%         Data Generation Interval & [1-4] s \\
%         Broadcast Interval & 10 s \\
%         Simulation Time & 400 s \\
%         Sensing Field & 100 x 100 meter square \\
%         Propagation Loss Model & Log distance \\
%         Node Failure Model & Random \\
%         Network Topology & Random \\
%         MAC Protocol & 802.15.4 \\
%         \hline\\
%         \end{tabular}
%         }
% }
% \caption{Main System Parameters}
% \label{tab:systemparams}
% \end{table}

\begin{table}[t]
    \centering
    \renewcommand{\arraystretch}{1.2}
        \caption{Main system parameters}\label{tab:systemparams}
    \begin{tabular}{| c | p{4.3cm} |}
\hline\hline	\textbf{Parameter} & \textbf{Value} \\
      		\hline\hline

Number of nodes & 100 \\ \hline
        %Buffer size per node & 100 data units \\
        Data Generation Interval & [1-4] s \\ \hline
        Broadcast Interval & 10 s \\ \hline
        Simulation Time & 400 s \\ \hline
        Sensing Field Area & 100 x 100 meter square \\ \hline
        Propagation Loss Model & Log distance \\ \hline
        Node Failure Model & Random \\ \hline
        Network Topology & Random \\ \hline
        MAC Protocol & 802.15.4 \\

        \hline\hline
        \end{tabular}
\end{table}

\subsection*{Performance Metrics:}

The portfolio of metrics used to analyze the comparative performance of the proposed algorithms includes:
\begin{itemize}

    \item {\textbf{Data Availability:} This metric represents the average percentage of unique data items available in the system (excluding copies) despite node failures. A data item is considered available if at least a single copy exists in the network; otherwise, its availability is recorded as 0.}
    
    \item{\textbf{Average Replicas Created:} This parameter measures the average number of copies created for each data item across the network. Although the replication degree is pre-set within the range [2-5], achieving the set replication degree for every data item may not always be possible due to the unavailability of suitable nodes for replication. Thus, calculating the average replicas created provides valuable insights into the effectiveness of the replication process.}
  
    \item {\textbf{Replica Spread:} Replica spread quantifies the distribution of replicas by calculating the average Euclidean distance traveled by replicas from their data owner nodes. It represents how well replicas spread across the network.}
   
    \item {\textbf{Data Collection Efficiency:} Data collection efficiency is defined as the percentage of total network data gathered by the mobile sink node after visiting a specific number of nodes. A higher efficiency indicates that the sink collects a larger proportion of network data by visiting fewer nodes. This metric depends on the nodes' data dissemination mechanism and the sink node's data collection strategy. Efficiency improves when each visited node provides a substantial amount of new, uncollected data to the sink.}
     
\end{itemize}

%%%%%%%%%%%%%%%%%%%%%%%%%%%%%%%%%%%%%%%%%%%%%%%%%%%%%%%%%%%%%%%%
\section{Data Replication}
%%%%%%%%%%%%%%%%%%%%%%%%%%%%%%%%%%%%%%%%%%%%%%%%%%%%%%%%%%%%%%%%
\label{sec4}

\textcolor{black}{In this section, we present details of the proposed data replication algorithm, a core component of the DRACO framework, designed to efficiently disseminate sensed data among peers in the network \cite{qaim2018draw}. The algorithm aims to enhance data availability and system robustness by distributing and storing redundant copies of generated data at multiple locations within the network.}

\subsection*{Algorithm Overview:}

\textcolor{black}{We propose a fully distributed hop-by-hop data replication algorithm to mitigate data loss in IoT networks caused by local memory shortages at the node level and to enhance data availability in the presence of node failures. A distinguishing feature of the proposed technique is its reliance on limited neighbor information to perform efficient in-network data replication, ensuring data preservation. Algorithm \ref{DR} is executed at each node either when it generates a new data item or when it receives a data item generated by another node in the network.}

\subsection*{Algorithm Inputs and Output:}
The algorithm takes several inputs, including the data item $D_{i}$, replication degree $R$, remaining memory $RM$, node IDs of previous replicas $PR$ for $D_{i}$, node IDs of previously visited nodes $PV$ by $D_{i}$, node IDs of 1-hop neighbors of previous replicas $CN$, hop count $H$, and the neighbors' attribute table $N_{Att}$. As output, the algorithm generates node ID of the best candidate $RC$ (among the neighbors of a node) to create a replica of data item $D_{i}$.

\begin{algorithm}[t]
{
%\KwIn{$D_{i}$, $R$, $RM$, $PR$, $PV$, $CN$, $HC$
\KwIn{Data item $D_{i}$, Replication Degree $R$,\\ Remaining Memory $RM$, Previous Replicas $PR$,\\Previously visited nodes $PV$, Common Neighbors $CN$, Hop Count $H$, and Neighbors' attribute table $N_{Att}$
%Threshold value $thresh$,\\
%Properties of neighboring nodes $T$,\\
%List of previously visited nodes $PN$
}
\BlankLine
\KwOut {Best Replica Candidate Node $RC$}
%\KwOut {Next replica node $RN$}
\BlankLine

Initialize $RC\gets 0$, $H \gets 0$;\\

\For{each Data item $D_{i}$}{
\eIf{RM $>$ 0}{
%save a local copy\;
Create replica\;
Decrement $R$\;
%append own id to the list of previously visited nodes\;
Append node ID to $PV$ and $PR$\;
%append own id to the list of previous replicas\;
%append node id to $PR$\;
Append neighbor node IDs to $CN$\;
}
{
Append node ID to $PV$\;
}
\eIf{Node is data owner}{
\uIf{R $>$ 0}{
Search for a node with max. $NoN$ and $RM$ $>$ 0 in $N_{Att}$\;
\uIf{node found}{
Select as $RC$\;
Increment $H$: $H \gets H + 1$\;
}
%\uElse{terminate;}
}
%{terminate\;}
}
{
%\If{replica node}{
\uIf{R $>$ 0}{
Search for a node with max. $NoN$, $RM$ $>$ 0, !$PV$, !$PR$, !$CN$ in $N_{Att}$\;
\uIf{node found}{
Select as $RC$\;
Increment $H$: $H \gets H + 1$\;
}
\uElse{Search for a node with max. $NoN$, $RM$ $>$ 0, !$PV$, !$PR$ in $N_{Att}$\;
\uIf{node found}{
Select as $RC$\;
Increment $H$: $H \gets H + 1$\;}
}
}
%{terminate\;}
}}
return $RC$;
%\Indm
\caption{\textcolor{black}{Data Replication}}
%\caption{DRAW Algorithm}
\label {DR}
}\end{algorithm}

\subsection*{Algorithm Description:}

In the beginning, the algorithm initializes the output variable, best replica candidate node $RC$ to 0. Additionally, the hop count $H$ is also set to 0 (line 1).

For each data item $D_{i}$, the algorithm begins by checking the node's available remaining memory $RM$. If sufficient memory space is available (line 3), the node performs a number of tasks, i.e., it creates a replica of the data item, decrements the replication degree $R$, appends its node ID to the lists of previously visited nodes $PV$ and previous replica nodes $PR$ (lines 4-6). Additionally, the node appends node IDs of its 1-hop neighbors to the list of common neighbors $CN$ (line 7). However, if the node does not have enough memory space to create a replica, it simply appends its node ID to the list of previously visited nodes $PV$ (lines 8-9) to avoid loops. 

Selection heuristic for the best candidate node to create the next replica slightly differs for the data owner than any other node in the network. If the node is a data owner (line 11), it checks the replication degree. If the replication degree is greater than zero (line 12), i.e., more copies need to be created in the network, it selects a node with the maximum number of neighbors $NoN$ with enough memory space to create a replica from the neighbors attribute table $N_{Att}$ and the hop count is incremented (lines 13-16). \par

When a node receives a data item from another node in the network, it performs the same steps to create a replica as defined above (lines 3-10). For the selection of the best replica candidate node, it first checks if the replication degree is greater than zero (line 18). If the condition holds true, it searches for a node that fulfills certain criteria, i.e., it has the maximum number of neighbors $NoN$, it has enough memory space available to create a replica, it's not a previous replica $PR$ of the received data item, not previously visited $PV$, and not a common neighbor $CN$ (line 19). If such a node is found it is selected as the next replica node and the hop count is incremented (lines 20-22). Otherwise, if no such node could be found among the uncommon neighbors, the algorithm expands its search to the common neighbors. If a suitable node is found in the common neighbors it is selected to be the next replica node and the hop count is incremented (lines 23-27). This process continues until either the replication degree is satisfied ($R=0$) or no suitable candidate node is found ($RC$ remains 0 after the algorithm’s execution). In the latter case, the node discards the data item $D_{i}$, marking it as a failed attempt to create a replica.

\subsection*{Algorithm Complexity:}
\textcolor{black}{Since every node in the system, whether a data owner or a replica node, must search its neighbors to determine the best replica candidate node, we estimate the number of neighbors per node in the system. Assuming an omnidirectional antenna, each sensor node's communication field is a circle around it with fixed radius $\alpha$ (determined by its transmit power). Therefore, the estimated number of neighbors per node in the system ($NoN/Node$) can be formalized as follows:}

\begin{figure*}[t]
\centering
%\scalebox{1}
\includegraphics[width=17.5cm, height=4cm,trim=1.8cm 0.7cm 0.6cm 0.7cm, clip=true]{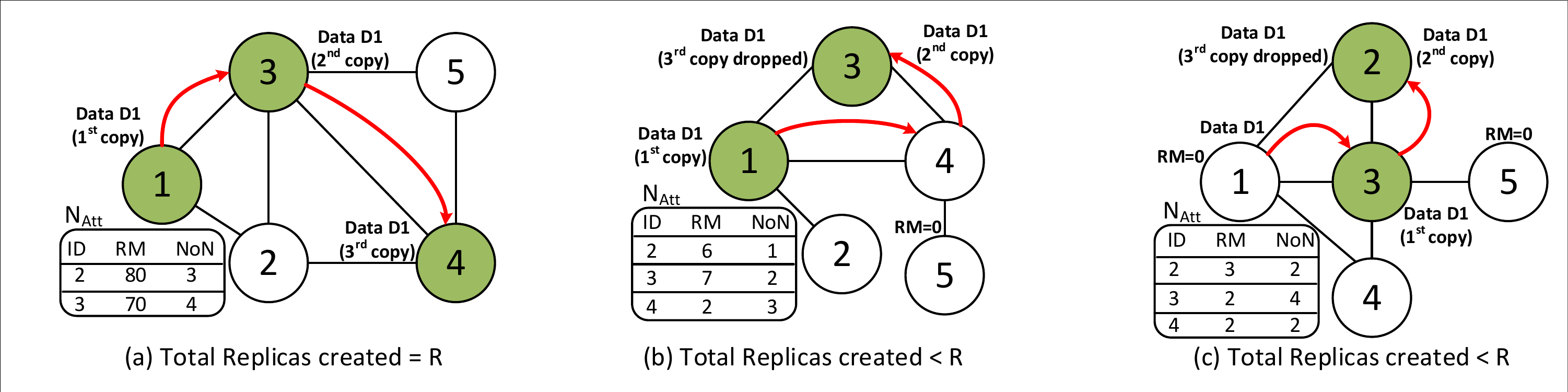}
\caption{Illustration for R=3 under several scenarios: (a) All the desired replicas are created including data owner and avoiding common neighbors (b) Replication degree not met because of unavailability of enough memory at intermediate node (c) Replication degree not met due to unavailability of enough memory at data owner node.}
\label{fig:DRAW}
\end{figure*}

\begin{equation}
\label{eq1}
NoN/Node = \frac{\pi \times \alpha^2 \times N}{A} - 1,
\end{equation}
\textcolor{black}{where $\pi \times \alpha^2$ gives the area of the circular communication field of a node, $N$ is the total number of nodes in the system, and $A$ is the area of the sensing field. For a given area of the field and fixed transmit power per node, the number of neighbors per node in equation \ref{eq1} depends on the system size, i.e., the total number of nodes $N$. Therefore, the worst-complexity of the proposed data replication algorithm is $O(n)$ which continues for at most $R$ times per data item.}

\subsection*{Illustration of Data Replication:}
Figure \ref{fig:DRAW} illustrates three scenarios demonstrating the operation of the proposed data replication algorithm for a replication degree of $R=3$. In this illustration, node 1 initiates the replication process upon generation of data item $D1$, though any node in the network can trigger replication when generating data. In figure \ref{fig:DRAW}(a), all the desired replicas are successfully created at nodes 1, 3, and 4. Each node selects the best replica candidate node among its neighbors based on the replica selection criteria explained above. When selecting the next replica candidate, node 3 has two possible options as nodes 2 and 4. However, it avoids node 2, since node 2 is a common neighbor with node 1 to ensure maximum replica spread. In figure \ref{fig:DRAW}(b), node 1 selects node 4 to create a replica. However, by the time this message reaches node 4, it has already consumed its memory. Hence, instead of creating a local replica, it simply forwards the data to node 3. It can be observed that avoiding common neighbors is not possible at node 4; since the only available option is node 3 which is a common neighbor with node 1. Thus, to create maximum replicas, it considers node 3 to create a replica. In this case, the replica spread is low. In figure \ref{fig:DRAW}(c), node 1 has no memory to create a local replica. Thus, it selects node 3 from its neighbors to create a replica instead of discarding it. Two replicas are created at nodes 3 and 2; however, the third copy is dropped at node 2, since it can not find any other suitable node to create a replica. In both cases depicted in figures \ref{fig:DRAW}(b) and \ref{fig:DRAW}(c), the replication degree is not met.

\subsection*{Comparison Algorithms:}

For performance comparison, we implemented two closely related data replication mechanisms as detailed below:
\subsubsection{Greedy)}
Greedy is a memory-based distributed data replication algorithm available in the literature that utilizes limited neighbors' information to create replicas of data items \cite{GONIZZI201516}. In the Greedy algorithm, each node creates a local copy of the data item it generates if sufficient memory is available. Additionally, it selects the best candidate among its neighbors, choosing the one with the highest remaining memory to create a replica. The replication process continues until the desired replication degree is reached or no suitable candidate nodes are available.
\subsubsection{Random replication)}
The Random Replication algorithm is another approach we implemented for comparison. In this mechanism, each node creates a local copy of the data item it generates if there is enough available memory. However, unlike the Greedy algorithm, the node randomly selects one of its neighbors to replicate the data item. Similar to Greedy, the replication process continues until the replication degree is met or no candidate node is available.

\subsection*{Performance Analysis:}
\setcounter{subsubsection}{0}

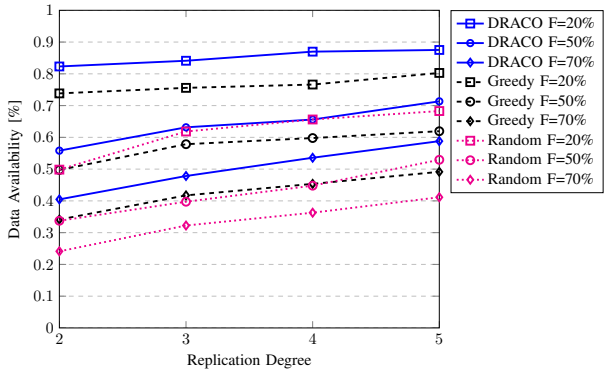
\begin{figure}[t]
\centering
    \scalebox{0.6}{\input{Graphs/DAvsR.tex}}
\caption{Effect of increasing replication degree on data availability in the system against node failure rates of 20\%, 50\% and 70\% with 100 nodes randomly deployed in 100 x 100 meter square region.}
\label{fig:DAvsR}
\end{figure}

\begin{figure}[t]
\centering
    \scalebox{0.6}{\input{Graphs/DAvsF.tex}}
\caption{Effect of increasing node failure rate on data availability in the system under replication degrees of 2, 3 and 5 with 100 nodes randomly deployed in 100 x 100 meter square region.}
\label{fig:DAvsF}
\end{figure}
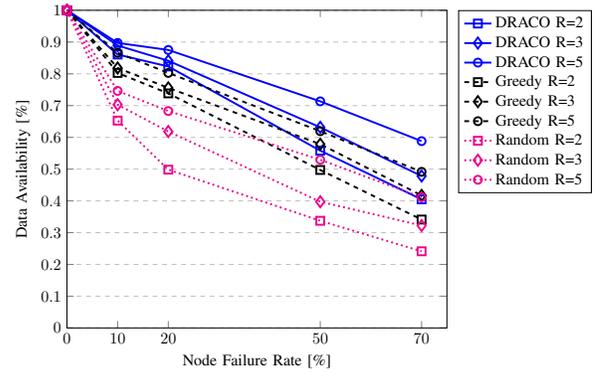

%Actual Replication Degree achieved
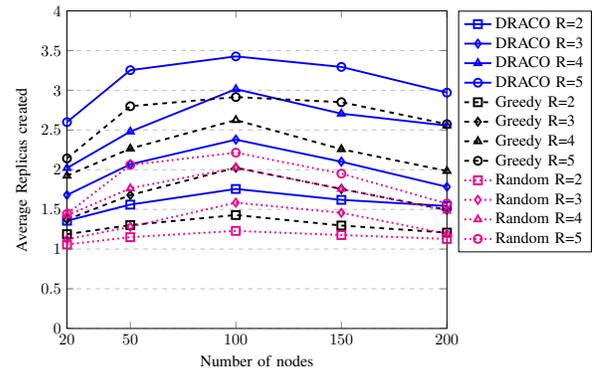
\begin{figure}
\centering
    \scalebox{0.6}{\input{Graphs/ARD.tex}}
    \caption{Actual Replication Degree achieved in the network for varying node densities. Number of nodes are varied in the range of 20-200 randomly deployed in 100 x 100 meter square region.}
\label{fig:ARD}
\end{figure}

\begin{figure}[t]
\centering
    \scalebox{0.6}{\input{Graphs/Replica_Distribution_Distance.tex}}
\caption{Replica distribution across the network in terms of average distance traveled by each successive replica from the data owner node as a function of the replica number.}
\label{fig:Dist}
\end{figure}
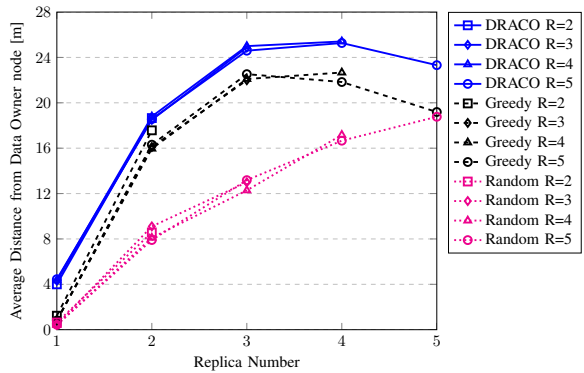

\subsubsection{Effect of increasing Replication Degree on Data Availability}

Figure \ref{fig:DAvsR} illustrates the impact of increasing the replication degree on data availability in the system. As the replication degree increases, data availability improves, as creating more copies of data increases the likelihood that at least one copy will survive node failures. The graph compares the performance for different replication degrees against various node failure rates. Our proposed approach consistently outperforms the other two methods, thanks to its efficient replica selection mechanism. While the Greedy approach performs better than the Random approach due to its memory-based selection heuristic, Random replication underperforms because its random selection cannot ensure optimal replication for each data item. Remarkably, even under the worst-case scenario of a 70\% node failure rate, our data replication algorithm achieves approximately 60\% data availability with a replication degree of 5. This results in an average improvement of around 15\% over Greedy and 34\% over Random replication, based on the average of corresponding data points.

\subsubsection{Effect of increasing Node Failure Rate on Data Availability}

Figure \ref{fig:DAvsF} shows the effect of node failures on data availability in the system. It can be observed that, as the node failures increase, data availability decreases. However, through data replication, a certain threshold of data availability can be maintained. The figure shows that even when half of the nodes fail, our proposed data replication mechanim can achieve a data availability of more than 70\%. Whereas, Greedy and Random approaches can provide data availability of slightly above 60\% and 50\%, respectively.

\subsubsection{Effect of Node Density on Average Replicas Created in the Network}

Figure \ref{fig:ARD} shows the average number of replicas created per data item as a function of the total number of nodes in the system. To determine the optimal node density, experiments were conducted with varying node counts ranging from 20 to 200. It can be observed that, generally, for all three algorithms, the performance initially improves as the number of nodes increases. This is because a higher node density increases the likelihood of finding suitable candidates among neighbors for replica creation, resulting in more replicas. However, increasing the number of nodes beyond a certain point degrades system performance. This is due to a trade-off, i.e. increasing the number of nodes may increase replica candidates, but at the same time, it also increases the total data items generated, which adversely affects system performance. Our proposed algorithm outperforms the other two due to its intelligent replica allocation decisions. The Greedy approach performs better than the Random approach, thanks to its memory-based replica selection, which ensures that selected nodes have sufficient memory to create replicas. In contrast, the Random approach underperforms because its random selection may lead to choosing nodes with inadequate memory, limiting effective replication. Compared to Greedy and Random, the proposed algorithm increases the average number of replicas created by up to 18\% and 40\%, respectively. Thus, while higher node density can initially boost performance, beyond a certain threshold, it can have declining returns.

\subsubsection{Replica Distribution across the Network}

Figure \ref{fig:Dist} illustrates the average Euclidean distance traveled by successive replicas from the data owner node. The distance curves are plotted for different replication degrees as a function of the replica number. It can be observed that the trend for all three algorithms generally overlaps and increases as the number of replicas grows. However, as the network saturates, the trend diverges for higher replica numbers, since it becomes increasingly difficult to find suitable candidate nodes at greater distances. A comparative analysis reveals that our proposed approach outperforms both the Greedy and Random techniques, as it ensures better replica distribution, which directly enhances the quality of data dissemination across the network.

%%%%%%%%%%%%%%%%%%%%%%%%%%%%%%%%%%%%%%%%%%%%%%%%%%%%%%%%%%%%%%%%
\section{Data Collection}
%%%%%%%%%%%%%%%%%%%%%%%%%%%%%%%%%%%%%%%%%%%%%%%%%%%%%%%%%%%%%%%%
\label{sec5}

\textcolor{black}{This section presents details of the proposed data collection algorithm of the DRACO framework, where a mobile sink node efficiently collects network data by randomly visiting nodes in the field. The objective is to enhance data collection efficiency, i.e., to collect a higher percentage of network data while visiting a significantly fewer number of nodes.}

\begin{algorithm}[t]
\KwIn{Previously Visited Sites $PVS$, Diagonal Coordinates of Sensor Field ($C1,C2$), Maximum Number of Sites to Visit $M$, Number of Sites Visited $S$, Communication Radius of Sink Node $CR$}
\BlankLine
\KwOut {Set of Collected Data Items $D_{c}$}
\BlankLine

\While {(S $<$ M)}{

$X \gets \text{generate random site coordinates within ($C1,C2$)\;}$

\If{($X \notin PVS$ and $X \notin CR$ at current site)}{
Move to $X$\;
Advertise presence with memory snapshot of sink node\;
Collect responses from nodes inside $CR$\;
Select node with highest number of new data items NOT already available at sink\;
Send a data transfer request to the selected node for new data items\;
Collect all new data items from the selected node\;
Append newly collected data items to $D_{c}$\;
Append $X$ to $PVS$\;
Increment $S: S \gets S + 1$\;
}
}
return $D_{c}$\;
\Indm
\caption{\textcolor{black}{Data Collection}}
\label {DC}
\end{algorithm}

\subsection*{Algorithm Overview:}
\textcolor{black}{Algorithm \ref{DC} is invoked at the mobile sink node by the end of each data dissemination phase to collect the generated network data. The mobile sink node is considered to be a more powerful node than other network nodes, with no significant resource constraints. However, it lacks knowledge of the sensor nodes' locations, as well as the amount and availability of network data. Therefore, by only knowing the diagonal coordinates of the field, the sink node follows a randomly chosen trajectory to visit nodes and collect the network data.}

\subsection*{Algorithm Inputs and Output:}
As inputs, the algorithm requires the list of previously visited sites $PVS$, diagonal coordinates of the field ($C1, C2$), the maximum number of sites to visit $M$, the number of sites already visited $S$, and the communication radius of the sink node $CR$. As output, the algorithm returns the set of collected data items $D_{c}$ in each round.

\subsection*{Algorithm Description:}

\textcolor{black}{The algorithm executes repeatedly until the predefined total number of sites $M$ have been visited (line 1). In each iteration, the sink node generates random coordinates for a potential site within the sensor field denoted as $X$ by utilizing the diagonal coordinates ($C1$ and $C2$) (line 2). If the randomly generated site has not been previously visited and is outside the communication radius of the sink at its present location(line 3), it proceeds to the site (line 4) and performs several tasks. First, the sink broadcasts its presence along with a memory snapshot of the data items it has already collected (line 5). Neighboring nodes within the sink’s communication radius receive this broadcast and respond with the number of new data items they have that are not already available at the sink (line 6). Upon receiving responses from neighboring nodes, the sink selects the node with the highest number of new data items and sends it a unicast data transfer request (lines 7-8). The selected node, upon receiving the request, transfers the requested data items to the sink (line 9). The newly collected data items are then appended to the sink's set of collected data items $D_{c}$ (line 10). The sink also updates its list of previously visited sites $PVS$ with the current site coordinates and increments the count of visited sites $S$ to prevent revisiting the same location (lines 11-12). This process repeats until the sink has visited the required number of sites $M$ to collect network data. Finally, the algorithm returns the complete set of collected data items $D_{c}$ (line 15).}

\subsection*{Algorithm Complexity:}
\textcolor{black}{At each randomly visited site, the sink node must determine the best node for data collection based on the availability of maximum new data items. This search is limited to the number of nodes inside the communication radius of the sink node. A sink node is also assumed to have a circular communication field just like any other sensor node in the system. Therefore, similar to equation \ref{eq1}, the estimated number of sensor nodes available around the sink per site ($NoN_{sink}/Site$) can be formalized as follows:}

\begin{equation}
\label{eq2}
NoN_{sink}/Site = \frac{\pi \times CR^2 \times N}{A},
\end{equation}
\textcolor{black}{where $\pi \times \alpha^2$ gives the area of the circular communication field of the sink node, $N$ is the total number of nodes in the system, and $A$ is the area of the field. For a fixed transmit power of the sink node and a given field area, the number of neighboring sensor nodes around the sink at each site, as shown in equation \ref{eq2}, is dependent on the system size, i.e., the total number of nodes $N$. Consequently, the worst-case computational complexity of the proposed data collection algorithm is $O(n)$, which persists for the total number of sites visited by the sink node to collect the required amount of network data.}

\begin{figure*}[t]
\centering
%\scalebox{1}
\includegraphics[width=14cm, height=7cm,trim=1.8cm 0.6cm 0.6cm 0.7cm, clip=true]{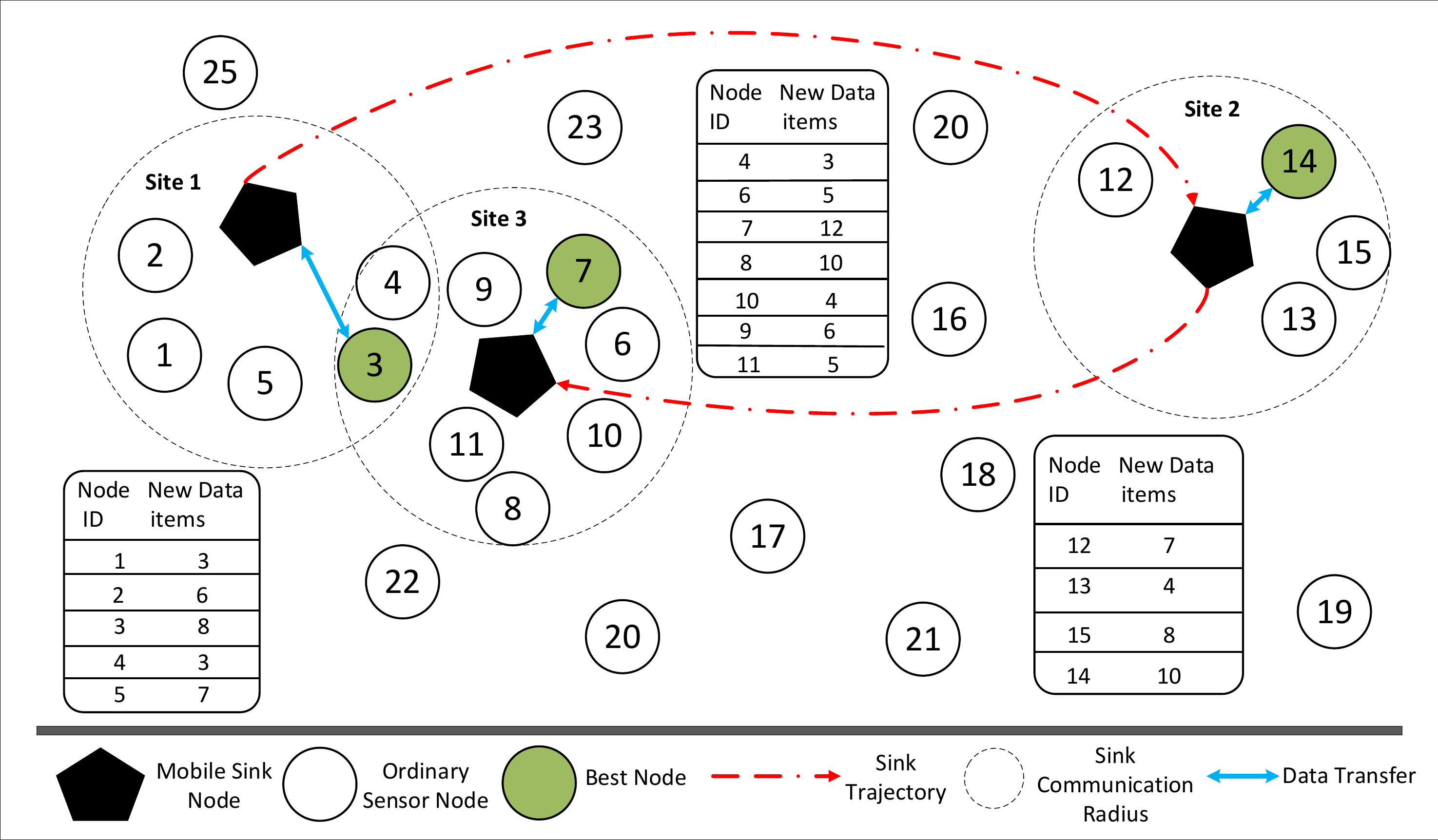}
\caption{Illustration for data collection using a mobile sink node visiting randomly generated sites inside the field and searching the best node based on the availability of new data items for data collection.}
\label{fig:MS}
\end{figure*}

\subsection*{Illustration of Data Collection:}

Figure \ref{fig:MS} illustrates the operation of the data collection mechanism using a mobile sink node. The sink node visits randomly generated sites within the field, as shown by three example sites. Upon reaching each site, it broadcasts its presence while also sharing its memory snapshot. Nodes within the sink’s communication radius receive the broadcast, become aware of the sink's presence, and respond with acknowledgment messages. These acknowledgments include the number of new data items each node holds that are not already collected by the sink. The sink node processes these acknowledgments, creating a table for the surrounding nodes and their respective new data items at each site. This process is repeated at every site. For instance, at site 1, node 3 holds the highest number of new data items unavailable at the sink, so the sink sends a unicast data transfer request to node 3. Similarly, at site 2, node 14 is selected, and at site 3, node 7 is identified as the optimal data source based on the sink’s corresponding data tables. When generating a new site, the mobile sink ensures that it is outside the communication radius of the current site to avoid redundant collection. However, overlap with previously visited sites’ communication radii may occur, as seen between sites 3 and 1. In such cases, overlapping nodes that were not previously visited are considered valid for data collection. For example, node 3, already visited at site 1, is excluded from the data table at site 3, as reflected in the figure.
If the sink is unable to identify any sensor nodes within its communication radius at a site, or if all nodes have already been visited, it moves to the next site without collecting data. This iterative process enables the sink to efficiently collect new data while minimizing redundant visits to previously accessed nodes.

\subsection*{Comparison Algorithms:}

For performance comparison of our proposed data collection algorithm, we implemented two widely used state-of-the-art data collection algorithms with uncontrolled sink mobility, as described below:

\subsubsection{Self Avoiding Random Walk (SA-RW)}
In the implementation of SA-RW, the mobile sink node visits a randomly generated site inside the field and collects data items from the nearest node. It keeps on visiting sites to collect network data until all or the desired number of nodes have been visited. SA-RW keeps track of the previously visited nodes to avoid revisiting the same nodes, thus the name self-avoiding. If at a given site, no node exists in the communication radius of the sink or the available nodes have been already visited, it simply moves to the next randomly generated site \cite{vecchio2010deep}. \par

\subsubsection{Random Walk (RW)}
In the implementation of RW, the mobile sink node visits a randomly generated site inside the field and collects data items from the nearest node. However, unlike SA-RW, it does not keep track of the previously visited nodes. Therefore, a node may be visited more than once, thus resulting in no data collection from that node since the data items from that have been already collected. It keeps on generating sites and visiting nodes to collect network data until all or the desired number of nodes have been visited. Similar to SA-RW, if no node exists in the communication radius of the mobile sink, it simply moves to the next randomly generated site \cite{maia2013distributed}.

\subsection*{Performance Analysis:}
\setcounter{subsubsection}{0}
We performed several experiments to validate and compare the performance of the proposed data collection algorithm coupled with the data replication mechanism (proposed in section \ref{sec4}) as detailed below:

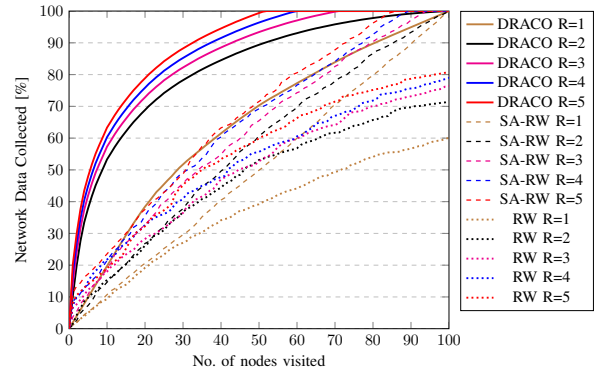
\begin{figure}[t]
\centering
    \scalebox{0.6}{\input{Graphs/DC-100vsR.tex}}
\caption{Effect of increasing replication degree (R=1-5) on data collection efficiency in the system with 100 nodes randomly deployed in 100 x 100 meter square region.}
\label{fig:DCvsR}
\end{figure}

%%%%%%%%%%%%%%%%%%%%%%%%%%%%%%%% trying minipage%%%%%%%%%%%%%%%%%%%%%%%%%%%%%%%
\begin{figure*}[!htb]
    \centering
    \begin{minipage}{0.5\textwidth}
        \centering
        \scalebox{0.6}{\input{Graphs/DC-20.tex}}
        \\(a) Data collection efficiency with 20 nodes
        %\label{fig:prob1_6_2}
        \vspace{1cm}
    \end{minipage}%
    \begin{minipage}{0.5\textwidth}
        \centering
        \scalebox{0.6}{\input{Graphs/DC-50.tex}}
        \\(b) Data collection efficiency with 50 nodes
        %\label{fig:prob1_6_1}
        \vspace{1cm}
    \end{minipage}
    \begin{minipage}{0.5\textwidth}
        \centering
        \scalebox{0.6}{\input{Graphs/DC-150.tex}}
        \\(c) Data collection efficiency with 150 nodes
        %\label{fig:prob1_6_2}
    \end{minipage}%
    \begin{minipage}{0.5\textwidth}
        \centering
        \scalebox{0.6}{\input{Graphs/DC-200.tex}}
        \\(d) Data collection efficiency with 200 nodes
        %\label{fig:prob1_6_1}
    \end{minipage}
    \vspace{0.5cm}
    
    \caption{Effect of increasing node density on data collection efficiency in the system. Number of nodes are varied in the range of 20-200 randomly deployed in 100 x 100 meter square region.}
    \label{fig:DCvsD}
\end{figure*}

\subsubsection{Effect of Data Replication on Data Collection Efficiency}

\textcolor{black}{Figure \ref{fig:DCvsR} shows the effect of increasing data replication on data collection efficiency. We conducted a comparative analysis of our proposed data collection algorithm against the SA-RW and RW algorithms (as described previously). Experiments were performed with replication degrees ranging from $R=1-5$, where $R=1$ represents no replication, meaning each data item is stored only once in the network.
%the node that generated it.
The results show that data collection efficiency improves with increasing replication degree. However, at higher replication degrees, the incremental gain in efficiency is minimal. The maximum efficiency is achieved at $R=5$, where the algorithms aim to create five replicas of each data item across the network, thereby increasing accessibility and enhancing data collection efficiency.
Our proposed data collection algorithm consistently outperforms the other two approaches, primarily due to its intelligent node selection heuristic. SA-RW eventually achieves full data collection after visiting all the nodes but with significantly lower efficiency. Conversely, RW exhibits the lowest performance because the mobile sink node visits random sites without avoiding previously visited nodes, leading to redundant visits and incomplete data collection even after visiting many nodes.
Interestingly, for the $R=1$ case, the data collection efficiency curve deviates slightly from a linear trend. One might expect a straight line since each data item is stored only on its generating node. However, this deviation is attributed to the randomness of sensing intervals among the nodes. Nodes generate data at random intervals within [1–4] seconds range. Therefore, nodes with shorter sensing intervals produce more data, meaning that when the mobile sink visits such nodes, it collects a larger share of the total network data.
For a clearer performance comparison, we examined the number of nodes each algorithm needed to visit to collect 80\% of the network data at $R=5$. Our proposed mechanism achieved this by visiting only 21 nodes, whereas SA-RW and RW required 60 and 96 nodes, respectively, in a network of 100 nodes.
Since all three algorithms achieve their best performance at $R=5$, the remaining experiments focus exclusively on the $R=5$ case for the sake of clarity and brevity.}

%%%%%%%%%%%%%%%%%5 with minipage %%%%%%%%%%%%%%%%%%%%%%%%%%%%%%%%

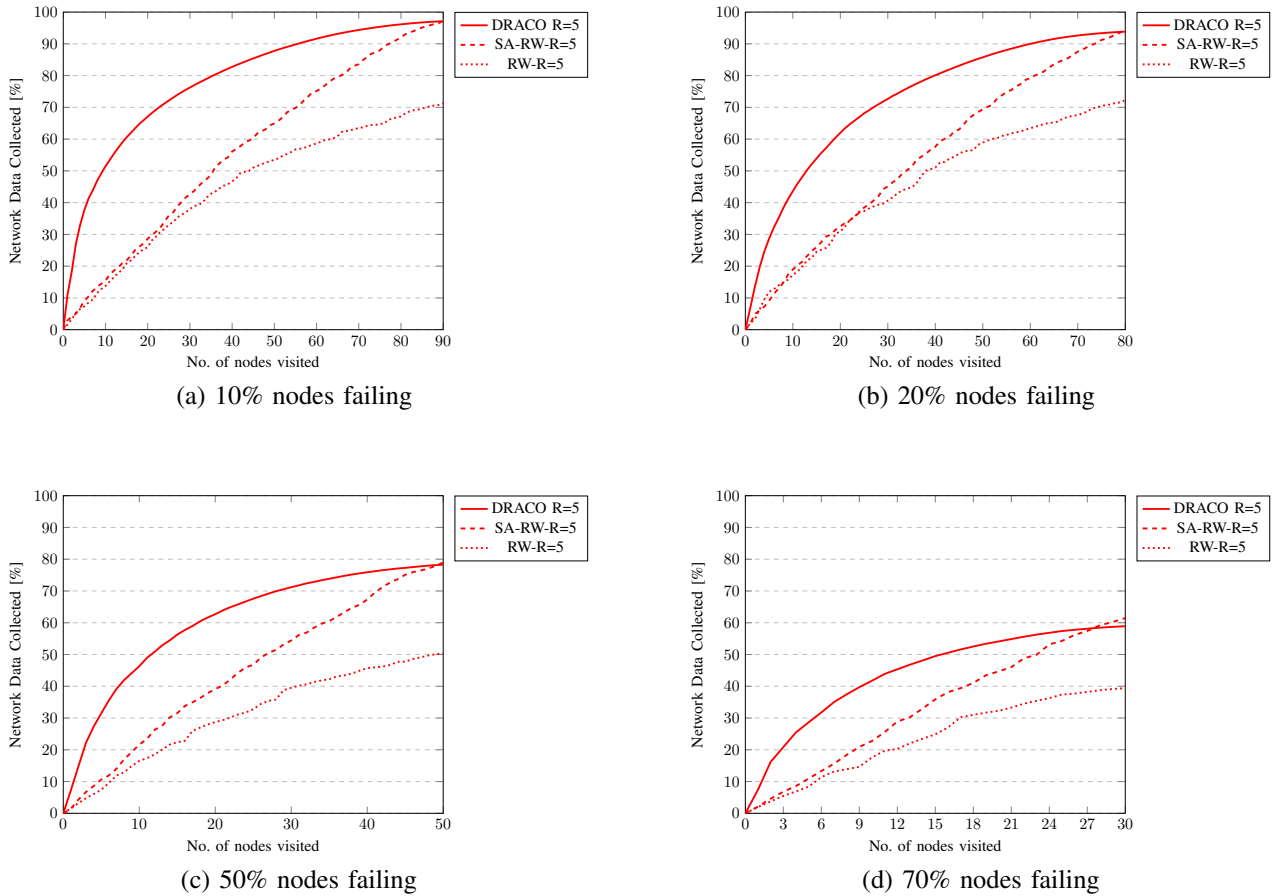
\begin{figure*}[!htb]
    \centering
    \begin{minipage}{0.5\textwidth}
        \centering
        \scalebox{0.6}{\input{Graphs/DC-100N-NF10.tex}}
        \\(a) 10\% nodes failing
        %\label{fig:prob1_6_2}
        \vspace{1cm}
    \end{minipage}%
    \begin{minipage}{0.5\textwidth}
        \centering
        \scalebox{0.6}{\input{Graphs/DC-100N-NF20.tex}}
        \\(b) 20\% nodes failing
        %\label{fig:prob1_6_1}
        \vspace{1cm}
    \end{minipage}
    \begin{minipage}{0.5\textwidth}
        \centering
        \scalebox{0.6}{\input{Graphs/DC-100N-NF50.tex}}
        \\(c) 50\% nodes failing
        %\label{fig:prob1_6_2}
    \end{minipage}%
    \begin{minipage}{0.5\textwidth}
        \centering
        \scalebox{0.6}{\input{Graphs/DC-100N-NF70.tex}}
        \\(d) 70\% nodes failing
        %\label{fig:prob1_6_1}
    \end{minipage}
    \vspace{0.5cm}
    
    \caption{Effect of increasing node failure rate (10\%, 20\%, 50\% and 70\%) on data collection efficiency in the system with 100 sensor nodes randomly deployed in 100 x 100 meter square region.}
    \label{fig:DCNF}
\end{figure*}

\subsubsection{Effect of Node Density on Data Collection Efficiency}

We also analyzed the effect of node density on the data collection efficiency of all three approaches. To compare their performance in both sparse and dense networks, we varied the number of nodes within a fixed area. Figure \ref{fig:DCvsD} shows the trends of data collection efficiency with varying numbers of nodes in the system ranging from 20 to 200 nodes randomly deployed in a 100 x 100-meter square region. The results indicate that our proposed data collection algorithm consistently outperforms SA-RW and RW across all network sizes, thanks to its intelligent data collection mechanism. While SA-RW is capable of collecting the entire network data by visiting all nodes, its data collection efficiency is noticeably lower compared to our proposed algorithm. In contrast, RW fails to ensure 100\% network data collection due to its lack of a mechanism to avoid revisiting previously visited nodes, leading to inefficiencies.
For a given threshold of network data collection, our proposed algorithm achieves the required data collection by visiting significantly fewer nodes than SA-RW and RW. This efficiency advantage is observed across all network sizes, highlighting the robustness of our approach. However, in denser networks, our algorithm requires visiting a higher percentage of nodes compared to sparse networks. This behavior arises because an increase in the number of nodes results in a corresponding increase in the total volume of generated data, necessitating visits to a higher number of nodes.
The best data collection efficiency is achieved with a system size of 100 nodes, as shown in Figure \ref{fig:DCvsR}, compared to other network sizes. This is primarily due to the high replication degree achieved at this system size, as also discussed in section \ref{sec3}.

\subsubsection{Effect of Node Failures on Data Collection Efficiency}

As a prime objective, we leverage data replication to enhance system robustness. To evaluate its impact, we conducted experiments to analyze the effect of node failures on data collection efficiency. Efficiency was measured for varying percentages of node failures, where the nodes followed a random failure model as described in Section \ref{sec2}. The comparative performance of the three data collection algorithms was evaluated for failure rates of 10\%, 20\%, 50\%, and 70\% in a network of 100 nodes with a replication degree of 5, as shown in Figure \ref{fig:DCNF}.
The results reveal that as the percentage of node failures increases, the number of nodes available for the mobile sink to visit decreases. This is because the sink can only visit active nodes to recover data items from the whole network. Under all failure scenarios, the mobile sink is unable to collect 100\% of the network data. This limitation arises because, although data replication ensures that failing nodes’ data items are preserved on neighboring nodes, some data items are inevitably lost. These losses occur when no suitable replica candidate nodes are available during the replication process.
Our proposed data collection algorithm demonstrates significantly higher efficiency in gathering network data as the failure rate increases, outperforming the other two approaches. For example, in the case of 50\% node failures, both our algorithm and SA-RW, manage to collect 80\% of the network data. However, our algorithm achieves this with far greater efficiency. Conversely, RW exhibits the worst performance under node failure conditions. In the 50\% failure scenario, RW struggles to collect even 50\% of the network data.
These findings highlight the resilience of the proposed algorithm in maintaining data collection efficiency even in the presence of high node failure rates, making it a robust solution for dynamic and failure-prone IoT networks.

\section{Conclusion and Future Work}
%%%%%%%%%%%%%%%%%%%%%%%%%%%%%%

\label{sec6}

\textcolor{black}{In the era of IoT-based smart systems, data generated by IoT devices is crucial for driving the functionality and effectiveness of modern applications. However, these devices are prone to failures due to their intrinsic properties and harsh operating conditions, leading to potential data loss and reduced system reliability. To address these challenges, we proposed DRACO, a fully distributed hop-by-hop data replication algorithm complemented by an efficient data collection mechanism using a mobile sink. The proposed framework enables IoT devices to redundantly store generated data within the network, ensuring data preservation even in the presence of node failures. Nodes make local decisions to select the best replica candidates based on available memory and neighbor information, communicated through periodic broadcasts. This eliminates the need for routing structures, reducing overhead and enhancing system simplicity. Additionally, the mobile sink node follows an uncontrolled random trajectory to collect data, making intelligent decisions based on the availability of new data items within its communication radius. Extensive simulations in ns-3 demonstrate the effectiveness of the proposed solution. Our data replication technique enhances data availability and replica spread, achieving gains of up to 15\% in availability and 18\% in replica creation compared to a similar state-of-the-art technique. The improved replica spread facilitates efficient data collection, outperforming other uncontrolled mobility approaches like random walk (RW) and self-avoiding random walk (SA-RW) in terms of data collection efficiency across various scenarios. While data replication enhances system robustness against node failures, the downside, however, is the decreased number of unique data items that can be stored in the network and incurs additional communication overhead.}

\subsection*{Future Work}

\textcolor{black}{Future work will focus on addressing energy consumption, which is a critical factor for battery-powered IoT devices. In particular, we aim to explore strategies that balance the trade-off between extending network lifetime and maintaining system robustness through data replication, while also considering storage–energy efficiency aspects. Moreover, we plan to investigate intelligent adaptive sink mobility mechanisms that dynamically adjust the sink’s movement based on network conditions such as node density, residual energy, and replica distribution. Unlike static or predefined mobility patterns, adaptive mobility decisions would enable the sink to prioritize high-value regions, avoid depleted areas, and opportunistically collect data from isolated nodes, thereby improving both energy efficiency and data collection reliability.}

\textcolor{black}{
Furthermore, while the current study relies on extensive network simulations to evaluate the proposed framework, we acknowledge the importance of real-world validation to further assess its practical feasibility, particularly with respect to energy consumption, mobility constraints, and heterogeneous IoT hardware. As part of our future research, we also intend to extend DRACO to experimental testbeds and field deployments. Such validation would capture practical dynamics beyond simulation models and further strengthen the applicability of the proposed framework, providing valuable insights into its performance under real-world operating conditions and guiding future optimizations.
}

% use section* for acknowledgment
\section*{Acknowledgment}

Waleed Bin Qaim’s work at Haltian Oy and Tampere University was supported by the Foundation for Economic Education, Finland (Grant No. 240031), through the Post Docs in Companies (PoDoCo) program.
Öznur Özkasap's work was supported in part by TUBITAK (The Scientific and Technological Research Council of Türkiye) 2247-A Award 121C338.

\bibliographystyle{IEEEtran}
\bibliography{bare_jrnl_new_sample4}

%\newpage

%\vspace{0.2cm}

%\newpage~%\newpage~
\section*{Biographies}
\vspace{-1cm}

\begin{IEEEbiography}[{\includegraphics[width=1in,height=1.25in,clip,keepaspectratio]{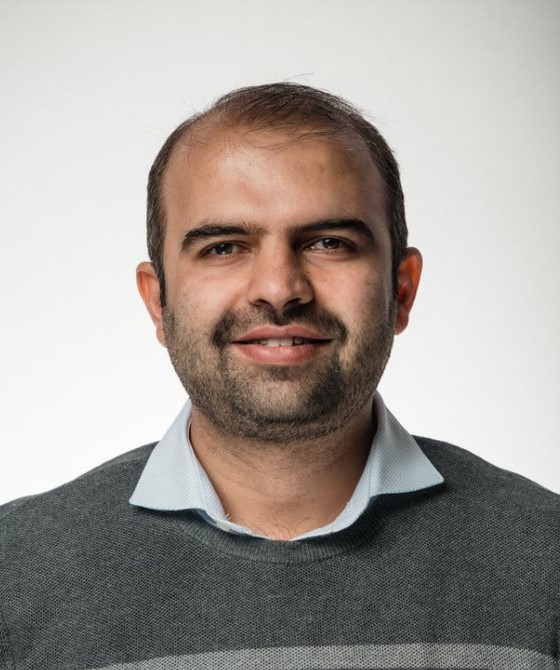}}]{Waleed Bin Qaim} received a double Ph.D. in 2024 from Tampere University (TAU), Finland, and the University Mediterranea of Reggio Calabria~(UNIRC), Italy, while working as an Early Stage Researcher~(ESR) in the European Union Horizon 2020 project A-WEAR, funded by a Marie Skłodowska-Curie (MSCA) grant. He completed his M.Sc. in Computer Science and Engineering at Koç University, Turkey, in 2018, and his B.Sc. in Telecommunication Engineering at the National University of Computer and Emerging Sciences (NUCES), Pakistan, in 2011. Currently, he is a Postdoctoral Researcher with the Signal Analysis and Machine Intelligence (SAMI) research group at TAU, Finland. His research interests include wireless communication, the Internet of Things, edge computing, wearable networks, reinforcement learning, and distributed computing systems.
	\end{IEEEbiography}
%	\vspace{-1cm}
 
\begin{IEEEbiography}
[{\includegraphics[width=1in,height=1.25in,clip,keepaspectratio]{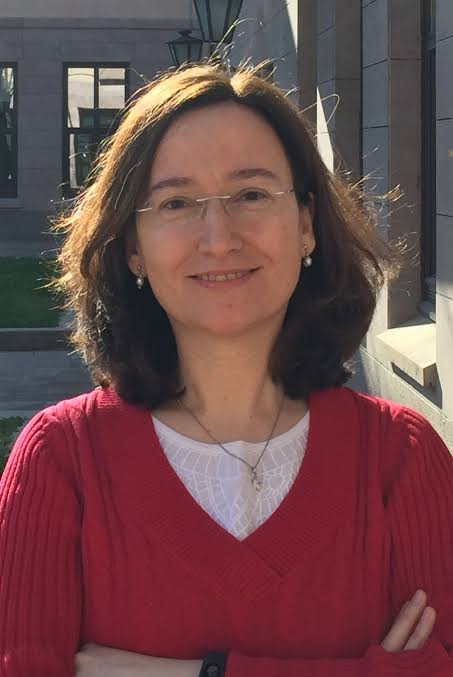}}]
{Öznur Özkasap} is a Professor with the Department of Computer Engineering, Koç University where she is leading the Distributed Systems and Reliable Networks (DISNET) Research Laboratory. She received the Ph.D. degree in Computer Engineering from Ege University. She was a Graduate Research Assistant with the Department of Computer Science, Cornell University where she completed her Ph.D. dissertation. Her research interests include distributed systems, peer-to-peer systems, energy efficiency, mobile and vehicular ad hoc networks, and artificial intelligence in distributed systems. She serves as an Area Editor for the IEEE Transactions on Cloud Computing, IEEE Transactions on Parallel and Distributed Systems, ACM Transactions on Internet Technology, Future Generation Computer Systems journal, Computer Communications journal, Elsevier Science and Cluster Computing journal, Springer. She is a recipient of the National Research Leaders Program Award 2021 of TUBITAK (The Scientific and Technological Research Council of Turkey) and The Informatics Association of Turkey 2019 Prof. Aydın Köksal Computer Engineering Science Award.
	\end{IEEEbiography}
%	\vspace{-1cm}

	\begin{IEEEbiography}
    [{\includegraphics[width=0.9in,height=1.25in]{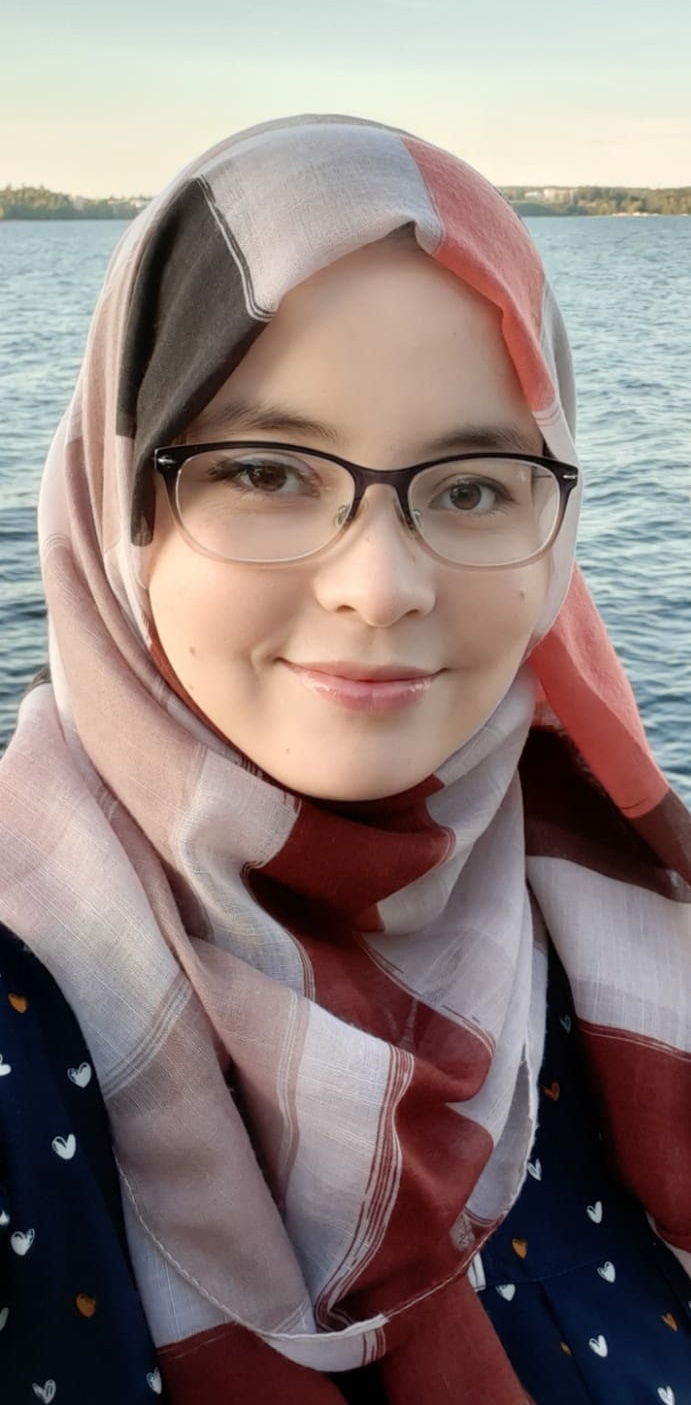}}]
    {Rabia Qadar} is a Doctoral Researcher with the Unit of Electrical Engineering at Tampere University, Finland, where she is pursuing a Ph.D. degree in Communications Engineering. She completed her Masters and Bachelors in Telecommunication Engineering from Balochistan University of IT, Engineering and Management Sciences (BUITEMS), Pakistan, in 2014 and 2016, respectively. She also served as a Lecture and later as an Assistant Professor at the department of Telecommunication Engineering, BUITEMS from 2015 to 2019. Her research interests include wireless communication, underwater networks, the Internet of Things, and reinforcement learning.
	\end{IEEEbiography}
 %\vspace{-1cm}

 	\begin{IEEEbiography}
[{\includegraphics[width=1in,height=1.25in,clip,keepaspectratio]{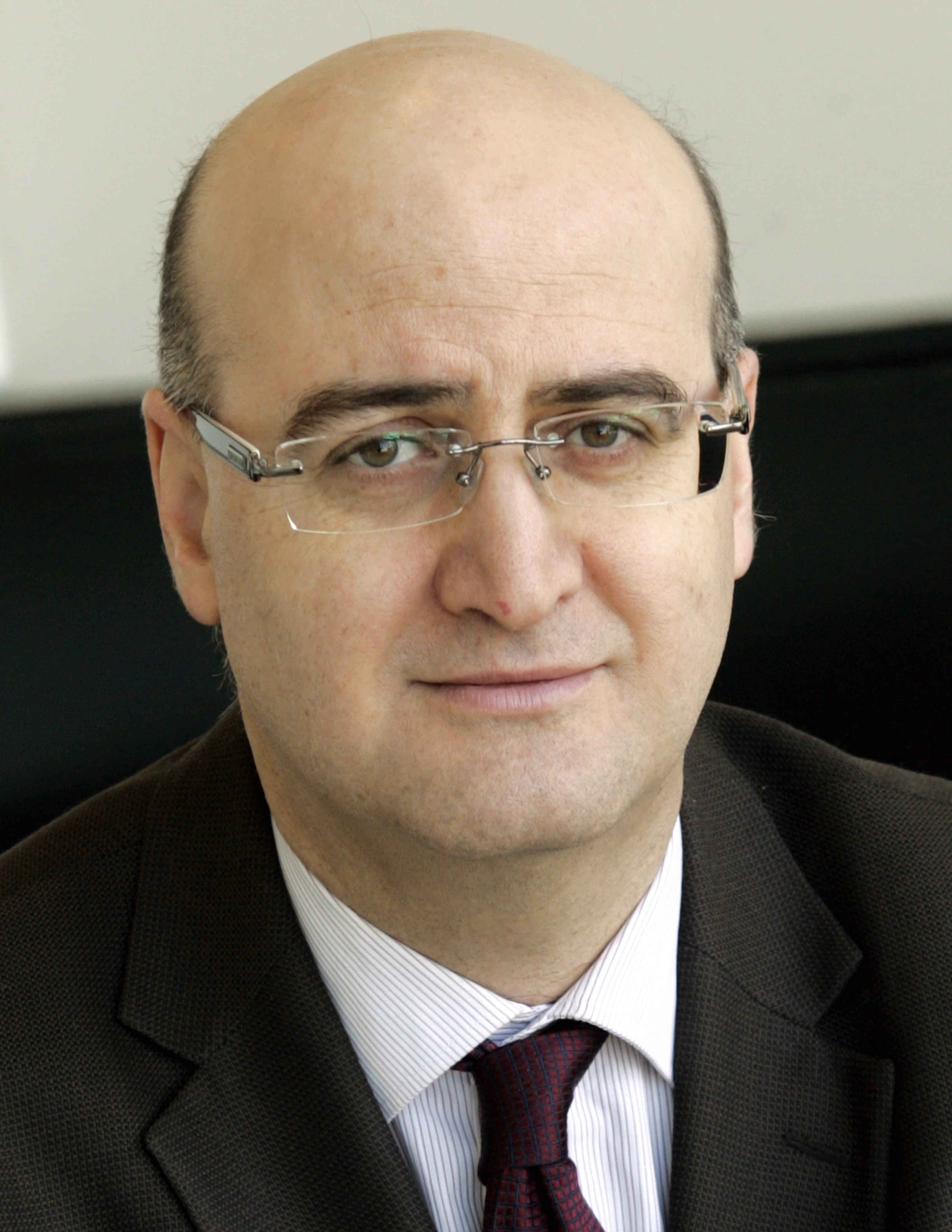}}] {Moncef Gabbouj} received his MS and PhD degrees in Electrical Engineering from Purdue University, in 1986 and 1989, respectively. Dr. Gabbouj is a Professor of Information Technology at the Department of Computing Sciences, Tampere University, Tampere, Finland. He was Academy of Finland Professor during 2011-2015. His research interests include Big Data analytics, multimedia content-based analysis, indexing and retrieval, artificial intelligence, machine learning, pattern recognition, nonlinear signal and image processing and analysis, voice conversion, and video processing and coding. Dr. Gabbouj is a Fellow of the IEEE and Asia-Pacific Artificial Intelligence Association. He is member of the Academia Europaea, the Finnish Academy of Science and Letters and the Finnish Academy of Engineering Sciences. He served as associate editor and guest editor of many IEEE, many international journals, and General Chair of IEEE SPS and CAS Flagship conferences, ICIP and ISCAS as well as ICME.
    
	\end{IEEEbiography}
	%\vspace{0.5cm}

% \bf{If you include a photo:}\vspace{-33pt}
% \begin{IEEEbiography}[{\includegraphics[width=1in,height=1.25in,clip,keepaspectratio]{fig1}}]{Michael Shell}
% Use $\backslash${\tt{begin\{IEEEbiography\}}} and then for the 1st argument use $\backslash${\tt{includegraphics}} to declare and link the author photo.
% Use the author name as the 3rd argument followed by the biography text.
% \end{IEEEbiography}

% \vspace{11pt}

% \bf{If you will not include a photo:}\vspace{-33pt}
% \begin{IEEEbiographynophoto}{John Doe}
% Use $\backslash${\tt{begin\{IEEEbiographynophoto\}}} and the author name as the argument followed by the biography text.
% \end{IEEEbiographynophoto}

\vfill

\end{document}

%% file: Graphs/DAvsR.tex
\begin{tikzpicture}
\begin{axis}[
    ylabel={Data Availability [\%]},
    xlabel={Replication Degree},
    xmin=2, xmax=5,
    ymin=0, ymax=1,
    xtick={2, 3, 4, 5},
    ytick={0, 0.1, 0.2, 0.3, 0.4, 0.5, 0.6, 0.7, 0.8, 0.9, 1},
    legend style={legend pos=outer north east, legend cell align = left}, 
    ymajorgrids=true,
    grid style=dashed,
]

    \addplot[color=blue, mark=square, mark size = 2.5, mark options={solid}, very thick]
    coordinates 
    {
    (2, 0.823182)
    (3, 0.84115)
    (4, 0.869826)
    (5, 0.87515)
    };
    \addlegendentry{DRACO F=20\%}
    
    \addplot[color=blue, mark=o, very thick]
    coordinates 
    {
    (2, 0.558178)
    (3, 0.631596)
    (4, 0.65607)
    (5, 0.713352)
    };
    \addlegendentry{DRACO F=50\%}
    
    \addplot[color=blue, mark=diamond, mark size = 2.5, mark options={solid}, very thick]
    coordinates 
    {
    (2, 0.404715)
    (3, 0.478319)
    (4, 0.535792)
    (5, 0.588091)
    };
    \addlegendentry{DRACO F=70\%}

    \addplot[color=black, dashed, mark= square, mark size = 2.5, mark options={solid}, very thick]
    coordinates 
    {
    (2, 0.7384)
    (3, 0.755823)
    (4, 0.76625)
    (5, 0.802817)
    };
    \addlegendentry{Greedy F=20\%}
    
    \addplot[color=black, dashed, mark=o, mark size = 2.5, mark options={solid}, very thick]
    coordinates 
    {
    (2, 0.497363)
    (3, 0.578464)
    (4, 0.597832)
    (5, 0.619358)
    };
    \addlegendentry{Greedy F=50\%}
   
    \addplot[color=black, dashed, mark=diamond, mark size = 2.5, mark options={solid}, very thick]
    coordinates 
    {
    (2, 0.341241)
    (3, 0.417327)
    (4, 0.453566)
    (5, 0.491463)
    };
    \addlegendentry{Greedy F=70\%}
    
    \addplot[color=magenta, dotted, mark= square, mark size = 2.5, mark options={solid}, very thick]
    coordinates 
    {
    (2, 0.4984)
    (3, 0.6184)
    (4, 0.65625)
    (5, 0.682817)
    };
    \addlegendentry{Random F=20\%}
    
    \addplot[color=magenta, dotted, mark=o, mark size = 2.5, mark options={solid}, very thick]
    coordinates 
    {
    (2, 0.337363)
    (3, 0.397363)
    (4, 0.447821)
    (5, 0.529358)
    };
    \addlegendentry{Random F=50\%}
   
    \addplot[color=magenta, dotted, mark=diamond, mark size = 2.5, mark options={solid}, very thick]
    coordinates 
    {
    (2, 0.241241)
    (3, 0.32241)
    (4, 0.362815)
    (5, 0.411463)
    };
    \addlegendentry{Random F=70\%}
   
\end{axis}
\end{tikzpicture}

%% file: Graphs/DAvsF.tex
\begin{tikzpicture}
\begin{axis}[
    ylabel={Data Availability [\%]},
    xlabel={Node Failure Rate [\%]},
    xmin=0, xmax=75,
    ymin=0, ymax=1,
    xtick={0, 10, 20, 50, 70},
    ytick={0, 0.1, 0.2, 0.3, 0.4, 0.5, 0.6, 0.7, 0.8, 0.9, 1},
    legend style={legend pos=outer north east, legend cell align = left}, 
    ymajorgrids=true,
    grid style=dashed,
]
    \addplot[color=blue, mark=square, mark size = 2.5, mark options={solid}, very thick]
    coordinates 
    {
    (0, 1)
    (10, 0.860746)
    (20, 0.823182)
    (50, 0.558178)
    (70, 0.404715)
    };
    \addlegendentry{DRACO R=2}
    
    \addplot[color=blue, mark=diamond, mark size = 3.5, mark options={solid}, very thick]
    coordinates 
    {
    (0, 1)
    (10, 0.889189)
    (20, 0.84115)
    (50, 0.631596)
    (70, 0.478319)
    };
    \addlegendentry{DRACO R=3}
    
    \addplot[color=blue, mark=o, mark size= 2.5, mark options={solid}, very thick]
    coordinates 
    {
    (0, 1)
    (10, 0.897291)
    (20, 0.87515)
    (50, 0.713352)
    (70, 0.588091)
    };
    \addlegendentry{DRACO R=5}

    \addplot[color=black, dashed, mark=square, mark size = 2.5, mark options={solid}, very thick]
    coordinates 
    {
    (0, 1)
    (10, 0.802817)
    (20, 0.7384)
    (50, 0.497363)
    (70, 0.341241)
    };
    \addlegendentry{Greedy R=2}
    
    \addplot[color=black, dashed, mark=diamond, mark size = 3.5, mark options={solid}, very thick]
    coordinates 
    {
    (0, 1)
    (10, 0.8184)
    (20, 0.755823)
    (50, 0.578464)
    (70, 0.417327)
    };
    \addlegendentry{Greedy R=3}
    
    \addplot[color=black, dashed, mark=o, mark size= 2.5, mark options={solid}, very thick]
    coordinates 
    {
    (0, 1)
    (10, 0.86625)
    (20, 0.802817)
    (50, 0.619358)
    (70, 0.491463)
    };
    \addlegendentry{Greedy R=5}
    
    \addplot[color=magenta, dotted, mark=square, mark size = 2.5, mark options={solid}, very thick]
    coordinates 
    {
    (0, 1)
    (10, 0.6528)
    (20, 0.4984)
    (50, 0.337363)
    (70, 0.241241)
    };
    \addlegendentry{Random R=2}
    
    \addplot[color=magenta, dotted, mark=diamond, mark size = 3.5, mark options={solid}, very thick]
    coordinates 
    {
    (0, 1)
    (10, 0.7028)
    (20, 0.6184)
    (50, 0.397363)
    (70, 0.32241)
    };
    \addlegendentry{Random R=3}
    
    \addplot[color=magenta, dotted, mark=o, mark size= 2.5, mark options={solid}, very thick]
    coordinates 
    {
    (0, 1)
    (10, 0.74625)
    (20, 0.682817)
    (50, 0.529358)
    (70, 0.411463)
    };
    \addlegendentry{Random R=5}
   
\end{axis}
\end{tikzpicture}

%% file: Graphs/ARD.tex
\begin{tikzpicture}
\begin{axis}[
    ylabel={Average Replicas created},
    xlabel={Number of nodes},
    xmin=20, xmax=200,
    ymin=0, ymax=4,
    xtick={20, 50, 100, 150, 200},
    ytick={0, 0.5, 1, 1.5, 2, 2.5, 3, 3.5, 4},
    legend style={legend pos=outer north east, legend cell align = left}, 
    ymajorgrids=true,
    grid style=dashed,
]

    \addplot[color=blue, mark= square, mark size=2.5pt, mark options={solid}, very thick]
    coordinates 
    {
    (20, 1.3569)
    (50, 1.56044)
    (100, 1.75725)
    (150, 1.62025)
    (200, 1.54239)
    };
    \addlegendentry{DRACO R=2}

    \addplot[color=blue, mark= diamond, mark size=2.5pt, mark options={solid}, very thick]
    coordinates 
    {
    (20, 1.68112)
    (50, 2.06758)
    (100, 2.37825)
    (150, 2.10023)
    (200, 1.78383)
    };
    \addlegendentry{DRACO R=3}

    \addplot[color=blue, mark= triangle, mark size=2.5pt, mark options={solid}, very thick]
    coordinates 
    {
    (20, 2.02077)
    (50, 2.47965)
    (100, 3.01423)
    (150, 2.70556)
    (200, 2.55543)
    };
    \addlegendentry{DRACO R=4}

    \addplot[color=blue, mark=o, mark size=2.5pt, mark options={solid}, very thick]
    coordinates 
    {
    (20, 2.59871)
    (50, 3.25306)
    (100, 3.42667)
    (150, 3.29401)
    (200, 2.97018)
    };
    \addlegendentry{DRACO R=5}

    \addplot[color=black, dashed, mark= square, mark size=2.5pt, mark options={solid}, very thick]
    coordinates 
    {
    (20, 1.18762)
    (50, 1.30185)
    (100, 1.42907)
    (150, 1.2969)
    (200, 1.2076)
    };
    \addlegendentry{Greedy R=2}

    \addplot[color=black, dashed, mark= diamond, mark size=2.5pt, mark options={solid}, very thick]
    coordinates 
    {
    (20, 1.37233)
    (50, 1.6802)
    (100, 2.0247)
    (150, 1.75747)
    (200, 1.49388)
    };
    \addlegendentry{Greedy R=3}

    \addplot[color=black, dashed, mark=triangle, mark size=2.5pt, mark options={solid}, very thick]
    coordinates 
    {
    (20, 1.92414)
    (50, 2.26526)
    (100, 2.62469)
    (150, 2.25768)
    (200, 1.98368)
    };
    \addlegendentry{Greedy R=4}

    \addplot[color=black, dashed, mark=o, mark size=2.5pt, mark options={solid}, very thick]
    coordinates 
    {
    (20, 2.14196)
    (50, 2.79908)
    (100, 2.91438)
    (150, 2.85025)
    (200, 2.57319)
    };
    \addlegendentry{Greedy R=5}

    \addplot[color=magenta, dotted, mark= square, mark size=2.5pt, mark options={solid}, very thick]
    coordinates 
    {
    (20, 1.05762)
    (50, 1.150185)
    (100, 1.22907)
    (150, 1.1769)
    (200, 1.1276)
    };
    \addlegendentry{Random R=2}
    
    \addplot[color=magenta, dotted, mark= diamond, mark size=2.5pt, mark options={solid}, very thick]
    coordinates 
    {
    (20, 1.12233)
    (50, 1.2802)
    (100, 1.5847)
    (150, 1.45747)
    (200, 1.19388)
    };
    \addlegendentry{Random R=3}
    
    \addplot[color=magenta, dotted, mark= triangle, mark size=2.5pt, mark options={solid}, very thick]
    coordinates 
    {
    (20, 1.42414)
    (50, 1.76526)
    (100, 2.02469)
    (150, 1.75768)
    (200, 1.48368)
    };
    \addlegendentry{Random R=4}
    
    \addplot[color=magenta, dotted, mark= o, mark size=2.5pt, mark options={solid}, very thick]
    coordinates 
    {
    (20, 1.44196)
    (50, 2.05908)
    (100, 2.21438)
    (150, 1.95025)
    (200, 1.57319)
    };
    \addlegendentry{Random R=5}

\end{axis}
\end{tikzpicture}

%% file: Graphs/Replica_Distribution_Distance.tex
\begin{tikzpicture}
\begin{axis}[
    ylabel={Average Distance from Data Owner node [m]},
    xlabel={Replica Number},
    xmin=1, xmax=5,
    ymin=0, ymax=28,
    xtick={1, 2, 3, 4, 5},
    ytick={0, 4, 8, 12, 16, 20, 24, 28},
    legend style={legend pos=outer north east, legend cell align = left}, 
    ymajorgrids=true,
    grid style=dashed,
]
    \addplot[color=blue, mark= square, mark size= 2.5, mark options={solid}, very thick]
    coordinates 
    {
    (1, 3.998059)
    (2, 18.65482)
    };
    \addlegendentry{DRACO R=2}
    
    \addplot[color=blue, mark= diamond, mark size= 2.5, mark options={solid}, very thick]
    coordinates 
    {
    (1,  4.315146)
    (2,  18.49879)
    (3,  24.8773)
    };
    \addlegendentry{DRACO R=3}
    
    \addplot[color=blue, mark= triangle, mark size= 2.5, mark options={solid}, very thick]
    coordinates 
    {
    (1,  4.360932)
    (2,  18.79261)
    (3,  24.99686)
    (4,  25.4094)
    };
    \addlegendentry{DRACO R=4}

    \addplot[color=blue, mark=o, mark size= 2.5, mark options={solid}, very thick]
    coordinates 
    {
    (1, 4.450952)
    (2, 18.5679)
    (3, 24.59528)
    (4, 25.26996)
    (5, 23.31895)
    };
    \addlegendentry{DRACO R=5}

    \addplot[color=black, dashed, mark= square, mark size= 2.5, mark options={solid}, very thick]
    coordinates 
    {
    (1, 1.23240)
    (2, 17.57447)
    };
    \addlegendentry{Greedy R=2}
    
    \addplot[color=black, dashed, mark= diamond, mark size= 2.5, mark options={solid}, very thick]
    coordinates 
    {
    (1,  0.83336)
    (2,  16.10452)
    (3,  21.98944)
    };
    \addlegendentry{Greedy R=3}
    
    \addplot[color=black, dashed, mark= triangle, mark size= 2.5, mark options={solid}, very thick]
    coordinates 
    {
    (1,  0.814253)
    (2,  15.95996)
    (3,  22.09985)
    (4,  22.66879)
    };
    \addlegendentry{Greedy R=4}

    \addplot[color=black, dashed, mark=o, mark size= 2.5, mark options={solid}, very thick]
    coordinates 
    {
    (1, 0.8110424)
    (2, 16.30803)
    (3, 22.52606)
    (4, 21.83293)
    (5, 19.18992)
    };
    \addlegendentry{Greedy R=5}
    
    \addplot[color=magenta, dotted, mark= square, mark size= 2.5, mark options={solid}, very thick]
    coordinates 
    {
    (1, 0.63240)
    (2, 8.57447)
    };
    \addlegendentry{Random R=2}
    
    \addplot[color=magenta, dotted, mark= diamond, mark size= 2.5, mark options={solid}, very thick]
    coordinates 
    {
    (1,  0.43336)
    (2,  9.10452)
    (3,  12.98944)
    };
    \addlegendentry{Random R=3}
    
    \addplot[color=magenta, dotted, mark= triangle, mark size= 2.5, mark options={solid}, very thick]
    coordinates 
    {
    (1,  0.41336)
    (2,  8.10361)
    (3,  12.28236)
    (4,  17.16932)
    };
    \addlegendentry{Random R=4}

    \addplot[color=magenta, dotted, mark=o, mark size= 2.5, mark options={solid}, very thick]
    coordinates 
    {
    (1,  0.43336)
    (2,  7.90452)
    (3,  13.18433)
    (4,  16.66141)
    (5,  18.769282)
    };
    \addlegendentry{Random R=5}

\end{axis}
\end{tikzpicture}

%% file: Graphs/DC-100vsR.tex
\begin{tikzpicture}

% \begin{axis}[%
% width=4.521in,
% height=3.566in,
% at={(0.758in,0.481in)},
% scale only axis,
% xmin=0,
% xmax=100,
% ymin=0,
% ymax=100,
% axis background/.style={fill=white}
% ]
\begin{axis}[
            legend pos=outer north east,
            ylabel={Network Data Collected [\%]},
            xlabel={No. of nodes visited},
            xmin=0, xmax=100,
            ymin=0, ymax=100,
            xtick={0, 10, 20, 30, 40, 50, 60, 70, 80, 90, 100},
            ytick={0, 10, 20, 30, 40, 50, 60, 70, 80, 90, 100},
            ymajorgrids=true,
            grid style=dashed,
          ]

\addplot [color=brown, very thick]
  table[row sep=crcr]{%
0	0\\
1	1.97412\\
2	3.94824\\
3	5.92235\\
4	7.89647\\
5	9.87059\\
6	11.8447\\
7	13.8188\\
8	15.7929\\
9	17.7671\\
10	19.7412\\
11	21.7153\\
12	23.6894\\
13	25.6635\\
14	27.5283\\
15	29.5024\\
16	31.3672\\
17	33.232\\
18	35\\
19	36.7644\\
20	38.436\\
21	39.9026\\
22	41.4755\\
23	43.0426\\
24	44.2237\\
25	45.5039\\
26	46.783\\
27	47.973\\
28	49.3479\\
29	50.6295\\
30	51.7231\\
31	52.8097\\
32	53.8065\\
33	54.7692\\
34	55.766\\
35	56.7287\\
36	57.7256\\
37	58.6511\\
38	59.6138\\
39	60.5389\\
40	61.5016\\
41	62.4295\\
42	63.3274\\
43	64.2896\\
44	65.181\\
45	66.0728\\
46	66.9324\\
47	67.758\\
48	68.6152\\
49	69.477\\
50	70.3039\\
51	71.0634\\
52	71.791\\
53	72.5198\\
54	73.1844\\
55	73.8489\\
56	74.5761\\
57	75.2407\\
58	75.9378\\
59	76.6024\\
60	77.2974\\
61	77.962\\
62	78.6265\\
63	79.2911\\
64	79.9557\\
65	80.6202\\
66	81.2848\\
67	81.9493\\
68	82.5964\\
69	83.2609\\
70	83.908\\
71	84.5402\\
72	85.1559\\
73	85.7571\\
74	86.3752\\
75	86.9436\\
76	87.5459\\
77	88.0999\\
78	88.6532\\
79	89.222\\
80	89.7457\\
81	90.2682\\
82	90.8073\\
83	91.3155\\
84	91.8535\\
85	92.3772\\
86	92.8854\\
87	93.3935\\
88	93.9017\\
89	94.4099\\
90	94.9181\\
91	95.4263\\
92	95.9345\\
93	96.4427\\
94	96.9509\\
95	97.4591\\
96	97.9672\\
97	98.4754\\
98	98.9836\\
99	99.4918\\
100	100\\
};
\addlegendentry{DRACO R=1}

\addplot [color=black, very thick]
  table[row sep=crcr]{%
0	0\\
1	10.3068\\
2	19.8451\\
3	27.5917\\
4	33.4944\\
5	37.8524\\
6	41.6938\\
7	45.1204\\
8	47.9472\\
9	50.6056\\
10	53.3133\\
11	55.2204\\
12	57.1702\\
13	58.9363\\
14	60.5814\\
15	62.1199\\
16	63.6476\\
17	65.0906\\
18	66.3673\\
19	67.6639\\
20	68.8531\\
21	70.0867\\
22	71.1373\\
23	72.2083\\
24	73.1723\\
25	74.2012\\
26	75.0416\\
27	75.9026\\
28	76.7075\\
29	77.5196\\
30	78.2341\\
31	78.939\\
32	79.6383\\
33	80.3219\\
34	81.0157\\
35	81.635\\
36	82.2494\\
37	82.8414\\
38	83.4184\\
39	83.981\\
40	84.5194\\
41	85.0616\\
42	85.5851\\
43	86.1034\\
44	86.6349\\
45	87.0973\\
46	87.5823\\
47	88.053\\
48	88.5084\\
49	88.9547\\
50	89.3869\\
51	89.7999\\
52	90.2005\\
53	90.5916\\
54	90.9589\\
55	91.3227\\
56	91.6966\\
57	92.0691\\
58	92.4111\\
59	92.7401\\
60	93.0438\\
61	93.3772\\
62	93.6868\\
63	93.9937\\
64	94.2787\\
65	94.5602\\
66	94.7993\\
67	95.0671\\
68	95.3133\\
69	95.5671\\
70	95.823\\
71	96.0385\\
72	96.2658\\
73	96.4685\\
74	96.696\\
75	96.878\\
76	97.0814\\
77	97.2846\\
78	97.4587\\
79	97.6329\\
80	97.8033\\
81	97.9694\\
82	98.125\\
83	98.2935\\
84	98.432\\
85	98.5873\\
86	98.725\\
87	98.853\\
88	98.9772\\
89	99.0985\\
90	99.2062\\
91	99.3092\\
92	99.4052\\
93	99.491\\
94	99.5897\\
95	99.6786\\
96	99.751\\
97	99.8312\\
98	99.8941\\
99	99.9553\\
100	100\\
};
\addlegendentry{DRACO R=2}

\addplot [color=magenta, very thick]
  table[row sep=crcr]{%
0	0\\
1	14.3068\\
2	23.8451\\
3	31.5917\\
4	37.4944\\
5	41.8524\\
6	45.6938\\
7	49.1204\\
8	51.9472\\
9	54.6056\\
10	57.3133\\
11	59.2204\\
12	61.1702\\
13	62.9363\\
14	64.5814\\
15	66.1199\\
16	67.6476\\
17	69.0906\\
18	70.3673\\
19	71.6639\\
20	72.8531\\
21	74.0867\\
22	75.1373\\
23	76.2083\\
24	77.1723\\
25	78.2012\\
26	79.0416\\
27	79.9026\\
28	80.7075\\
29	81.5196\\
30	82.2341\\
31	82.939\\
32	83.6383\\
33	84.3219\\
34	85.0157\\
35	85.635\\
36	86.2494\\
37	86.8414\\
38	87.4184\\
39	87.981\\
40	88.5194\\
41	89.0616\\
42	89.5851\\
43	90.1034\\
44	90.6349\\
45	91.0973\\
46	91.5823\\
47	92.053\\
48	92.5084\\
49	92.9547\\
50	93.3869\\
51	93.7999\\
52	94.2005\\
53	94.5916\\
54	94.9589\\
55	95.3227\\
56	95.6966\\
57	96.0691\\
58	96.4111\\
59	96.7401\\
60	97.0438\\
61	97.3772\\
62	97.6868\\
63	97.9937\\
64	98.2787\\
65	98.5602\\
66	98.7993\\
67	99.0671\\
68	99.3133\\
69	99.5671\\
70	99.823\\
71	100\\
72	100\\
73	100\\
74	100\\
75	100\\
76	100\\
77	100\\
78	100\\
79	100\\
80	100\\
81	100\\
82	100\\
83	100\\
84	100\\
85	100\\
86	100\\
87	100\\
88	100\\
89	100\\
90	100\\
91	100\\
92	100\\
93	100\\
94	100\\
95	100\\
96	100\\
97	100\\
98	100\\
99	100\\
100	100\\
};

\addlegendentry{DRACO R=3}

\addplot [color=blue, very thick]
  table[row sep=crcr]{%
0	0\\
1	17.3068\\
2	26.8451\\
3	34.5917\\
4	40.4944\\
5	44.8524\\
6	48.6938\\
7	52.1204\\
8	54.9472\\
9	57.6056\\
10	60.3133\\
11	62.2204\\
12	64.1702\\
13	65.9363\\
14	67.5814\\
15	69.1199\\
16	70.6476\\
17	72.0906\\
18	73.3673\\
19	74.6639\\
20	75.8531\\
21	77.0867\\
22	78.1373\\
23	79.2083\\
24	80.1723\\
25	81.2012\\
26	82.0416\\
27	82.9026\\
28	83.7075\\
29	84.5196\\
30	85.2341\\
31	85.939\\
32	86.6383\\
33	87.3219\\
34	88.0157\\
35	88.635\\
36	89.2494\\
37	89.8414\\
38	90.4184\\
39	90.981\\
40	91.5194\\
41	92.0616\\
42	92.5851\\
43	93.1034\\
44	93.6349\\
45	94.0973\\
46	94.5823\\
47	95.053\\
48	95.5084\\
49	95.9547\\
50	96.3869\\
51	96.7999\\
52	97.2005\\
53	97.5916\\
54	97.9589\\
55	98.3227\\
56	98.6966\\
57	99.0691\\
58	99.4111\\
59	99.7401\\
60	100\\
61	100\\
62	100\\
63	100\\
64	100\\
65	100\\
66	100\\
67	100\\
68	100\\
69	100\\
70	100\\
71	100\\
72	100\\
73	100\\
74	100\\
75	100\\
76	100\\
77	100\\
78	100\\
79	100\\
80	100\\
81	100\\
82	100\\
83	100\\
84	100\\
85	100\\
86	100\\
87	100\\
88	100\\
89	100\\
90	100\\
91	100\\
92	100\\
93	100\\
94	100\\
95	100\\
96	100\\
97	100\\
98	100\\
99	100\\
100	100\\
};
\addlegendentry{DRACO R=4}

\addplot [color=red, very thick]
  table[row sep=crcr]{%
0	0\\
1	20.3068\\
2	29.8451\\
3	37.5917\\
4	43.4944\\
5	47.8524\\
6	51.6938\\
7	55.1204\\
8	57.9472\\
9	60.6056\\
10	63.3133\\
11	65.2204\\
12	67.1702\\
13	68.9363\\
14	70.5814\\
15	72.1199\\
16	73.6476\\
17	75.0906\\
18	76.3673\\
19	77.6639\\
20	78.8531\\
21	80.0867\\
22	81.1373\\
23	82.2083\\
24	83.1723\\
25	84.2012\\
26	85.0416\\
27	85.9026\\
28	86.7075\\
29	87.5196\\
30	88.2341\\
31	88.939\\
32	89.6383\\
33	90.3219\\
34	91.0157\\
35	91.635\\
36	92.2494\\
37	92.8414\\
38	93.4184\\
39	93.981\\
40	94.5194\\
41	95.0616\\
42	95.5851\\
43	96.1034\\
44	96.6349\\
45	97.0973\\
46	97.5823\\
47	98.053\\
48	98.5084\\
49	98.9547\\
50	99.3869\\
51	99.7999\\
52	100\\
53	100\\
54	100\\
55	100\\
56	100\\
57	100\\
58	100\\
59	100\\
60	100\\
61	100\\
62	100\\
63	100\\
64	100\\
65	100\\
66	100\\
67	100\\
68	100\\
69	100\\
70	100\\
71	100\\
72	100\\
73	100\\
74	100\\
75	100\\
76	100\\
77	100\\
78	100\\
79	100\\
80	100\\
81	100\\
82	100\\
83	100\\
84	100\\
85	100\\
86	100\\
87	100\\
88	100\\
89	100\\
90	100\\
91	100\\
92	100\\
93	100\\
94	100\\
95	100\\
96	100\\
97	100\\
98	100\\
99	100\\
100	100\\
};
\addlegendentry{DRACO R=5}

\addplot [color=brown, dashed, thick]
  table[row sep=crcr]{%
0   0\\
1	1.19924\\
2	1.95055\\
3	3.08658\\
4	4.41153\\
5	5.33822\\
6	6.16681\\
7	7.29342\\
8	8.23204\\
9	9.51097\\
10	10.7202\\
11	11.9255\\
12	12.5895\\
13	13.6762\\
14	14.6316\\
15	15.6787\\
16	16.7701\\
17	17.5762\\
18	18.6571\\
19	19.732\\
20	20.5761\\
21	21.4175\\
22	22.351\\
23	23.2296\\
24	24.1669\\
25	25.2737\\
26	26.0848\\
27	27.0226\\
28	28.0227\\
29	28.701\\
30	29.6175\\
31	30.4104\\
32	31.4016\\
33	32.4999\\
34	33.7341\\
35	35.2079\\
36	36.2771\\
37	37.2771\\
38	38.2617\\
39	39.4944\\
40	40.499\\
41	41.221\\
42	42.5447\\
43	43.4095\\
44	44.2334\\
45	45.1093\\
46	46.1886\\
47	47.0973\\
48	47.9379\\
49	48.8733\\
50	49.9087\\
51	50.7296\\
52	51.8671\\
53	52.818\\
54	53.6125\\
55	54.7533\\
56	55.8952\\
57	56.7713\\
58	57.79\\
59	58.9092\\
60	59.9739\\
61	60.8651\\
62	61.829\\
63	62.6827\\
64	63.8215\\
65	64.6696\\
66	65.7981\\
67	67.089\\
68	68.1578\\
69	69.6129\\
70	70.5721\\
71	71.3349\\
72	72.4417\\
73	73.2145\\
74	74.425\\
75	75.1991\\
76	76.1413\\
77	76.9968\\
78	77.9332\\
79	78.9436\\
80	79.7806\\
81	80.7298\\
82	81.679\\
83	82.8328\\
84	84.1087\\
85	84.9128\\
86	85.5834\\
87	86.6639\\
88	87.45\\
89	88.5237\\
90	89.6523\\
91	90.8131\\
92	91.8418\\
93	92.8044\\
94	93.8218\\
95	94.8538\\
96	95.8151\\
97	96.7906\\
98	98.0538\\
99	99.0741\\
100	100\\
};
\addlegendentry{SA-RW R=1}

\addplot [color=black, dashed, thick]
  table[row sep=crcr]{%
0   0\\
1	1.20421\\
2	2.59042\\
3	4.87986\\
4	6.20618\\
5	7.27397\\
6	8.73841\\
7	10.7875\\
8	12.491\\
9	13.289\\
10	14.6301\\
11	15.804\\
12	17.2558\\
13	18.1951\\
14	19.3274\\
15	20.3375\\
16	21.8656\\
17	23.2148\\
18	24.526\\
19	25.5556\\
20	26.8446\\
21	27.4689\\
22	28.8301\\
23	30.2391\\
24	31.1875\\
25	32.1037\\
26	33.9068\\
27	34.7244\\
28	35.9903\\
29	36.8477\\
30	37.7586\\
31	39.3343\\
32	40.7359\\
33	42.4492\\
34	43.6692\\
35	44.6355\\
36	45.5667\\
37	46.4762\\
38	47.2541\\
39	48.353\\
40	49.6397\\
41	50.4235\\
42	51.7676\\
43	52.9956\\
44	54.2487\\
45	55.7783\\
46	56.6553\\
47	57.6016\\
48	58.5297\\
49	59.3246\\
50	60.7313\\
51	61.6008\\
52	62.6991\\
53	63.6125\\
54	64.4626\\
55	65.2412\\
56	66.0522\\
57	66.9613\\
58	68.0845\\
59	68.7\\
60	69.6532\\
61	70.964\\
62	72.0818\\
63	72.725\\
64	73.2923\\
65	73.8691\\
66	74.5958\\
67	75.7881\\
68	76.3541\\
69	76.9213\\
70	77.9035\\
71	78.7954\\
72	79.4571\\
73	79.9049\\
74	80.8785\\
75	81.78\\
76	82.4326\\
77	83.652\\
78	84.8004\\
79	85.9156\\
80	86.6186\\
81	87.4836\\
82	87.9317\\
83	88.5175\\
84	89.0988\\
85	89.7384\\
86	90.4994\\
87	91.1124\\
88	91.6212\\
89	92.5776\\
90	93.3037\\
91	93.9652\\
92	94.8816\\
93	95.6842\\
94	96.2337\\
95	96.8474\\
96	97.6532\\
97	98.3364\\
98	98.6954\\
99	99.3372\\
100	100\\
};
\addlegendentry{SA-RW R=2}

\addplot [color=magenta, dashed, thick]
  table[row sep=crcr]{%
0   0\\
1	3.735903\\
2	5.36902\\
3	7.76306\\
4	9.34643\\
5	10.76471\\
6	13.0836\\
7	14.4452\\
8	15.3936\\
9	16.8188\\
10	19.2147\\
11	20.5237\\
12	23.3021\\
13	24.2737\\
14	25.9497\\
15	27.0231\\
16	28.0468\\
17	29.342\\
18	30.1212\\
19	31.643\\
20	32.7562\\
21	33.9911\\
22	35.4137\\
23	36.7794\\
24	38.827\\
25	39.784\\
26	40.7131\\
27	42.5045\\
28	43.6031\\
29	44.691\\
30	45.7236\\
31	47.0572\\
32	47.795\\
33	48.8288\\
34	49.9937\\
35	51.0423\\
36	51.4858\\
37	52.2932\\
38	53.023\\
39	53.902\\
40	54.8618\\
41	56.2919\\
42	57.5208\\
43	58.3279\\
44	58.8328\\
45	59.8956\\
46	60.7714\\
47	62.1901\\
48	64.023\\
49	64.7116\\
50	65.4821\\
51	66.4805\\
52	67.4797\\
53	68.6036\\
54	69.3345\\
55	70.5249\\
56	71.5178\\
57	72.1655\\
58	72.7454\\
59	73.4741\\
60	74.292\\
61	74.7542\\
62	75.8048\\
63	76.3292\\
64	77.1764\\
65	77.9173\\
66	78.8528\\
67	79.8531\\
68	80.4608\\
69	81.2312\\
70	82.5763\\
71	83.1782\\
72	84.6083\\
73	85.8367\\
74	86.3578\\
75	87.1651\\
76	87.6776\\
77	88.2979\\
78	88.7201\\
79	89.6805\\
80	90.3766\\
81	91.1808\\
82	91.7978\\
83	92.1853\\
84	92.922\\
85	93.605\\
86	94.0832\\
87	94.9756\\
88	95.8346\\
89	96.1271\\
90	97.1589\\
91	97.8403\\
92	98.5758\\
93	99.2758\\
94	100\\
95	100\\
96	100\\
97	100\\
98	100\\
99	100\\
100	100\\
};
\addlegendentry{SA-RW R=3}

\addplot [color=blue, dashed, thick]
  table[row sep=crcr]{%
0   0\\
1	8.48886\\
2	9.95608\\
3	11.14472\\
4	12.9983\\
5	13.80491\\
6	15.1876\\
7	16.30422\\
8	17.4324\\
9	19.6768\\
10	21.4739\\
11	23.2112\\
12	24.6438\\
13	25.5632\\
14	27.5915\\
15	28.7588\\
16	29.7508\\
17	31.931\\
18	32.6998\\
19	33.5711\\
20	35.5645\\
21	37.429\\
22	38.3512\\
23	39.4921\\
24	41.4077\\
25	42.9967\\
26	44.0105\\
27	45.1293\\
28	47.0877\\
29	48.2482\\
30	49.2347\\
31	49.886\\
32	50.7123\\
33	51.2129\\
34	52.3181\\
35	53.6601\\
36	54.6819\\
37	56.5327\\
38	58.2404\\
39	59.6584\\
40	60.3492\\
41	61.6089\\
42	62.4989\\
43	63.2279\\
44	64.1268\\
45	65.0739\\
46	65.9047\\
47	66.6208\\
48	67.9196\\
49	68.5793\\
50	69.2402\\
51	69.9382\\
52	70.4859\\
53	71.3955\\
54	72.1327\\
55	72.829\\
56	73.613\\
57	74.3263\\
58	74.8915\\
59	75.5745\\
60	76.7189\\
61	77.2893\\
62	78.1628\\
63	78.8694\\
64	79.5684\\
65	80.4043\\
66	81.1923\\
67	82.2001\\
68	82.734\\
69	83.4131\\
70	84.0664\\
71	85.1337\\
72	85.7433\\
73	86.8663\\
74	87.5673\\
75	88.3232\\
76	89.2798\\
77	89.786\\
78	91.143\\
79	91.7549\\
80	92.6698\\
81	93.4991\\
82	94.9707\\
83	95.7828\\
84	96.2875\\
85	96.9939\\
86	97.7475\\
87	98.6669\\
88	99.2054\\
89	99.7559\\
90	100\\
91	100\\
92	100\\
93	100\\
94	100\\
95	100\\
96	100\\
97	100\\
98	100\\
99	100\\
100	100\\
};
\addlegendentry{SA-RW R=4}

\addplot [color=red, dashed, thick]
  table[row sep=crcr]{%
0   0\\
1	11.17627\\
2	13.46082\\
3	14.43133\\
4	16.19481\\
5	17.21269\\
6	18.30003\\
7	19.6139\\
8	20.6972\\
9	22.2393\\
10	23.5399\\
11	24.9291\\
12	25.9524\\
13	27.5057\\
14	28.6157\\
15	30.0792\\
16	31.2304\\
17	33.0778\\
18	34.8465\\
19	36.541\\
20	37.538\\
21	38.8877\\
22	40.1522\\
23	40.8792\\
24	41.7737\\
25	42.9499\\
26	43.8702\\
27	45.1913\\
28	46.3787\\
29	47.6768\\
30	48.6915\\
31	50.6532\\
32	53.1225\\
33	54.4646\\
34	55.8031\\
35	57.399\\
36	58.194\\
37	59.8339\\
38	61.1205\\
39	62.2534\\
40	63.2303\\
41	63.7651\\
42	64.1806\\
43	64.8355\\
44	66.2662\\
45	67.1253\\
46	67.8462\\
47	69.1231\\
48	69.6727\\
49	70.7629\\
50	71.5377\\
51	72.3514\\
52	73.7254\\
53	74.0668\\
54	74.7034\\
55	75.6469\\
56	77.749\\
57	78.2985\\
58	79.1893\\
59	80.0867\\
60	80.5715\\
61	81.1919\\
62	81.968\\
63	82.808\\
64	83.3296\\
65	83.9275\\
66	84.5801\\
67	85.5223\\
68	86.1968\\
69	86.8951\\
70	87.7997\\
71	88.6403\\
72	89.6272\\
73	90.1351\\
74	91.2699\\
75	91.7557\\
76	92.3301\\
77	93.1737\\
78	94.3752\\
79	95.3764\\
80	96.24\\
81	96.9296\\
82	97.533\\
83	98.4734\\
84	99.2967\\
85	99.6179\\
86	100\\
87	100\\
88	100\\
89	100\\
90	100\\
91	100\\
92	100\\
93	100\\
94	100\\
95	100\\
96	100\\
97	100\\
98	100\\
99	100\\
100	100\\
};
\addlegendentry{SA-RW R=5}

\addplot [color=brown, dotted, very thick]
  table[row sep=crcr]{%
0   0\\
1	1.22065\\
2	2.04511\\
3	3.12658\\
4	3.90251\\
5	4.98276\\
6	5.95982\\
7	6.87994\\
8	8.01409\\
9	8.76041\\
10	9.29872\\
11	10.6908\\
12	11.8111\\
13	12.6889\\
14	13.282\\
15	13.8791\\
16	14.9593\\
17	16.0645\\
18	17.2163\\
19	18.0678\\
20	19.195\\
21	20.0076\\
22	21.1671\\
23	21.8341\\
24	22.5758\\
25	23.2747\\
26	24.5078\\
27	25.14\\
28	26.1015\\
29	26.5509\\
30	27.0473\\
31	27.6555\\
32	28.7409\\
33	29.2534\\
34	29.7673\\
35	30.4531\\
36	31.0259\\
37	32.0383\\
38	32.6713\\
39	33.2243\\
40	34.1573\\
41	34.53\\
42	34.809\\
43	35.4668\\
44	36.4261\\
45	37.1418\\
46	37.3334\\
47	37.8068\\
48	38.4017\\
49	38.864\\
50	39.1763\\
51	39.8613\\
52	40.4255\\
53	40.7729\\
54	41.5576\\
55	42.1071\\
56	42.6448\\
57	43.1331\\
58	43.4907\\
59	43.9061\\
60	44.1113\\
61	44.7801\\
62	45.225\\
63	45.8937\\
64	46.517\\
65	46.9469\\
66	47.8539\\
67	48.0838\\
68	48.2431\\
69	48.9251\\
70	49.3154\\
71	49.661\\
72	50.3018\\
73	50.4513\\
74	51.4126\\
75	51.8082\\
76	52.1854\\
77	52.8032\\
78	53.3887\\
79	53.7589\\
80	54.0177\\
81	54.8054\\
82	54.8054\\
83	55.229\\
84	55.4883\\
85	55.4883\\
86	55.6538\\
87	55.7211\\
88	56.2153\\
89	56.4133\\
90	56.7507\\
91	57.0301\\
92	57.46\\
93	57.5798\\
94	58.1367\\
95	58.6273\\
96	58.9565\\
97	59.1579\\
98	59.3216\\
99	59.7219\\
100	60.1304\\
};
\addlegendentry{RW R=1}

\addplot [color=black, dotted, very thick]
  table[row sep=crcr]{%
0   0\\
1	1.56535\\
2	4.2432\\
3	6.47174\\
4	8.61447\\
5	9.76107\\
6	10.9153\\
7	11.737\\
8	13.045\\
9	14.448\\
10	15.307\\
11	16.2064\\
12	16.9716\\
13	18.1978\\
14	19.3524\\
15	20.2905\\
16	21.4712\\
17	22.6635\\
18	23.3932\\
19	24.6153\\
20	26.2724\\
21	27.0335\\
22	28.9911\\
23	29.4959\\
24	30.1858\\
25	31.0745\\
26	32.4794\\
27	33.6821\\
28	34.5162\\
29	35.3935\\
30	36.8203\\
31	37.6909\\
32	38.2922\\
33	39.0507\\
34	39.8655\\
35	40.8131\\
36	41.2944\\
37	42.1832\\
38	42.8823\\
39	43.8156\\
40	44.3142\\
41	45.6345\\
42	46.1358\\
43	46.8085\\
44	48.1536\\
45	48.448\\
46	49.9016\\
47	51.1249\\
48	51.4421\\
49	52.1971\\
50	53.1231\\
51	53.4095\\
52	54.0357\\
53	54.2691\\
54	54.6447\\
55	55.0586\\
56	55.5209\\
57	55.7388\\
58	56.2481\\
59	56.4328\\
60	56.8056\\
61	57.8651\\
62	58.4352\\
63	58.8514\\
64	59.1677\\
65	59.6277\\
66	60.2527\\
67	60.6991\\
68	61.4144\\
69	61.5296\\
70	61.6879\\
71	61.7437\\
72	62.3432\\
73	62.9029\\
74	63.2592\\
75	63.9048\\
76	64.0922\\
77	64.3602\\
78	64.8488\\
79	65.3297\\
80	65.7179\\
81	65.8761\\
82	66.5549\\
83	66.8679\\
84	66.9318\\
85	68.3163\\
86	68.7445\\
87	69.2167\\
88	69.3797\\
89	69.6796\\
90	69.6796\\
91	70.0526\\
92	70.0526\\
93	70.2734\\
94	70.4512\\
95	70.6464\\
96	70.8253\\
97	70.8662\\
98	71.1678\\
99	71.2473\\
100	71.6444\\
};
\addlegendentry{RW R=2}

\addplot [color=magenta, dotted, very thick]
  table[row sep=crcr]{%
0   0\\
1	6.63162\\
2	7.69423\\
3	9.16433\\
4	10.5876\\
5	12.31436\\
6	13.06908\\
7	14.27439\\
8	15.6003\\
9	17.0131\\
10	17.973\\
11	19.3878\\
12	20.1272\\
13	21.4488\\
14	22.5476\\
15	23.3879\\
16	24.2493\\
17	24.819\\
18	26.4024\\
19	27.4547\\
20	28.2432\\
21	29.1842\\
22	29.6223\\
23	30.0992\\
24	30.9795\\
25	31.8572\\
26	32.684\\
27	33.7011\\
28	34.915\\
29	35.3688\\
30	36.4624\\
31	38.3994\\
32	38.9617\\
33	40.5194\\
34	40.988\\
35	41.639\\
36	42.3023\\
37	43.4058\\
38	44.0589\\
39	45.6664\\
40	47.0234\\
41	47.6472\\
42	48.3476\\
43	48.7868\\
44	49.5471\\
45	50.2877\\
46	50.5623\\
47	51.0465\\
48	51.3133\\
49	51.6548\\
50	52.716\\
51	53.429\\
52	54.2686\\
53	54.5805\\
54	55.8194\\
55	57.2724\\
56	58.0465\\
57	58.6864\\
58	59.2051\\
59	59.462\\
60	59.8102\\
61	60.5257\\
62	60.8491\\
63	61.6652\\
64	62.1368\\
65	62.4555\\
66	62.9516\\
67	63.3928\\
68	63.7106\\
69	63.9534\\
70	64.168\\
71	64.6103\\
72	65.8343\\
73	66.7787\\
74	68.0467\\
75	68.1986\\
76	68.3729\\
77	68.5898\\
78	69.1212\\
79	69.8198\\
80	70.0907\\
81	70.2322\\
82	70.6962\\
83	70.9143\\
84	71.1842\\
85	71.7283\\
86	72.1623\\
87	72.3255\\
88	72.617\\
89	72.9855\\
90	73.1661\\
91	73.4253\\
92	73.7966\\
93	73.8804\\
94	74.297\\
95	74.7469\\
96	75.0689\\
97	75.3076\\
98	75.4754\\
99	76.2401\\
100	76.4169\\
};
\addlegendentry{RW R=3}

\addplot [color=blue, dotted, very thick]
  table[row sep=crcr]{%
0   0\\
1	7.34627\\
2	9.96225\\
3	11.09781\\
4	12.4601\\
5	14.81696\\
6	16.8697\\
7	18.2476\\
8	19.3107\\
9	20.877\\
10	21.7868\\
11	22.5808\\
12	23.0773\\
13	24.9083\\
14	26.1047\\
15	26.9046\\
16	28.3465\\
17	28.9706\\
18	30.6211\\
19	31.4035\\
20	32.6766\\
21	33.5305\\
22	34.573\\
23	35.1742\\
24	35.7065\\
25	36.1158\\
26	36.6949\\
27	37.6058\\
28	39.5011\\
29	40.097\\
30	40.8417\\
31	42.4561\\
32	43.1206\\
33	43.448\\
34	44.3248\\
35	45.0031\\
36	45.3218\\
37	45.8549\\
38	46.8951\\
39	47.254\\
40	47.7754\\
41	48.1006\\
42	50.3956\\
43	50.9565\\
44	52.3551\\
45	52.8113\\
46	53.2614\\
47	54.1476\\
48	55.1333\\
49	55.4203\\
50	55.6331\\
51	56.41\\
52	56.6308\\
53	57.4906\\
54	57.8949\\
55	58.3189\\
56	58.6758\\
57	59.0273\\
58	59.2643\\
59	59.6464\\
60	60.5727\\
61	60.7918\\
62	61.6312\\
63	62.6235\\
64	63.0582\\
65	64.6067\\
66	64.8645\\
67	64.9426\\
68	65.3463\\
69	66.5633\\
70	67.0323\\
71	67.3279\\
72	67.916\\
73	68.375\\
74	68.5022\\
75	69.1558\\
76	69.7354\\
77	70.9467\\
78	71.3222\\
79	71.5291\\
80	71.8235\\
81	72.4246\\
82	73.6894\\
83	73.9417\\
84	74.0413\\
85	74.2018\\
86	74.4217\\
87	74.6488\\
88	75.0579\\
89	75.5332\\
90	75.6312\\
91	75.7729\\
92	76.0509\\
93	76.605\\
94	77.0011\\
95	77.1094\\
96	77.6441\\
97	77.663\\
98	78.341\\
99	78.5501\\
100	78.9949\\
};
\addlegendentry{RW R=4}

\addplot [color=red, dotted, very thick]
  table[row sep=crcr]{%
0   0\\
1	8.03801\\
2	10.55341\\
3	11.67699\\
4	12.87871\\
5	14.07923\\
6	14.87341\\
7	15.72461\\
8	16.995\\
9	17.9957\\
10	19.0415\\
11	20.0695\\
12	21.2883\\
13	21.9997\\
14	22.9416\\
15	25.2744\\
16	27.3527\\
17	29.3805\\
18	30.6364\\
19	31.325\\
20	32.2552\\
21	33.8191\\
22	35.2897\\
23	35.8198\\
24	36.65\\
25	37.2864\\
26	38.2354\\
27	39.3218\\
28	42.5942\\
29	43.9831\\
30	45.0979\\
31	46.6101\\
32	47.1497\\
33	47.8706\\
34	49.2918\\
35	49.9693\\
36	50.9075\\
37	51.2616\\
38	51.7758\\
39	52.2711\\
40	53.4045\\
41	54.0781\\
42	54.56\\
43	55.1087\\
44	56.1112\\
45	56.7231\\
46	57.3328\\
47	57.906\\
48	58.2726\\
49	59.1621\\
50	59.9249\\
51	60.5727\\
52	60.9364\\
53	61.7434\\
54	62.2088\\
55	62.6768\\
56	63.1742\\
57	63.7161\\
58	64.8601\\
59	65.4747\\
60	66.2944\\
61	67.1512\\
62	67.4948\\
63	67.6838\\
64	68.3083\\
65	68.7513\\
66	69.0818\\
67	70.1527\\
68	70.7143\\
69	71.0115\\
70	71.5004\\
71	71.7948\\
72	73.0179\\
73	73.3259\\
74	73.4153\\
75	73.6328\\
76	74.1168\\
77	74.2854\\
78	74.4628\\
79	74.9552\\
80	75.1646\\
81	75.4426\\
82	75.6281\\
83	75.9524\\
84	76.3956\\
85	76.7016\\
86	76.9794\\
87	77.0238\\
88	78.322\\
89	78.4652\\
90	78.6896\\
91	78.976\\
92	78.9922\\
93	79.2446\\
94	79.6334\\
95	79.6334\\
96	80.0069\\
97	80.0952\\
98	80.3972\\
99	80.5709\\
100	80.8859\\
};
\addlegendentry{RW R=5}
\end{axis}
\end{tikzpicture}%

%% file: Graphs/DC-20.tex
% %updated correct results

\begin{tikzpicture}
        \begin{axis}[
            legend pos=outer north east,
            ylabel={Network Data Collected [\%]},
            xlabel={No. of nodes visited},
            xmin=0, xmax=20,
            ymin=0, ymax=100,
            %symbolic x coords={R=2, R=3, R=4, R=5},
            %xtick=data,
            %xtick={0, 10, 20, 30, 40, 50, 60, 70, 80, 90, 100},
            xtick={0, 2, 4, 6, 8, 10, 12, 14, 16, 18, 20},
            ytick={0, 10, 20, 30, 40, 50, 60, 70, 80, 90, 100},
            ymajorgrids=true,
            grid style=dashed,
            %legend style={legend pos=outer north east, legend cell align = left},
            %symbolic x coords={a small bar, a medium bar, a large bar},
            %xtick=data
          ]

    \addplot[color=red , mark size=2.5pt, mark options={solid}, very thick]
    coordinates 
    {
% (1, 13.6916)
% (2, 27.1402)
% (3, 40.4025)
% (4, 50.4768)
% (5, 60.1051)
% (6, 69.7408)
% (7, 75.3641)
% (8, 82.0804)
% (9, 85.7124)
% (10, 89.5107)
% (11, 90.7725)
% (12, 93.1287)
% (13, 95.0205)
% (14, 96.3685)
% (15, 98.42)
% (16, 99.0803)
% (17, 99.3775)
% (18, 99.5116)
% (19, 99.7023)
% (20, 99.8227)
% (21, 99.8788)
% (22, 99.9029)
% (23, 99.9297)
% (24, 99.9593)
% (25, 99.9593)
% (26, 99.9726)
% (27, 99.9726)
% (28, 99.9726)
% (29, 99.988)
% (30, 99.988)
% (31, 100)
% (32, 100)
% (33, 100)
% (34, 100)
% (35, 100)
% (36, 100)
% (37, 100)
% (38, 100)
% (39, 100)
% (40, 100)
% (41, 100)
% (42, 100)
% (43, 100)
% (44, 100)
% (45, 100)
% (46, 100)
% (47, 100)
% (48, 100)
% (49, 100)
% (50, 100)
% (51, 100)
% (52, 100)
% (53, 100)
% (54, 100)
% (55, 100)
% (56, 100)
% (57, 100)
% (58, 100)
% (59, 100)
% (60, 100)
% (61, 100)
% (62, 100)
% (63, 100)
% (64, 100)
% (65, 100)
% (66, 100)
% (67, 100)
% (68, 100)
% (69, 100)
% (70, 100)
% (71, 100)
% (72, 100)
% (73, 100)
% (74, 100)
% (75, 100)
% (76, 100)
% (77, 100)
% (78, 100)
% (79, 100)
% (80, 100)
% (81, 100)
% (82, 100)
% (83, 100)
% (84, 100)
% (85, 100)
% (86, 100)
% (87, 100)
% (88, 100)
% (89, 100)
% (90, 100)
% (91, 100)
% (92, 100)
% (93, 100)
% (94, 100)
% (95, 100)
% (96, 100)
% (97, 100)
% (98, 100)
% (99, 100)
% (100, 100)

% (1, 9.86093)
% (2, 19.7219)
% (3, 29.5828)
% (4, 39.4437)
% (5, 49.2726)
% (6, 58.2361)
% (7, 68.0324)
% (8, 71.7439)
% (9, 81.0885)
% (10, 87.7498)
% (11, 91.8504)
% (12, 95.8548)
% (13, 97.2133)
% (14, 98.4392)
% (15, 98.733)
% (16, 99.3487)
% (17, 99.4796)
% (18, 99.5782)
% (19, 99.8691)
% (20, 100)

(0, 0)
(1, 33.9876)
(2, 61.3846)
(3, 72.8655)
(4, 81.0885)
(5, 86.131)
(6, 91.0894)
(7, 93.3291)
(8, 95.1226)
(9, 95.9971)
(10, 96.4436)
(11, 97.1095)
(12, 98.2807)
(13, 99.0509)
(14, 99.202)
(15, 99.4182)
(16, 99.5628)
(17, 99.6798)
(18, 99.8089)
(19, 99.9006)
(20, 100)
    };
    \addlegendentry{DRACO R=5}

    %%%%%%%%%%%%%%%%%%%%%%%%%%%%%%%%%%%%%%%% SA-RW results %%%%%%%%%%%%%%%%%%%%%%%%%%%%%%%%%%%
    
%     \addplot[color=blue, dashed, mark size=2.5pt, mark options={solid}, very thick]
%     coordinates 
%     {
% (1, 9.50538)
% (2, 13.3199)
% (3, 17.1526)
% (4, 26.5989)
% (5, 32.0813)
% (6, 36.659)
% (7, 41.8999)
% (8, 47.8208)
% (9, 54.8002)
% (10, 59.6278)
% (11, 62.4392)
% (12, 65.3998)
% (13, 68.7904)
% (14, 72.9468)
% (15, 75.7815)
% (16, 78.1797)
% (17, 79.6871)
% (18, 81.6256)
% (19, 82.126)
% (20, 84.4822)
%     };
%     \addlegendentry{SA-RW-R=2}
    
%     \addplot[color=black, dashed, mark size=2.5pt, mark options={solid}, very thick]
%     coordinates 
%     {
% (1, 7.11267)
% (2, 14.0913)
% (3, 21.2932)
% (4, 30.6391)
% (5, 36.7854)
% (6, 39.3227)
% (7, 44.7706)
% (8, 51.8921)
% (9, 55.7045)
% (10, 55.7501)
% (11, 59.5955)
% (12, 62.943)
% (13, 62.967)
% (14, 64.8291)
% (15, 69.9171)
% (16, 73.4068)
% (17, 75.444)
% (18, 76.7196)
% (19, 81.8109)
% (20, 85.0867)
%     };
%     \addlegendentry{SA-RW-R=3}
    
%     \addplot[color=brown, dashed, mark size=2.5pt, mark options={solid}, very thick]
%     coordinates 
%     {
% (1, 8.18576)
% (2, 17.822)
% (3, 24.3128)
% (4, 34.0121)
% (5, 41.2137)
% (6, 46.4855)
% (7, 51.2427)
% (8, 56.2911)
% (9, 59.7283)
% (10, 63.1268)
% (11, 64.5607)
% (12, 66.7857)
% (13, 67.2033)
% (14, 67.3545)
% (15, 71.6821)
% (16, 78.4487)
% (17, 82.8687)
% (18, 86.2744)
% (19, 86.2991)
% (20, 86.3756)
%     };
%     \addlegendentry{SA-RW-R=4}
    
    \addplot[color=red, dashed, mark size=2.5pt, mark options={solid}, very thick]
    coordinates 
    {
% (1, 12.8545)
% (2, 23.4099)
% (3, 30.9709)
% (4, 40.0815)
% (5, 44.4752)
% (6, 50.9973)
% (7, 56.8903)
% (8, 59.3543)
% (9, 64.1063)
% (10, 67.9389)
% (11, 68.8907)
% (12, 73.8578)
% (13, 76.7923)
% (14, 76.9721)
% (15, 79.8676)
% (16, 81.4762)
% (17, 82.2407)
% (18, 85.0475)
% (19, 87.6996)
% (20, 88.4916)

(0, 0)
(1, 15.8139)
(2, 28.9822)
(3, 40.1052)
(4, 46.9448)
(5, 52.4342)
(6, 58.2613)
(7, 68.9003)
(8, 75.877)
(9, 78.642)
(10, 81.4804)
(11, 88.9618)
(12, 92.3236)
(13, 94.2688)
(14, 95.0337)
(15, 95.9663)
(16, 96.7004)
(17, 98.5583)
(18, 99.1561)
(19, 99.5514)
(20, 100)
    };
    \addlegendentry{SA-RW-R=5}
    
\addplot[color=red, dotted, mark size=2.5pt, mark options={solid}, very thick]
    coordinates 
    {
% (1, 14.1243)
% (2, 28.2486)
% (3, 42.3729)
% (4, 51.6949)
% (5, 65.5367)
% (6, 65.9605)
% (7, 75.7062)
% (8, 75.7062)
% (9, 75.9887)
% (10, 76.5537)
% (11, 76.5537)
% (12, 76.8362)
% (13, 88.2768)
% (14, 88.2768)
% (15, 89.6893)
% (16, 94.9153)
% (17, 94.9153)
% (18, 94.9153)
% (19, 95.0565)
% (20, 95.0565)

(0, 0)
(1, 18.4363)
(2, 25.4032)
(3, 32.3478)
(4, 46.0981)
(5, 56.399)
(6, 61.8299)
(7, 64.8419)
(8, 67.0818)
(9, 71.477)
(10, 73.7919)
(11, 74.8182)
(12, 77.7734)
(13, 81.0093)
(14, 83.7263)
(15, 84.4429)
(16, 84.6278)
(17, 87.726)
(18, 88.4677)
(19, 89.0954)
(20, 89.5176)
    };
    \addlegendentry{RW-R=5}

% \addplot[color=black, dashed, mark size=2.5pt, mark options={solid}, very thick]
%     coordinates 
%     {
% (0, 0)
% (1, 5)
% (2, 10)
% (3, 15)
% (4, 20)
% (5, 25)
% (6, 30)
% (7, 35)
% (8, 40)
% (9, 45)
% (10, 50)
% (11, 55)
% (12, 60)
% (13, 65)
% (14, 70)
% (15, 75)
% (16, 80)
% (17, 85)
% (18, 90)
% (19, 95)
% (20, 100)
%     };
%     \addlegendentry{No replication}

        \end{axis}
    \end{tikzpicture}

%% file: Graphs/DC-50.tex
% %updated correct results

\begin{tikzpicture}
        \begin{axis}[
            legend pos=outer north east,
            ylabel={Network Data Collected [\%]},
            xlabel={No. of nodes visited},
            xmin=0, xmax=50,
            ymin=0, ymax=100,
            %symbolic x coords={R=2, R=3, R=4, R=5},
            %xtick=data,
            xtick={0, 5, 10, 15, 20, 25, 30, 35, 40, 45, 50},
            ytick={0, 10, 20, 30, 40, 50, 60, 70, 80, 90, 100},
            ymajorgrids=true,
            grid style=dashed,
            %legend style={legend pos=outer north east, legend cell align = left},
            %symbolic x coords={a small bar, a medium bar, a large bar},
            %xtick=data
          ]

    \addplot[color=red , mark size=2.5pt, mark options={solid}, very thick]
    coordinates 
    {
(0, 0)
(1, 26.1352)
(2, 48.6286)
(3, 59.0034)
(4, 67.2843)
(5, 72.1068)
(6, 75.5577)
(7, 77.9602)
(8, 81.3487)
(9, 83.2301)
(10, 84.7613)
(11, 85.8954)
(12, 86.9382)
(13, 87.829)
(14, 88.9696)
(15, 89.7017)
(16, 90.4506)
(17, 91.1276)
(18, 91.9127)
(19, 92.6087)
(20, 93.1825)
(21, 93.7415)
(22, 94.3818)
(23, 94.7744)
(24, 95.2308)
(25, 95.672)
(26, 96.0584)
(27, 96.3998)
(28, 96.666)
(29, 96.9941)
(30, 97.2804)
(31, 97.5197)
(32, 97.7609)
(33, 98.0169)
(34, 98.2241)
(35, 98.4165)
(36, 98.6229)
(37, 98.7685)
(38, 98.9179)
(39, 99.062)
(40, 99.1776)
(41, 99.3031)
(42, 99.4103)
(43, 99.5228)
(44, 99.6189)
(45, 99.692)
(46, 99.7612)
(47, 99.8388)
(48, 99.8962)
(49, 99.9465)
(50, 100)
    };
    \addlegendentry{DRACO R=5}
    
        \addplot[color=red, dashed, mark size=2.5pt, mark options={solid}, very thick]
    coordinates 
    {
% (1, 4.61467)
% (2, 9.22935)
% (3, 12.9672)
% (4, 17.5819)
% (5, 22.1043)
% (6, 26.6267)
% (7, 31.1491)
% (8, 35.6714)
% (9, 39.6862)
% (10, 42.0397)
% (11, 46.1467)
% (12, 50.0692)
% (13, 53.8533)
% (14, 57.3604)
% (15, 61.329)
% (16, 61.329)
% (17, 64.5593)
% (18, 64.5593)
% (19, 67.4665)
% (20, 70.7891)
% (21, 72.4042)
% (22, 74.2501)
% (23, 76.3267)
% (24, 77.9419)
% (25, 80.0185)
% (26, 81.3106)
% (27, 81.3106)
% (28, 82.9257)
% (29, 82.9257)
% (30, 84.1717)
% (31, 84.587)
% (32, 86.3406)
% (33, 87.6788)
% (34, 87.6788)
% (35, 88.9709)
% (36, 90.0323)
% (37, 91.9705)
% (38, 92.3396)
% (39, 92.3396)
% (40, 92.3396)
% (41, 93.5856)
% (42, 93.9086)
% (43, 94.6008)
% (44, 94.7393)
% (45, 96.0314)
% (46, 96.4006)
% (47, 96.4006)
% (48, 97.3235)
% (49, 97.3235)
% (50, 97.3235)

(0, 0)
(1, 4.82544)
(2, 8.10562)
(3, 13.9024)
(4, 16.6945)
(5, 22.6037)
(6, 23.4681)
(7, 26.1105)
(8, 34.0054)
(9, 36.6104)
(10, 37.6708)
(11, 39.2752)
(12, 43.6461)
(13, 44.9949)
(14, 51.229)
(15, 53.1433)
(16, 54.984)
(17, 57.0413)
(18, 62.9114)
(19, 68.2097)
(20, 70.6785)
(21, 75.3709)
(22, 76.6974)
(23, 77.6553)
(24, 79.5299)
(25, 80.293)
(26, 80.7477)
(27, 81.6568)
(28, 82.8724)
(29, 83.6875)
(30, 83.9434)
(31, 85.1284)
(32, 85.9387)
(33, 87.8532)
(34, 88.7544)
(35, 89.4824)
(36, 89.9332)
(37, 90.275)
(38, 91.1311)
(39, 91.3407)
(40, 92.4384)
(41, 94.0537)
(42, 94.9854)
(43, 95.9769)
(44, 97.0825)
(45, 97.5428)
(46, 98.3105)
(47, 98.8451)
(48, 99.3677)
(49, 99.5994)
(50, 100)

    };
    \addlegendentry{SA-RW-R=5}
    
\addplot[color=red, dotted, mark size=2.5pt, mark options={solid}, very thick]
    coordinates 
    {
% (1, 4.9505)
% (2, 9.90099)
% (3, 14.8515)
% (4, 19.0594)
% (5, 23.3663)
% (6, 26.2871)
% (7, 31.2376)
% (8, 34.802)
% (9, 37.7228)
% (10, 42.6733)
% (11, 47.4257)
% (12, 50.6436)
% (13, 52.0792)
% (14, 53.9109)
% (15, 55.198)
% (16, 59.0594)
% (17, 63.0198)
% (18, 67.3267)
% (19, 67.4752)
% (20, 70)
% (21, 71.3366)
% (22, 71.4356)
% (23, 74.9505)
% (24, 74.9505)
% (25, 77.0792)
% (26, 79.6535)
% (27, 82.5743)
% (28, 82.5743)
% (29, 84.3069)
% (30, 86.0396)
% (31, 88.8119)
% (32, 89.1584)
% (33, 90.198)
% (34, 90.3465)
% (35, 90.3465)
% (36, 90.9406)
% (37, 92.0792)
% (38, 92.0792)
% (39, 92.2277)
% (40, 93.0198)
% (41, 93.7624)
% (42, 95.5446)
% (43, 95.6436)
% (44, 96.5347)
% (45, 96.5347)
% (46, 96.5347)
% (47, 96.5347)
% (48, 96.5347)
% (49, 96.5347)
% (50, 96.5347)

(0, 0)
(1, 5.5558)
(2, 10.6642)
(3, 18.3209)
(4, 24.8346)
(5, 27.9095)
(6, 29.8025)
(7, 31.7013)
(8, 36.0336)
(9, 38.4463)
(10, 42.1065)
(11, 42.9366)
(12, 52.0987)
(13, 53.7522)
(14, 55.9405)
(15, 57.4913)
(16, 58.1421)
(17, 59.2722)
(18, 63.0918)
(19, 64.7906)
(20, 65.8006)
(21, 67.0668)
(22, 67.6281)
(23, 69.1019)
(24, 70.3759)
(25, 71.4653)
(26, 71.9521)
(27, 72.4693)
(28, 73.3734)
(29, 74.5824)
(30, 75.381)
(31, 76.0436)
(32, 76.6954)
(33, 77.3194)
(34, 77.5558)
(35, 77.9676)
(36, 80.1255)
(37, 80.3793)
(38, 80.6135)
(39, 81.1237)
(40, 81.5314)
(41, 82.2892)
(42, 82.7817)
(43, 83.0701)
(44, 83.3027)
(45, 83.475)
(46, 83.8194)
(47, 84.0434)
(48, 84.2099)
(49, 84.6997)
(50, 85.3221)
    };
    \addlegendentry{RW-R=5}

        \end{axis}
    \end{tikzpicture}

%% file: Graphs/DC-150.tex
% %updated correct results

\begin{tikzpicture}
        \begin{axis}[
            legend pos=outer north east,
            ylabel={Network Data Collected [\%]},
            xlabel={No. of nodes visited},
            xmin=0, xmax=150,
            ymin=0, ymax=100,
            %symbolic x coords={R=2, R=3, R=4, R=5},
            %xtick=data,
            xtick={0, 15, 30, 45, 60, 75, 90, 105, 120, 135, 150},
            ytick={0, 10, 20, 30, 40, 50, 60, 70, 80, 90, 100},
            ymajorgrids=true,
            grid style=dashed,
            %legend style={legend pos=outer north east, legend cell align = left},
            %symbolic x coords={a small bar, a medium bar, a large bar},
            %xtick=data
          ]

    \addplot[color=red , mark size=2.5pt, mark options={solid}, very thick]
    coordinates 
    {

(0,0)
(1, 3.26716)
(2, 6.55178)
(3, 8.83153)
(4, 10.9704)
(5, 12.7009)
(6, 14.4924)
(7, 16.1382)
(8, 17.7633)
(9, 19.1648)
(10, 20.6011)
(11, 21.9832)
(12, 23.2913)
(13, 24.5902)
(14, 25.8602)
(15, 27.1649)
(16, 28.4267)
(17, 29.6934)
(18, 30.9504)
(19, 32.2098)
(20, 33.4705)
(21, 34.7275)
(22, 35.9845)
(23, 37.2)
(24, 38.4583)
(25, 39.7045)
(26, 40.9602)
(27, 42.1761)
(28, 43.3684)
(29, 44.5304)
(30, 45.7182)
(31, 46.865)
(32, 47.9454)
(33, 49.0707)
(34, 50.0781)
(35, 51.1445)
(36, 52.1483)
(37, 53.1025)
(38, 53.9849)
(39, 54.9895)
(40, 55.8922)
(41, 56.7842)
(42, 57.6729)
(43, 58.5003)
(44, 59.3183)
(45, 60.0797)
(46, 60.8515)
(47, 61.6207)
(48, 62.3111)
(49, 62.9565)
(50, 63.6192)
(51, 64.3047)
(52, 64.9525)
(53, 65.6663)
(54, 66.2875)
(55, 66.9221)
(56, 67.5334)
(57, 68.1885)
(58, 68.7915)
(59, 69.4051)
(60, 69.9958)
(61, 70.6046)
(62, 71.1971)
(63, 71.7875)
(64, 72.3381)
(65, 72.8969)
(66, 73.4545)
(67, 73.9908)
(68, 74.5163)
(69, 75.0343)
(70, 75.5313)
(71, 76.0254)
(72, 76.4983)
(73, 76.9927)
(74, 77.4453)
(75, 77.9105)
(76, 78.355)
(77, 78.8044)
(78, 79.2672)
(79, 79.7149)
(80, 80.1607)
(81, 80.6211)
(82, 81.0342)
(83, 81.4734)
(84, 81.904)
(85, 82.3178)
(86, 82.7432)
(87, 83.141)
(88, 83.5584)
(89, 83.9624)
(90, 84.3677)
(91, 84.756)
(92, 85.1629)
(93, 85.5679)
(94, 85.9584)
(95, 86.3485)
(96, 86.7158)
(97, 87.0917)
(98, 87.4544)
(99, 87.8094)
(100, 88.1659)
(101, 88.5131)
(102, 88.8633)
(103, 89.1976)
(104, 89.543)
(105, 89.8786)
(106, 90.2166)
(107, 90.5426)
(108, 90.8697)
(109, 91.1992)
(110, 91.5239)
(111, 91.8521)
(112, 92.1745)
(113, 92.4981)
(114, 92.8145)
(115, 93.1357)
(116, 93.4509)
(117, 93.7672)
(118, 94.0776)
(119, 94.3865)
(120, 94.6875)
(121, 94.9925)
(122, 95.2946)
(123, 95.5882)
(124, 95.8781)
(125, 96.167)
(126, 96.432)
(127, 96.6777)
(128, 96.9353)
(129, 97.1789)
(130, 97.4294)
(131, 97.6545)
(132, 97.8585)
(133, 98.0496)
(134, 98.2176)
(135, 98.3981)
(136, 98.6004)
(137, 98.7564)
(138, 98.8848)
(139, 99.0366)
(140, 99.1706)
(141, 99.2771)
(142, 99.3943)
(143, 99.5156)
(144, 99.6032)
(145, 99.6825)
(146, 99.7628)
(147, 99.8367)
(148, 99.8916)
(149, 99.9509)
(150, 100)

    };
    \addlegendentry{DRACO R=5}
    
        \addplot[color=red, dashed, mark size=2.5pt, mark options={solid}, very thick]
    coordinates 
    {

(0, 0)
(1, 0.712975)
(2, 1.599)
(3, 2.7314)
(4, 3.45192)
(5, 4.10769)
(6, 4.91793)
(7, 5.8109)
(8, 6.47678)
(9, 7.77526)
(10, 8.45244)
(11, 9.25189)
(12, 10.2036)
(13, 10.9168)
(14, 11.6401)
(15, 12.5227)
(16, 13.1583)
(17, 13.8444)
(18, 14.4414)
(19, 15.5649)
(20, 16.1246)
(21, 16.6093)
(22, 17.3461)
(23, 17.988)
(24, 18.6178)
(25, 19.3925)
(26, 20.116)
(27, 21.0543)
(28, 21.4801)
(29, 22.2093)
(30, 22.8688)
(31, 23.5655)
(32, 24.2747)
(33, 24.783)
(34, 25.4713)
(35, 25.9484)
(36, 26.6403)
(37, 27.4225)
(38, 28.2149)
(39, 29.1285)
(40, 29.7666)
(41, 30.5023)
(42, 31.0619)
(43, 31.9543)
(44, 32.7115)
(45, 34.1582)
(46, 34.8363)
(47, 35.7146)
(48, 36.2704)
(49, 36.9628)
(50, 37.7666)
(51, 38.344)
(52, 38.8492)
(53, 39.4541)
(54, 40.2064)
(55, 40.807)
(56, 41.5684)
(57, 42.3376)
(58, 43.0165)
(59, 43.7687)
(60, 44.4533)
(61, 45.3093)
(62, 46.0444)
(63, 46.7314)
(64, 47.1847)
(65, 47.8773)
(66, 48.7127)
(67, 49.4538)
(68, 49.9479)
(69, 50.728)
(70, 51.2402)
(71, 51.7088)
(72, 52.4246)
(73, 53.1078)
(74, 53.7458)
(75, 54.4123)
(76, 55.1846)
(77, 55.8729)
(78, 56.446)
(79, 57.1518)
(80, 57.8527)
(81, 58.674)
(82, 59.5608)
(83, 59.9219)
(84, 60.5407)
(85, 61.1665)
(86, 61.8201)
(87, 62.4449)
(88, 63.1155)
(89, 63.8542)
(90, 64.5376)
(91, 65.3849)
(92, 66.3395)
(93, 66.8599)
(94, 67.3357)
(95, 67.7773)
(96, 68.2694)
(97, 69.0109)
(98, 69.7139)
(99, 70.2344)
(100, 70.9876)
(101, 71.4717)
(102, 71.9623)
(103, 72.6493)
(104, 73.4113)
(105, 74.1717)
(106, 74.9164)
(107, 75.4991)
(108, 76.1595)
(109, 76.7167)
(110, 77.4283)
(111, 77.9671)
(112, 78.677)
(113, 79.2519)
(114, 79.7217)
(115, 80.3844)
(116, 81.0177)
(117, 81.6176)
(118, 82.2643)
(119, 82.6792)
(120, 83.2506)
(121, 83.7116)
(122, 84.1525)
(123, 84.6142)
(124, 85.2145)
(125, 85.7942)
(126, 86.4321)
(127, 87.1065)
(128, 87.7943)
(129, 88.2892)
(130, 88.6789)
(131, 89.2247)
(132, 90.0364)
(133, 90.6332)
(134, 91.2535)
(135, 91.9526)
(136, 92.6999)
(137, 93.2512)
(138, 93.8009)
(139, 94.3874)
(140, 94.9811)
(141, 95.5977)
(142, 96.1281)
(143, 96.7068)
(144, 97.2168)
(145, 97.685)
(146, 98.146)
(147, 98.4795)
(148, 98.991)
(149, 99.5265)
(150, 100)

    };
    \addlegendentry{SA-RW-R=5}
    
\addplot[color=red, dotted, mark size=2.5pt, mark options={solid}, very thick]
    coordinates 
    {

(0, 0)
(1, 0.500657)
(2, 1.24738)
(3, 2.43732)
(4, 3.24667)
(5, 3.70841)
(6, 4.37923)
(7, 5.11681)
(8, 5.54457)
(9, 6.25595)
(10, 8.29012)
(11, 8.74973)
(12, 9.31671)
(13, 9.87156)
(14, 10.675)
(15, 11.5413)
(16, 12.342)
(17, 13.0646)
(18, 13.5006)
(19, 13.9252)
(20, 14.4322)
(21, 15.2074)
(22, 15.8251)
(23, 16.2185)
(24, 17.5368)
(25, 18.3853)
(26, 18.938)
(27, 19.6713)
(28, 20.0942)
(29, 20.4582)
(30, 20.8654)
(31, 22.0038)
(32, 22.8442)
(33, 23.4965)
(34, 23.8857)
(35, 24.4154)
(36, 25.2451)
(37, 26.2766)
(38, 26.704)
(39, 27.2514)
(40, 27.8022)
(41, 28.1093)
(42, 28.2334)
(43, 29.2787)
(44, 30.593)
(45, 31.2421)
(46, 32.0156)
(47, 32.4414)
(48, 33.1528)
(49, 33.4035)
(50, 34.5551)
(51, 34.8989)
(52, 35.5246)
(53, 36.2749)
(54, 37.513)
(55, 37.6082)
(56, 38.7678)
(57, 38.9789)
(58, 39.2924)
(59, 39.4385)
(60, 39.8685)
(61, 40.1855)
(62, 40.3013)
(63, 41.3697)
(64, 41.5887)
(65, 41.5887)
(66, 41.5887)
(67, 41.8994)
(68, 41.8994)
(69, 42.5385)
(70, 42.6792)
(71, 42.8199)
(72, 43.3734)
(73, 44.1151)
(74, 44.252)
(75, 44.6256)
(76, 44.831)
(77, 45.1059)
(78, 45.522)
(79, 45.6589)
(80, 46.0669)
(81, 46.3276)
(82, 46.6669)
(83, 47.0335)
(84, 47.4647)
(85, 48.0921)
(86, 48.3443)
(87, 48.3443)
(88, 48.9711)
(89, 49.5381)
(90, 50.3254)
(91, 50.7498)
(92, 50.7498)
(93, 51.7405)
(94, 51.8481)
(95, 52.1434)
(96, 52.4178)
(97, 52.8244)
(98, 53.0298)
(99, 53.4364)
(100, 53.763)
(101, 54.2975)
(102, 55.2076)
(103, 55.7971)
(104, 56.2913)
(105, 56.4459)
(106, 57.0804)
(107, 57.5142)
(108, 57.6149)
(109, 57.6149)
(110, 58.2591)
(111, 58.6617)
(112, 58.6617)
(113, 58.6617)
(114, 58.8807)
(115, 59.1717)
(116, 59.1717)
(117, 59.1717)
(118, 59.357)
(119, 60.2046)
(120, 60.6225)
(121, 60.6225)
(122, 60.6225)
(123, 61.0529)
(124, 61.1536)
(125, 61.823)
(126, 62.2037)
(127, 62.4142)
(128, 62.6253)
(129, 62.6253)
(130, 62.6983)
(131, 62.9176)
(132, 63.0223)
(133, 63.6614)
(134, 64.0099)
(135, 64.0099)
(136, 64.4165)
(137, 64.5212)
(138, 64.5212)
(139, 64.6716)
(140, 65.0564)
(141, 65.1971)
(142, 65.1971)
(143, 65.1971)
(144, 65.4161)
(145, 65.4161)
(146, 65.6215)
(147, 65.6215)
(148, 65.6215)
(149, 65.6215)
(150, 66.1688)

    };
    \addlegendentry{RW-R=5}
            
        \end{axis}
    \end{tikzpicture}

%% file: Graphs/DC-200.tex
% %updated correct results

\begin{tikzpicture}
        \begin{axis}[
            legend pos=outer north east,
            ylabel={Network Data Collected [\%]},
            xlabel={No. of nodes visited},
            xmin=0, xmax=200,
            ymin=0, ymax=100,
            %symbolic x coords={R=2, R=3, R=4, R=5},
            %xtick=data,
            xtick={0, 20, 40, 60, 80, 100, 120, 140, 160, 180, 200},
            ytick={0, 10, 20, 30, 40, 50, 60, 70, 80, 90, 100},
            ymajorgrids=true,
            grid style=dashed,
            %legend style={legend pos=outer north east, legend cell align = left},
            %symbolic x coords={a small bar, a medium bar, a large bar},
            %xtick=data
          ]

    \addplot[color=red , mark size=2.5pt, mark options={solid}, very thick]
    coordinates 
    {

(0, 0)
(1, 1.51377)
(2, 2.95195)
(3, 4.19614)
(4, 5.38018)
(5, 6.38517)
(6, 7.54754)
(7, 8.6414)
(8, 9.62606)
(9, 10.6488)
(10, 11.6223)
(11, 12.5921)
(12, 13.5628)
(13, 14.5317)
(14, 15.4996)
(15, 16.4675)
(16, 17.4345)
(17, 18.4024)
(18, 19.3694)
(19, 20.3373)
(20, 21.3043)
(21, 22.2713)
(22, 23.2382)
(23, 24.2052)
(24, 25.1722)
(25, 26.1392)
(26, 27.1062)
(27, 28.0732)
(28, 29.0402)
(29, 30.0071)
(30, 30.9741)
(31, 31.9411)
(32, 32.9081)
(33, 33.8751)
(34, 34.8421)
(35, 35.809)
(36, 36.773)
(37, 37.6996)
(38, 38.6666)
(39, 39.5892)
(40, 40.5391)
(41, 41.3645)
(42, 42.2777)
(43, 43.1389)
(44, 43.9998)
(45, 44.7969)
(46, 45.5082)
(47, 46.3269)
(48, 47.0713)
(49, 47.7608)
(50, 48.3933)
(51, 49.0381)
(52, 49.7043)
(53, 50.2962)
(54, 50.8488)
(55, 51.445)
(56, 52.0207)
(57, 52.5279)
(58, 53.0408)
(59, 53.5551)
(60, 54.0818)
(61, 54.5904)
(62, 55.0787)
(63, 55.5679)
(64, 56.0609)
(65, 56.5492)
(66, 57.0375)
(67, 57.5257)
(68, 58.014)
(69, 58.5023)
(70, 58.9906)
(71, 59.4788)
(72, 59.9671)
(73, 60.4554)
(74, 60.9437)
(75, 61.432)
(76, 61.9202)
(77, 62.4085)
(78, 62.8968)
(79, 63.3851)
(80, 63.8734)
(81, 64.3616)
(82, 64.8469)
(83, 65.3352)
(84, 65.8235)
(85, 66.2956)
(86, 66.766)
(87, 67.2244)
(88, 67.682)
(89, 68.1521)
(90, 68.6196)
(91, 69.1018)
(92, 69.5504)
(93, 70.0139)
(94, 70.433)
(95, 70.8797)
(96, 71.2914)
(97, 71.6893)
(98, 72.0901)
(99, 72.4616)
(100, 72.8355)
(101, 73.1909)
(102, 73.5373)
(103, 73.8739)
(104, 74.2122)
(105, 74.5587)
(106, 74.8931)
(107, 75.2304)
(108, 75.5597)
(109, 75.889)
(110, 76.2172)
(111, 76.5438)
(112, 76.8693)
(113, 77.1948)
(114, 77.5203)
(115, 77.8458)
(116, 78.1723)
(117, 78.4978)
(118, 78.8233)
(119, 79.1488)
(120, 79.4725)
(121, 79.798)
(122, 80.1236)
(123, 80.4491)
(124, 80.7746)
(125, 81.1001)
(126, 81.4256)
(127, 81.7512)
(128, 82.0767)
(129, 82.4012)
(130, 82.7267)
(131, 83.0522)
(132, 83.3778)
(133, 83.7014)
(134, 84.0251)
(135, 84.3497)
(136, 84.6733)
(137, 84.9877)
(138, 85.2989)
(139, 85.5979)
(140, 85.8964)
(141, 86.1922)
(142, 86.4897)
(143, 86.7685)
(144, 87.0359)
(145, 87.2955)
(146, 87.5625)
(147, 87.8232)
(148, 88.0869)
(149, 88.3359)
(150, 88.6009)
(151, 88.8499)
(152, 89.1045)
(153, 89.3534)
(154, 89.6023)
(155, 89.8513)
(156, 90.1002)
(157, 90.3491)
(158, 90.598)
(159, 90.847)
(160, 91.0959)
(161, 91.3448)
(162, 91.5938)
(163, 91.8427)
(164, 92.0916)
(165, 92.3405)
(166, 92.5895)
(167, 92.8384)
(168, 93.0873)
(169, 93.3362)
(170, 93.5852)
(171, 93.8341)
(172, 94.083)
(173, 94.3319)
(174, 94.5809)
(175, 94.8298)
(176, 95.0787)
(177, 95.3277)
(178, 95.5766)
(179, 95.8255)
(180, 96.0678)
(181, 96.3168)
(182, 96.5648)
(183, 96.8137)
(184, 97.0617)
(185, 97.3097)
(186, 97.5503)
(187, 97.7955)
(188, 98.0273)
(189, 98.2632)
(190, 98.4969)
(191, 98.699)
(192, 98.8786)
(193, 99.0824)
(194, 99.2984)
(195, 99.4311)
(196, 99.5753)
(197, 99.6797)
(198, 99.8394)
(199, 99.9361)
(200, 100)
    };
    \addlegendentry{DRACO R=5}

        \addplot[color=red, dashed, mark size=2.5pt, mark options={solid}, very thick]
    coordinates 
    {

(0, 0)
(1, 0.39189)
(2, 0.75092)
(3, 1.28913)
(4, 1.86763)
(5, 2.22966)
(6, 2.77243)
(7, 3.0878)
(8, 3.62601)
(9, 3.96812)
(10, 4.89583)
(11, 5.60377)
(12, 5.97119)
(13, 6.5454)
(14, 6.80798)
(15, 7.18765)
(16, 7.45097)
(17, 8.02814)
(18, 8.30163)
(19, 8.86506)
(20, 9.15359)
(21, 9.85162)
(22, 10.1119)
(23, 10.4555)
(24, 10.8735)
(25, 11.5648)
(26, 12.1485)
(27, 12.5665)
(28, 12.9071)
(29, 13.3817)
(30, 13.9312)
(31, 14.6208)
(32, 15.0893)
(33, 15.8728)
(34, 16.1331)
(35, 16.6231)
(36, 16.9354)
(37, 17.4425)
(38, 18.0596)
(39, 18.3043)
(40, 18.6539)
(41, 19.3941)
(42, 20.1197)
(43, 20.6066)
(44, 21.3115)
(45, 21.5755)
(46, 22.1137)
(47, 22.4551)
(48, 23.2465)
(49, 23.6325)
(50, 23.9448)
(51, 24.6537)
(52, 25.2912)
(53, 25.6318)
(54, 26.2517)
(55, 26.515)
(56, 26.8021)
(57, 27.1574)
(58, 27.6191)
(59, 28.1627)
(60, 29.1975)
(61, 29.8166)
(62, 30.911)
(63, 31.3037)
(64, 31.8748)
(65, 32.4243)
(66, 32.9892)
(67, 33.4577)
(68, 33.8556)
(69, 34.5605)
(70, 35.1366)
(71, 35.7106)
(72, 35.9166)
(73, 36.258)
(74, 36.8229)
(75, 37.0869)
(76, 38.0057)
(77, 38.2691)
(78, 38.7559)
(79, 39.0735)
(80, 39.6844)
(81, 40.3163)
(82, 40.6041)
(83, 40.8667)
(84, 41.4436)
(85, 41.9538)
(86, 42.2171)
(87, 42.7566)
(88, 43.0437)
(89, 43.3322)
(90, 43.6326)
(91, 43.9457)
(92, 44.7249)
(93, 45.2186)
(94, 45.6803)
(95, 46.5735)
(96, 47.1415)
(97, 47.8702)
(98, 48.6448)
(99, 48.983)
(100, 49.3994)
(101, 49.9583)
(102, 50.4268)
(103, 50.8004)
(104, 51.3811)
(105, 51.7454)
(106, 52.0785)
(107, 52.547)
(108, 53.3058)
(109, 53.8553)
(110, 54.3904)
(111, 55.165)
(112, 55.7221)
(113, 56.4878)
(114, 57.0558)
(115, 57.6862)
(116, 57.9733)
(117, 58.6646)
(118, 59.2028)
(119, 59.4202)
(120, 60.1399)
(121, 60.3876)
(122, 60.8782)
(123, 61.6479)
(124, 62.2001)
(125, 62.9671)
(126, 63.541)
(127, 64.0645)
(128, 64.6118)
(129, 65.0934)
(130, 65.4833)
(131, 65.8209)
(132, 66.2098)
(133, 67.0362)
(134, 67.6953)
(135, 68.1547)
(136, 68.6691)
(137, 69.1375)
(138, 69.5524)
(139, 70.1234)
(140, 70.8283)
(141, 71.1911)
(142, 71.678)
(143, 71.9934)
(144, 72.4083)
(145, 73.0303)
(146, 73.4988)
(147, 73.8364)
(148, 74.4561)
(149, 74.6595)
(150, 75.1978)
(151, 75.4804)
(152, 76.1717)
(153, 76.8221)
(154, 77.5855)
(155, 78.135)
(156, 78.8851)
(157, 79.1887)
(158, 79.6824)
(159, 80.0185)
(160, 80.7886)
(161, 81.0772)
(162, 81.3406)
(163, 81.8274)
(164, 82.1152)
(165, 82.403)
(166, 82.9249)
(167, 83.4776)
(168, 84.2447)
(169, 84.5779)
(170, 85.1274)
(171, 85.5313)
(172, 86.0182)
(173, 86.7409)
(174, 87.2093)
(175, 87.7521)
(176, 88.1815)
(177, 88.4969)
(178, 88.7342)
(179, 89.3657)
(180, 89.6811)
(181, 90.249)
(182, 90.8008)
(183, 91.0137)
(184, 91.7154)
(185, 92.1096)
(186, 92.5559)
(187, 92.835)
(188, 93.5417)
(189, 94.2703)
(190, 94.8873)
(191, 95.4507)
(192, 96.1421)
(193, 96.5585)
(194, 97.0446)
(195, 97.4535)
(196, 98.0291)
(197, 98.5152)
(198, 99.0534)
(199, 99.5286)
(200, 100)

    };
    \addlegendentry{SA-RW-R=5}
    
\addplot[color=red, dotted, mark size=2.5pt, mark options={solid}, very thick]
    coordinates 
    {

(0, 0)
(1, 0.422009)
(2, 0.796446)
(3, 1.37283)
(4, 1.81646)
(5, 2.24407)
(6, 2.77179)
(7, 2.96136)
(8, 3.43109)
(9, 4.00402)
(10, 4.74495)
(11, 5.26508)
(12, 5.70679)
(13, 6.21866)
(14, 7.0436)
(15, 7.86526)
(16, 8.22183)
(17, 8.4686)
(18, 8.8296)
(19, 9.34973)
(20, 9.91972)
(21, 10.1665)
(22, 10.7394)
(23, 11.2981)
(24, 11.5056)
(25, 11.9493)
(26, 12.3513)
(27, 12.795)
(28, 13.0203)
(29, 13.351)
(30, 13.4608)
(31, 13.7411)
(32, 13.9952)
(33, 14.5098)
(34, 14.79)
(35, 15.2287)
(36, 15.6775)
(37, 16.1636)
(38, 16.4438)
(39, 16.7871)
(40, 17.2188)
(41, 17.5527)
(42, 18.0728)
(43, 18.4885)
(44, 18.9042)
(45, 18.9925)
(46, 19.1081)
(47, 19.2728)
(48, 19.6525)
(49, 19.9957)
(50, 20.2746)
(51, 20.9264)
(52, 21.1718)
(53, 21.4769)
(54, 21.7892)
(55, 22.0574)
(56, 22.6015)
(57, 22.7997)
(58, 23.143)
(59, 23.2585)
(60, 23.5847)
(61, 24.1048)
(62, 24.769)
(63, 25.107)
(64, 25.2654)
(65, 25.7535)
(66, 26.2529)
(67, 26.6769)
(68, 27.0031)
(69, 27.8163)
(70, 28.1729)
(71, 28.5783)
(72, 28.5783)
(73, 29.0653)
(74, 29.4563)
(75, 30.0365)
(76, 30.2098)
(77, 30.2098)
(78, 30.8234)
(79, 31.1551)
(80, 31.3135)
(81, 31.6474)
(82, 31.9611)
(83, 32.5612)
(84, 33.389)
(85, 33.8739)
(86, 34.443)
(87, 34.9301)
(88, 35.2)
(89, 35.5746)
(90, 35.6844)
(91, 36.4398)
(92, 37.1062)
(93, 37.6083)
(94, 38.0103)
(95, 38.3536)
(96, 38.3536)
(97, 38.4592)
(98, 38.6753)
(99, 39.073)
(100, 39.2871)
(101, 39.8262)
(102, 40.1399)
(103, 40.4178)
(104, 40.6074)
(105, 40.7979)
(106, 41.08)
(107, 41.1955)
(108, 41.5422)
(109, 41.6326)
(110, 41.721)
(111, 42.3904)
(112, 42.5593)
(113, 43.3095)
(114, 43.8486)
(115, 44.0635)
(116, 44.0635)
(117, 44.2219)
(118, 44.5356)
(119, 44.6454)
(120, 44.8101)
(121, 45.1356)
(122, 45.7212)
(123, 46.0791)
(124, 46.0791)
(125, 46.0791)
(126, 46.1675)
(127, 46.3638)
(128, 46.6211)
(129, 47.0628)
(130, 47.406)
(131, 47.5707)
(132, 47.914)
(133, 47.914)
(134, 48.4103)
(135, 48.829)
(136, 48.9346)
(137, 49.2102)
(138, 49.2102)
(139, 49.7118)
(140, 50.3743)
(141, 50.3743)
(142, 50.4899)
(143, 50.6632)
(144, 50.7753)
(145, 50.8637)
(146, 50.9477)
(147, 51.0316)
(148, 51.0316)
(149, 51.12)
(150, 51.3731)
(151, 51.8003)
(152, 51.8003)
(153, 51.8003)
(154, 51.8842)
(155, 51.9898)
(156, 52.1234)
(157, 52.2106)
(158, 52.4011)
(159, 52.6265)
(160, 53.1098)
(161, 53.2185)
(162, 53.3283)
(163, 53.4868)
(164, 53.5707)
(165, 54.1647)
(166, 54.9925)
(167, 55.4217)
(168, 55.5407)
(169, 55.6505)
(170, 55.7345)
(171, 56.5061)
(172, 56.5868)
(173, 56.5868)
(174, 57.0453)
(175, 57.359)
(176, 57.6727)
(177, 57.6727)
(178, 57.7535)
(179, 57.8995)
(180, 58.4196)
(181, 58.4196)
(182, 58.4196)
(183, 58.578)
(184, 58.7427)
(185, 59.3095)
(186, 59.4679)
(187, 59.9797)
(188, 60.0681)
(189, 60.517)
(190, 60.8171)
(191, 61.6449)
(192, 61.6449)
(193, 61.6449)
(194, 61.7505)
(195, 61.7505)
(196, 61.8603)
(197, 62.5002)
(198, 62.61)
(199, 62.61)
(200, 63.0743)

    };
    \addlegendentry{RW-R=5}

        \end{axis}
    \end{tikzpicture}

%% file: Graphs/DC-100N-NF10.tex
% %updated correct results

\begin{tikzpicture}
        \begin{axis}[
            legend pos=outer north east,
            ylabel={Network Data Collected [\%]},
            xlabel={No. of nodes visited},
            xmin=0, xmax=90,
            ymin=0, ymax=100,
            %symbolic x coords={R=2, R=3, R=4, R=5},
            %xtick=data,
            xtick={0, 10, 20, 30, 40, 50, 60, 70, 80, 90},
            ytick={0, 10, 20, 30, 40, 50, 60, 70, 80, 90, 100},
            ymajorgrids=true,
            grid style=dashed,
            %legend style={legend pos=outer north east, legend cell align = left},
            %symbolic x coords={a small bar, a medium bar, a large bar},
            %xtick=data
          ]

    \addplot[color=red , mark size=2.5pt, mark options={solid}, very thick]
    coordinates 
    {
(0, 0)
(1, 10.9561)
(2, 18.2564)
(3, 27.1153)
(4, 33.0039)
(5, 37.6725)
(6, 41.335)
(7, 43.8893)
(8, 46.8166)
(9, 49.2034)
(10, 51.466)
(11, 53.4199)
(12, 55.3975)
(13, 57.2517)
(14, 58.9791)
(15, 60.5541)
(16, 61.9585)
(17, 63.3855)
(18, 64.7634)
(19, 65.9053)
(20, 67.1081)
(21, 68.2208)
(22, 69.2721)
(23, 70.3022)
(24, 71.2347)
(25, 72.1566)
(26, 73.0429)
(27, 73.9103)
(28, 74.706)
(29, 75.5188)
(30, 76.246)
(31, 76.9801)
(32, 77.6766)
(33, 78.3336)
(34, 79.0061)
(35, 79.6781)
(36, 80.3423)
(37, 80.9403)
(38, 81.5777)
(39, 82.1525)
(40, 82.7183)
(41, 83.3037)
(42, 83.8281)
(43, 84.3649)
(44, 84.8869)
(45, 85.4012)
(46, 85.865)
(47, 86.3866)
(48, 86.8463)
(49, 87.3229)
(50, 87.7799)
(51, 88.2081)
(52, 88.6295)
(53, 89.0255)
(54, 89.4036)
(55, 89.7891)
(56, 90.1765)
(57, 90.5569)
(58, 90.9051)
(59, 91.26)
(60, 91.6076)
(61, 91.9451)
(62, 92.2654)
(63, 92.5541)
(64, 92.8698)
(65, 93.1515)
(66, 93.4122)
(67, 93.6565)
(68, 93.9002)
(69, 94.1407)
(70, 94.3574)
(71, 94.5795)
(72, 94.7663)
(73, 94.9766)
(74, 95.1711)
(75, 95.3458)
(76, 95.5226)
(77, 95.6919)
(78, 95.8567)
(79, 96.0093)
(80, 96.126)
(81, 96.265)
(82, 96.3987)
(83, 96.5237)
(84, 96.6285)
(85, 96.734)
(86, 96.8291)
(87, 96.9219)
(88, 96.9896)
(89, 97.0687)
(90, 97.1225)

    };
    \addlegendentry{DRACO R=5}

    \addplot[color=red, dashed, mark size=2.5pt, mark options={solid}, very thick]
    coordinates 
    {

(0, 0)
(1, 2.9513)
(2, 3.99504)
(3, 5.04697)
(4, 6.87739)
(5, 9.09451)
(6, 10.4879)
(7, 12.2175)
(8, 13.4568)
(9, 14.4654)
(10, 15.4616)
(11, 17.3654)
(12, 18.5826)
(13, 19.6159)
(14, 20.9677)
(15, 21.8613)
(16, 23.2202)
(17, 25.0629)
(18, 26.081)
(19, 27.1807)
(20, 28.6179)
(21, 29.7489)
(22, 30.9058)
(23, 32.1341)
(24, 34.0677)
(25, 35.9497)
(26, 36.895)
(27, 38.7228)
(28, 40.219)
(29, 41.6292)
(30, 42.6862)
(31, 43.6967)
(32, 45.1301)
(33, 46.4183)
(34, 47.4942)
(35, 48.6877)
(36, 51.1798)
(37, 52.3713)
(38, 53.4815)
(39, 54.2671)
(40, 56.0805)
(41, 56.9308)
(42, 57.8157)
(43, 58.9921)
(44, 59.55)
(45, 60.3596)
(46, 61.8305)
(47, 62.4717)
(48, 63.184)
(49, 64.3488)
(50, 64.9677)
(51, 65.7972)
(52, 66.9103)
(53, 68.4488)
(54, 69.1633)
(55, 69.8021)
(56, 70.7452)
(57, 72.1905)
(58, 73.4136)
(59, 74.7008)
(60, 75.1413)
(61, 75.8662)
(62, 76.8929)
(63, 78.0292)
(64, 78.9999)
(65, 79.4254)
(66, 80.2296)
(67, 80.9387)
(68, 82.4008)
(69, 83.1057)
(70, 83.586)
(71, 84.8798)
(72, 85.8821)
(73, 86.9224)
(74, 87.253)
(75, 87.9649)
(76, 89.1812)
(77, 89.9251)
(78, 90.5519)
(79, 91.2696)
(80, 92.0457)
(81, 92.9912)
(82, 93.5178)
(83, 94.0183)
(84, 94.5952)
(85, 95.1203)
(86, 95.5419)
(87, 95.888)
(88, 96.287)
(89, 96.5825)
(90, 97.1679)

    };
    \addlegendentry{SA-RW-R=5}
    
\addplot[color=red, dotted, mark size=2.5pt, mark options={solid}, very thick]
    coordinates 
    {

(0, 0)
(1, 1.70143)
(2, 3.27677)
(3, 5.33668)
(4, 6.48851)
(5, 7.34273)
(6, 8.55091)
(7, 9.47539)
(8, 11.5347)
(9, 12.7908)
(10, 13.7713)
(11, 15.2003)
(12, 16.5548)
(13, 17.8531)
(14, 18.8652)
(15, 20.8542)
(16, 21.8052)
(17, 23.1284)
(18, 24.394)
(19, 25.2936)
(20, 25.9801)
(21, 28.1457)
(22, 29.3307)
(23, 30.8365)
(24, 31.7428)
(25, 33.058)
(26, 33.8408)
(27, 35.3079)
(28, 36.1717)
(29, 36.9238)
(30, 37.9752)
(31, 38.9159)
(32, 39.2868)
(33, 40.0307)
(34, 41.62)
(35, 42.8025)
(36, 43.4869)
(37, 44.5071)
(38, 45.3071)
(39, 45.7834)
(40, 46.6722)
(41, 47.7337)
(42, 49.3993)
(43, 49.6956)
(44, 49.9807)
(45, 51.0376)
(46, 51.5491)
(47, 52.084)
(48, 52.6948)
(49, 53.0122)
(50, 53.531)
(51, 53.9198)
(52, 54.7928)
(53, 55.3188)
(54, 55.9687)
(55, 56.7458)
(56, 56.9446)
(57, 57.2196)
(58, 57.7478)
(59, 58.1751)
(60, 58.6904)
(61, 59.1384)
(62, 59.7072)
(63, 59.935)
(64, 60.2507)
(65, 60.9325)
(66, 62.2703)
(67, 62.4965)
(68, 62.7297)
(69, 63.2455)
(70, 63.3769)
(71, 63.7616)
(72, 64.1313)
(73, 64.4357)
(74, 64.484)
(75, 64.7721)
(76, 64.9115)
(77, 66.0609)
(78, 66.5321)
(79, 66.7754)
(80, 67.1418)
(81, 68.0798)
(82, 68.7145)
(83, 69.0667)
(84, 69.2726)
(85, 69.6967)
(86, 69.9252)
(87, 70.4284)
(88, 70.5475)
(89, 70.7902)
(90, 71.2503)

    };
    \addlegendentry{RW-R=5}

        \end{axis}
    \end{tikzpicture}

%% file: Graphs/DC-100N-NF20.tex
% %updated correct results

\begin{tikzpicture}
        \begin{axis}[
            legend pos=outer north east,
            ylabel={Network Data Collected [\%]},
            xlabel={No. of nodes visited},
            xmin=0, xmax=80,
            ymin=0, ymax=100,
            %symbolic x coords={R=2, R=3, R=4, R=5},
            %xtick=data,
            xtick={0, 10, 20, 30, 40, 50, 60, 70, 80},
            ytick={0, 10, 20, 30, 40, 50, 60, 70, 80, 90, 100},
            ymajorgrids=true,
            grid style=dashed,
            %legend style={legend pos=outer north east, legend cell align = left},
            %symbolic x coords={a small bar, a medium bar, a large bar},
            %xtick=data
          ]

    \addplot[color=red , mark size=2.5pt, mark options={solid}, very thick]
    coordinates 
    {

(0, 0)
(1, 6.57076)
(2, 13.4426)
(3, 19.5663)
(4, 24.6121)
(5, 28.6419)
(6, 32.1361)
(7, 35.1346)
(8, 38.3268)
(9, 41.1234)
(10, 43.6199)
(11, 46.0317)
(12, 48.1227)
(13, 50.2682)
(14, 52.1598)
(15, 53.9212)
(16, 55.6778)
(17, 57.1685)
(18, 58.8724)
(19, 60.4665)
(20, 61.9991)
(21, 63.4391)
(22, 64.6854)
(23, 65.8218)
(24, 66.9489)
(25, 68.1119)
(26, 69.0434)
(27, 70.0388)
(28, 70.9302)
(29, 71.8167)
(30, 72.6376)
(31, 73.5795)
(32, 74.316)
(33, 75.1525)
(34, 75.9426)
(35, 76.6904)
(36, 77.4264)
(37, 78.1168)
(38, 78.8081)
(39, 79.4803)
(40, 80.0943)
(41, 80.7291)
(42, 81.3191)
(43, 81.9256)
(44, 82.5145)
(45, 83.1087)
(46, 83.6865)
(47, 84.221)
(48, 84.7566)
(49, 85.2712)
(50, 85.7516)
(51, 86.2642)
(52, 86.7559)
(53, 87.1965)
(54, 87.6419)
(55, 88.0493)
(56, 88.4695)
(57, 88.8509)
(58, 89.2487)
(59, 89.6407)
(60, 89.9928)
(61, 90.317)
(62, 90.6579)
(63, 90.9573)
(64, 91.2612)
(65, 91.5431)
(66, 91.7845)
(67, 92.0184)
(68, 92.2343)
(69, 92.4315)
(70, 92.6294)
(71, 92.7902)
(72, 92.9439)
(73, 93.099)
(74, 93.2414)
(75, 93.3463)
(76, 93.4618)
(77, 93.5794)
(78, 93.6693)
(79, 93.7577)
(80, 93.8329)

    };
    \addlegendentry{DRACO R=5}

    \addplot[color=red, dashed, mark size=2.5pt, mark options={solid}, very thick]
    coordinates 
    {

(0, 0)
(1, 1.77516)
(2, 4.92919)
(3, 6.05257)
(4, 7.28377)
(5, 9.1179)
(6, 11.2408)
(7, 12.9774)
(8, 14.768)
(9, 17.3656)
(10, 18.9489)
(11, 20.3042)
(12, 21.432)
(13, 23.2022)
(14, 24.8092)
(15, 26.0022)
(16, 27.5642)
(17, 29.3172)
(18, 29.9762)
(19, 31.4899)
(20, 32.4857)
(21, 33.6039)
(22, 34.4104)
(23, 35.7038)
(24, 36.8786)
(25, 38.4424)
(26, 39.4444)
(27, 40.637)
(28, 41.9494)
(29, 44.2175)
(30, 45.1292)
(31, 46.3917)
(32, 47.3437)
(33, 49.0735)
(34, 50.1605)
(35, 51.2675)
(36, 53.3582)
(37, 54.3946)
(38, 55.3986)
(39, 56.6384)
(40, 57.5945)
(41, 59.3194)
(42, 60.1361)
(43, 60.7817)
(44, 62.4647)
(45, 63.015)
(46, 64.682)
(47, 66.4846)
(48, 67.5153)
(49, 68.4833)
(50, 69.4245)
(51, 70.1744)
(52, 70.7585)
(53, 72.8061)
(54, 73.7652)
(55, 74.8178)
(56, 75.4875)
(57, 76.5357)
(58, 77.381)
(59, 78.6423)
(60, 79.1869)
(61, 80.0603)
(62, 80.5186)
(63, 81.4592)
(64, 82.5951)
(65, 83.5486)
(66, 84.303)
(67, 84.967)
(68, 85.5254)
(69, 86.5819)
(70, 87.5346)
(71, 88.128)
(72, 88.9912)
(73, 89.7)
(74, 90.274)
(75, 91.2253)
(76, 91.5999)
(77, 92.2304)
(78, 92.9216)
(79, 93.6061)
(80, 94.6681)

    };
    \addlegendentry{SA-RW-R=5}
    
\addplot[color=red, dotted, mark size=2.5pt, mark options={solid}, very thick]
    coordinates 
    {

(0, 0)
(1, 1.93797)
(2, 3.30085)
(3, 6.348)
(4, 9.57574)
(5, 11.9258)
(6, 12.76)
(7, 13.9823)
(8, 14.9337)
(9, 15.7698)
(10, 17.2641)
(11, 18.2836)
(12, 19.9509)
(13, 21.7619)
(14, 22.6883)
(15, 24.5288)
(16, 25.1656)
(17, 25.7742)
(18, 27.2901)
(19, 30.1851)
(20, 30.9265)
(21, 32.515)
(22, 34.534)
(23, 35.5201)
(24, 37.1853)
(25, 37.342)
(26, 38.1509)
(27, 38.8175)
(28, 39.343)
(29, 39.6791)
(30, 40.6479)
(31, 41.6348)
(32, 42.8755)
(33, 43.8621)
(34, 44.3596)
(35, 44.8063)
(36, 45.9699)
(37, 48.3089)
(38, 50.0153)
(39, 50.5506)
(40, 51.0258)
(41, 52.6886)
(42, 52.7925)
(43, 53.8727)
(44, 54.4679)
(45, 55.3842)
(46, 56.1609)
(47, 56.3661)
(48, 56.5155)
(49, 58.2206)
(50, 58.913)
(51, 59.5294)
(52, 60.0625)
(53, 60.3203)
(54, 60.9872)
(55, 61.3077)
(56, 61.8158)
(57, 62.1435)
(58, 62.43)
(59, 62.9678)
(60, 63.4226)
(61, 63.8699)
(62, 64.2936)
(63, 64.8374)
(64, 65.0619)
(65, 65.2544)
(66, 65.5816)
(67, 66.6309)
(68, 66.9224)
(69, 67.3741)
(70, 67.5424)
(71, 68.1093)
(72, 68.6468)
(73, 69.6539)
(74, 69.9483)
(75, 70.4589)
(76, 70.6869)
(77, 70.9611)
(78, 71.4266)
(79, 71.5421)
(80, 72.297)

    };
    \addlegendentry{RW-R=5}

        \end{axis}
    \end{tikzpicture}

%% file: Graphs/DC-100N-NF50.tex
% %updated correct results

\begin{tikzpicture}
        \begin{axis}[
            legend pos=outer north east,
            ylabel={Network Data Collected [\%]},
            xlabel={No. of nodes visited},
            xmin=0, xmax=50,
            ymin=0, ymax=100,
            %symbolic x coords={R=2, R=3, R=4, R=5},
            %xtick=data,
            xtick={0, 10, 20, 30, 40, 50},
            ytick={0, 10, 20, 30, 40, 50, 60, 70, 80, 90, 100},
            ymajorgrids=true,
            grid style=dashed,
            %legend style={legend pos=outer north east, legend cell align = left},
            %symbolic x coords={a small bar, a medium bar, a large bar},
            %xtick=data
          ]

    \addplot[color=red , mark size=2.5pt, mark options={solid}, very thick]
    coordinates 
    {

(0, 0)
(1, 7.23128)
(2, 14.8544)
(3, 22.2617)
(4, 27.3059)
(5, 31.489)
(6, 35.5373)
(7, 39.1649)
(8, 41.9206)
(9, 44.1333)
(10, 46.2897)
(11, 48.8987)
(12, 50.8041)
(13, 52.8371)
(14, 54.4312)
(15, 56.2522)
(16, 57.7025)
(17, 58.9539)
(18, 60.4469)
(19, 61.6275)
(20, 62.6855)
(21, 63.8781)
(22, 64.9146)
(23, 65.7741)
(24, 66.6709)
(25, 67.5634)
(26, 68.3688)
(27, 69.1699)
(28, 69.9274)
(29, 70.5399)
(30, 71.1909)
(31, 71.7687)
(32, 72.3831)
(33, 72.8552)
(34, 73.3844)
(35, 73.8368)
(36, 74.2953)
(37, 74.77)
(38, 75.1791)
(39, 75.5485)
(40, 75.8894)
(41, 76.2053)
(42, 76.5371)
(43, 76.7961)
(44, 77.0751)
(45, 77.269)
(46, 77.5163)
(47, 77.7526)
(48, 77.9681)
(49, 78.14)
(50, 78.2659)

    };
    \addlegendentry{DRACO R=5}

    \addplot[color=red, dashed, mark size=2.5pt, mark options={solid}, very thick]
    coordinates 
    {

(0, 0)
(1, 1.40393)
(2, 4.32637)
(3, 6.73704)
(4, 8.6016)
(5, 10.6363)
(6, 11.9169)
(7, 14.014)
(8, 16.9834)
(9, 19.2452)
(10, 21.6473)
(11, 23.3947)
(12, 26.2542)
(13, 27.5641)
(14, 30.1762)
(15, 31.5222)
(16, 33.7826)
(17, 34.9329)
(18, 36.2568)
(19, 37.8546)
(20, 39.0744)
(21, 40.2785)
(22, 42.1568)
(23, 44.2536)
(24, 45.9143)
(25, 46.7895)
(26, 48.9871)
(27, 50.4861)
(28, 51.4872)
(29, 53.1652)
(30, 54.4142)
(31, 56.134)
(32, 56.8737)
(33, 58.4286)
(34, 59.6812)
(35, 60.5156)
(36, 61.843)
(37, 63.3348)
(38, 64.7186)
(39, 65.658)
(40, 67.3395)
(41, 69.4533)
(42, 71.2141)
(43, 72.6217)
(44, 73.7302)
(45, 75.1007)
(46, 75.937)
(47, 76.4113)
(48, 77.0582)
(49, 77.9219)
(50, 78.9716)

    };
    \addlegendentry{SA-RW-R=5}
    
\addplot[color=red, dotted, mark size=2.5pt, mark options={solid}, very thick]
    coordinates 
    {

(0, 0)
(1, 1.59294)
(2, 3.20126)
(3, 4.78153)
(4, 6.06886)
(5, 7.40477)
(6, 9.57935)
(7, 11.9001)
(8, 12.9146)
(9, 14.8745)
(10, 16.4891)
(11, 17.3481)
(12, 18.5088)
(13, 20.0462)
(14, 21.6232)
(15, 22.3966)
(16, 22.8193)
(17, 25.9195)
(18, 26.9753)
(19, 27.936)
(20, 28.7283)
(21, 29.3379)
(22, 30.1712)
(23, 31.1103)
(24, 31.7341)
(25, 32.7696)
(26, 34.4414)
(27, 35.3935)
(28, 35.855)
(29, 38.8063)
(30, 39.5212)
(31, 40.186)
(32, 40.5891)
(33, 41.3191)
(34, 41.843)
(35, 42.2342)
(36, 43.0966)
(37, 43.448)
(38, 44.0284)
(39, 44.8984)
(40, 45.7228)
(41, 45.9482)
(42, 46.1602)
(43, 46.7098)
(44, 47.6858)
(45, 47.7529)
(46, 48.5452)
(47, 49.0947)
(48, 49.6833)
(49, 50.0815)
(50, 50.3329)

    };
    \addlegendentry{RW-R=5}

        \end{axis}
    \end{tikzpicture}

%% file: Graphs/DC-100N-NF70.tex
% %updated correct results

\begin{tikzpicture}
        \begin{axis}[
            legend pos=outer north east,
            ylabel={Network Data Collected [\%]},
            xlabel={No. of nodes visited},
            xmin=0, xmax=30,
            ymin=0, ymax=100,
            %symbolic x coords={R=2, R=3, R=4, R=5},
            %xtick=data,
            xtick={0, 3, 6, 9, 12, 15, 18, 21, 24, 27, 30},
            ytick={0, 10, 20, 30, 40, 50, 60, 70, 80, 90, 100},
            ymajorgrids=true,
            grid style=dashed,
            %legend style={legend pos=outer north east, legend cell align = left},
            %symbolic x coords={a small bar, a medium bar, a large bar},
            %xtick=data
          ]

    \addplot[color=red , mark size=2.5pt, mark options={solid}, very thick]
    coordinates 
    {

(0, 0)
(1, 7.42841)
(2, 16.2024)
(3, 20.8536)
(4, 25.484)
(5, 28.6654)
(6, 31.7541)
(7, 35.0032)
(8, 37.4585)
(9, 39.7081)
(10, 41.7387)
(11, 43.8777)
(12, 45.3341)
(13, 46.7975)
(14, 48.1271)
(15, 49.4931)
(16, 50.5353)
(17, 51.5789)
(18, 52.5131)
(19, 53.3798)
(20, 54.1122)
(21, 54.8578)
(22, 55.5942)
(23, 56.2681)
(24, 56.8282)
(25, 57.3804)
(26, 57.7715)
(27, 58.125)
(28, 58.4411)
(29, 58.702)
(30, 58.9111)
    };
    \addlegendentry{DRACO R=5}

    \addplot[color=red, dashed, mark size=2.5pt, mark options={solid}, very thick]
    coordinates 
    {

(0, 0)
(1, 2.11739)
(2, 4.61191)
(3, 6.79938)
(4, 8.59908)
(5, 10.9483)
(6, 13.3227)
(7, 15.6084)
(8, 18.3796)
(9, 20.8733)
(10, 22.7666)
(11, 25.5505)
(12, 28.8875)
(13, 30.2798)
(14, 33.0081)
(15, 35.8717)
(16, 38.1249)
(17, 39.415)
(18, 41.0475)
(19, 43.4656)
(20, 44.6832)
(21, 46.0223)
(22, 48.7988)
(23, 49.9258)
(24, 53.1794)
(25, 54.2693)
(26, 56.1729)
(27, 57.3938)
(28, 59.2211)
(29, 60.2155)
(30, 61.4697)
    };
    \addlegendentry{SA-RW-R=5}
    
\addplot[color=red, dotted, mark size=2.5pt, mark options={solid}, very thick]
    coordinates 
    {

(0, 0)
(1, 1.99185)
(2, 3.75199)
(3, 5.4716)
(4, 6.90198)
(5, 8.38269)
(6, 11.3855)
(7, 13.1509)
(8, 13.892)
(9, 14.6427)
(10, 17.6158)
(11, 19.7137)
(12, 20.2608)
(13, 22.0343)
(14, 23.4925)
(15, 24.8776)
(16, 27.0305)
(17, 30.2431)
(18, 31.0146)
(19, 31.7023)
(20, 32.2712)
(21, 33.2707)
(22, 34.5166)
(23, 35.4252)
(24, 36.2798)
(25, 37.3937)
(26, 37.7073)
(27, 38.2174)
(28, 38.7618)
(29, 39.148)
(30, 39.3877)

    };
    \addlegendentry{RW-R=5}

        \end{axis}
    \end{tikzpicture}